\documentstyle[preprint,tighten,aps,epsf,floats,psfig,epsfig,rotate]{revtex}

\begin{document}

\newcommand{\gaprox}{$ {\raisebox{-.6ex}{{$\stackrel{\textstyle >}{\sim}$}}} $}
\newcommand{\saprox}{$ {\raisebox{-.6ex}{{$\stackrel{\textstyle <}{\sim}$}}} $}
\newcommand{\boldpi}{\mbox{\boldmath $\pi$}}
\newcommand{\boldT}{\mbox{\boldmath $T$}}
\newcommand{\boldD}{\mbox{\boldmath $D$}}
\newcommand{\boldE}{\mbox{\boldmath $E$}}
\newcommand{\boldt}{\mbox{\boldmath $t$}}
\newcommand{\boldphi}{\mbox{\boldmath $\varphi$}} 
\newcommand{\boldepsilon}{\mbox{\boldmath $\varepsilon$}}
\newcommand{\boldi}{\mbox{\boldmath $i$}}
\newcommand{\boldj}{\mbox{\boldmath $j$}}
\newcommand{\boldtau}{\mbox{\boldmath $\tau$}}
\newcommand{\boldvarphi}{\mbox{\boldmath $\varphi$}}
\newcommand{\pronabla}{\stackrel{\rightarrow}{\nabla}}
\newcommand{\conabla}{\stackrel{\leftarrow}{\nabla}}
\newcommand{\bothnabla}{(\pronabla-\conabla)} 
\def\L{\Lambda}

\thispagestyle{empty}

\hfill{\small KRL MAP-247}

\vspace*{60pt}
\begin{center}
{\large\bf EFFECTIVE FIELD THEORY OF NUCLEAR FORCES}
\footnote{Commissioned for Prog. Part. Nucl. Phys.} 

\vspace{90pt}

{\bf U. van Kolck} 

\vspace*{24pt}
{\sl Kellogg Radiation Laboratory, 106-38}\\
{\sl  California Institute of Technology}\\ 
{\sl  Pasadena, CA 91125, USA}\\
\vspace*{12pt}
{\tt vankolck@krl.caltech.edu}\\

\vspace{30pt}
\end{center}

\begin{abstract}
The application of the effective field theory (EFT) method 
to nuclear systems is reviewed. 
The roles of degrees of freedom, QCD symmetries, power
counting, renormalization, and potentials are discussed.
EFTs are constructed for the various energy regimes
of relevance in nuclear physics,
and are used in systematic expansions to derive 
nuclear forces in terms of a number of parameters that 
embody information about QCD dynamics.
Two-, three-, and many-nucleon
systems, including external probes, are considered.
\end{abstract}

\vspace*{12pt}
\vfill\eject

\tableofcontents

\vfill\eject

\setcounter{page}{1}
\setcounter{section}{0}

\section{Introduction}

There is little doubt that QCD is the correct theory
of the strong forces that bind quarks into baryons and mesons and yet,
despite many efforts, it has not been possible to derive
the interaction among these composite states from first principles. 
At the root of the problem is our inability to solve the dynamics of QCD 
at hadronic scales because of a large coupling constant.
Much of the description of hadronic properties and interactions
has been limited to phenomenological models whose connections
with QCD are unclear.

The strong-coupled dynamics among baryons is particularly rich:
it gives rises to bound states, of
which ordinary nuclei are particular interesting.
A superficial look at hadronic and nuclear tables conveys complexity,
but a closer look reveals some regularities that can be translated
into different mass scales.
At least two scales influence the bulk of the 
dynamics in a nucleus.
The masses of the main constituents (nucleons) and their
excitations (chiefly the delta isobar), 
the masses of most mesons that can be exchanged among
them (the rho, omega, {\it etc.}), 
and the chiral symmetry breaking scale; they all 
cluster near what we can
call the characteristic QCD scale, $M_{QCD} \sim 1$ GeV.
On the other hand, the reciprocal of the sizes of light nuclei,
the Fermi momentum of equilibrium nuclear matter,
the pion mass and decay constant, 
and the delta-nucleon mass difference
take values closer to a lighter scale 
$M_{nuc} \sim 100$ MeV.
{}From these two scales,
a third, $M_{nuc}^2/M_{QCD} \sim 10$ MeV, manifests itself in 
the order of magnitude of nuclear binding energies.

In nuclear physics there is thus a natural separation
of scales, and one can hope to use it to simplify 
the theoretical analysis.
Such an approach has proved useful in other areas
of physics, and is at the heart of the use of renormalization
group techniques.
The method of effective field theory (EFT) evolved from these
techniques, and it was designed
to incorporate and benefit from the existence of separate scales, even
when couplings are not small compared to 1.
It seems to constitute an
ideal tool for low-energy nuclear physics, a tool
which has similarities but also significant differences compared to
the more traditional phenomenological approach.
The method is  firmly based on the known (approximate) symmetries
of QCD, but parametrizes our ignorance of dynamical details 
in a number of undetermined constants.   
This article is a review of the program of research
that attempts to describe nuclei and processes involving them
using the method of EFT.

The scales described above are consistent with a picture of
nuclei as composed of 
non-relativistic nucleons of typical momenta $Q\sim  M_{nuc}$
and energies $Q^2/M_{QCD}$.
Now, slow particles make for poor microscopes.
A nucleon in the nucleus most of the time cannot
resolve features with a spatial extent $\saprox 1/Q$.
In particular, particles that live for short times and distances
can in first approximation
be treated as points.
Only those particles that propagate over distances longer than 
$\sim 1/Q$ need to be retained in the theory as explicit degrees
of freedom. They will supply the correct low-energy analytic structure 
(poles and cuts) of physical amplitudes. 
This is a significant simplification,
as it means that a large number of states (massive conventional
hadrons or exotic quark states) can be
accounted for just through their
contributions to local interactions.
Further simplification follows from the fact that we have been
able to identify a number of symmetries of QCD,
as well as the pattern of their breaking.
Low-energy $S$-matrix elements have to exhibit the same symmetry pattern,
which markedly constrains the possible interactions 
between the relevant degrees of freedom.
This still leaves us with infinitely many local interactions, as
arbitrary numbers of derivatives appear to account for corrections
to the point approximation.
Worse still, parameters characterizing the interaction strengths
need not be small. 
The crux of the EFT method is to find a way to organize
this infinite set of interactions using 
an expansion in the small parameter $Q$.

The method can be illustrated with a very familiar example.
Let us suppose we want to study a system whose overall
size is $R_l$ but that contains smaller structures of size
$R_s\ll R_l$. We can do so by shining light 
of wavelength $\lambda$ onto the system,
but for my purposes here
it does not really matter if this ``light'' comes from a laser source
or from electron scattering or whatever; in fact I will consider just 
a static field, the only important thing being that there is 
a scale $\lambda$
associated with its (spatial) variation.
We can learn about the system
because its charge density $\rho (\vec{r})$
affects  
the interaction energy between the system and light.

Let us consider first the case when $\lambda \gg R_l$. 
In classical electrodynamics
we argue that the interaction energy can under these
circumstances be written as a multipole
expansion. In first approximation, we need only to know the overall
charge $q$
of the system and the electromagnetic potential at its center.
In second approximation we need also the system's dipole moment and
the gradient of the potential, and so on.
Schematically,
\begin{equation}
\int d^3r\rho (\vec{r}) \phi (\vec{r}) = 
q \sum_{n=0}^{\infty} c_n \left(\frac{R_l}{\lambda} \right)^n.
\label{multipoles}
\end{equation}
This expansion is justified by dimensional analysis: except
for very pathological charge distributions, the interaction energy
is analytic in $1/\lambda$ and successive terms
must come in powers of $R_l/\lambda$ times dimensionless factors 
$c_n$ of $O(1)$. The latter encode all we need to know about the system;
they can be calculated from $\rho (\vec{r})$, but if $\rho (\vec{r})$
is not known, we do not have to weigh on every single dimple
of the charge distribution. To a given accuracy, we can truncate the
multipole
expansion and parametrize the electromagnetic properties of the system
by a finite number of parameters instead of a continuous $\rho (\vec{r})$. 
The role of an exact (approximate) symmetry is that it 
forbids (suppresses) certain terms in the
expansion.
Now, all this can be translated to QED. All we need to do is to write the 
most general Lagrangian allowed by gauge invariance that involves
fields for the system (and perhaps its excited states)
and the photon. This Lagrangian will
contain not only minimal couplings governed by charge but also
operators with more derivatives, such as a Pauli term.
The rules of field theory then can be used to derive the
electromagnetic form factors of the target ---or more generally
any scattering amplitude---
with contributions ordered by $R_l/\lambda$. 
There is a difference, however.
Quantum fluctuations from loops can probe small distances.
The procedure of renormalization is designed to
ensure that the physics of these small distances
contribute to the physics of large distances
only in an average sense, and that the multipole expansion remain
valid. Left behind are some calculable cuts.
Because (point) charge interactions are
renormalizable, the non-analytic terms
are suppressed only by powers of the 
fine-structure constant, $\alpha$.
Loops involving higher-derivative operators
are further suppressed by powers of $R_l/\lambda$.

Consider now an increase in resolution: $\lambda \sim R_l$.
Again, because $ \lambda \gg R_s$ the contributions
of the small structures can be treated in an expansion in
$R_s/\lambda$ as in Eq. (\ref{multipoles}), except that the multipole
parameters now refer to the inner charge distribution 
of charge $q_s$. 
Contributions from the outer cloud, on the other hand,
can only be accounted for by the full knowledge of the 
outer distribution $\rho_l (\vec{r})$.
The interaction energy is now
\begin{equation}
\int d^3r\rho (\vec{r}) \phi (\vec{r}) = 
q_s \sum_{n=0}^{\infty} c_n^{(s)} \left(\frac{R_s}{\lambda} \right)^n
+\int d^3r\rho_l (\vec{r}) \phi (\vec{r}).
\label{multipolesplusnonanalytic}
\end{equation}
In the quantum context these outer contributions that are
non-analytic in $1/\lambda$ come from loops generated
by the light particles flying within  the target at the time light strikes.
The magnitudes of loop diagrams will depend crucially
on the interactions among the structures that make up the system.
In general, they are {\em not} small compared to tree-level graphs. 

We can translate this simple example to a typical nucleus, where
$M_{QCD}\sim 1/R_s$, 
\linebreak
$M_{nuc}\sim 1/R_l$, and $Q\sim 1/\lambda$.
It turns out that in this case we can 
justify not only the ``multipole'' or derivative expansion,
but we can control the loop expansion as well.

The first (and simplest) situation, $Q\saprox M_{nuc}$,
corresponds to nucleons of
momenta sufficiently small that {\it all}
interactions among nucleons can be described as contact interactions.
One might be tempted to dismiss this as an uninteresting case,
but it is not so, as some information on the two-nucleon system reveals.
At low momenta the scattering amplitude in both isospin channels
($I=0$ and $I=1$) can be very well-described by the effective range
expansion.
One finds that, as expected on the basis of dimensional analysis,
most of the $S$-wave effective range parameters
scale with $M_{nuc}^{-1}$.
However, the scattering lengths $a_2$ are large and reflect the existence
of a new, smaller mass scale which I denote by $\aleph\sim 1/a_2$. 
In the $^3S_1$ channel,  $\aleph^{(^3S_1)} \sim 40$ MeV,
somewhat smaller than $M_{nuc}$.
In the  
$^1S_0$ channel,  $\aleph^{(^1S_0)} \sim 8$ MeV $\ll M_{nuc}$.
For simplicity, I take an average $\aleph\sim 30$ MeV.
The existence of a new scale requires some amount
of fine-tuning.
whatever its origins, 
the consequence is that these channels have anomalously shallow
bound states
with binding energies $\sim \aleph^2/M_{QCD} \ll M_{nuc}^2/M_{QCD}$
In the singlet channel this bound state is virtual, but
in the triplet it is real ---the deuteron.

The idea of a low-energy expansion is in fact as old as nuclear physics: 
already in the 30's Bethe and Peierls \cite{Peierls}
reasoned that, because the deuteron has a large size $\lambda$ compared to
the range $R_l$ of nuclear forces,
up to an error $O(R_l/\lambda)$ the sole 
effect of the potential is to provide an energy-independent 
boundary condition at $r\sim R_l$ to an otherwise free
Schr\"odinger equation.
This can be recast in terms of an EFT
of non-relativistic nucleons ---non-relativistic
because $Q$ is small compared to the nucleon mass.
As we will see in detail below, 
interactions contain $\aleph$ besides $M_{nuc}$ and 
the appearance of $Q^2/M_{QCD}$ 
as the only scale in energy denominators of loop
diagrams generates enhancements that
require a resummation of a large class of diagrams.
After resummation, the EFT includes terms of all orders in $Q/\aleph$
but is perturbative in $Q/M_{nuc}$.
Although the EFT in the two-body case is little more than a generalization
of the technique used by Bethe and Peierls,
it allows an easy extension of these arguments
to processes where low-energy
photons, electrons and neutrinos probe the deuteron.
It also allows the study of nucleon-deuteron scattering at low-energies,
and perhaps of the triton and
other nuclei as well.
I will refer to  momenta 
\linebreak
$Q\saprox  M_{nuc}$ as ``very low''.

As we increase momenta and $Q\sim M_{nuc}$,
pions can no longer be considered heavy, as $Q$
is comparable to their mass.
The EFT has to be augmented to include them as explicit
degrees of freedom. 
(Because the delta-nucleon mass difference
is of the same order of magnitude, explicit inclusion of
the delta might also be called for.)
Explicit inclusion of the pion dynamics gives
rise to non-analytic terms analogous to those in our 
simple electrodynamics example.
All short-range dynamics from massive states can be
treated in an expansion in $Q/M_{QCD}$, and is
parametrized in analogy to the
$c_n^{(s)}$ in Eq. (\ref{multipolesplusnonanalytic}).
What is crucial for the success of the approach is
that one can show that ---in a sense to be made
more precise below--- pion loops can {\it also} be
ordered in terms of an expansion in $Q/M_{QCD}$.
In terms of Eq. (\ref{multipolesplusnonanalytic}), this would mean
that, apart from logarithms and dimensionless factors of
$O(1)$, contributions from the outer cloud would come
with a small factor $(R_l/4\pi \lambda)^{2}$ for each loop,
namely
\begin{equation}
\int d^3r\rho_l (\vec{r}) \phi (\vec{r}) = 
q_l \sum_{L=0}^{\infty} \gamma_L \left(\frac{R_l}{4\pi\lambda} \right)^{2L},
\label{nonanalytic}
\end{equation}
where $\gamma_L$ are calculable functions of the Lagrangian parameters.
It is chiral symmetry that effects this magic
by requiring that pion interactions be 
proportional to $Q$, and therefore weak at low energies.

The EFT appropriate to these momenta has been extensively applied
to systems containing one or no nucleon, and in that context 
it is called Chiral Perturbation Theory ($\chi$PT).
The application of $\chi$PT
to nuclear systems is complicated by the 
resummation needed to accommodate the infrared enhancement.
What class of diagrams is to be re-summed? The answer
to this question depends on the relative magnitude of
pion and shorter-range effects. 
It is reasonable that at sufficiently small $Q$, pions can be treated 
perturbatively, while as $Q$ reaches some scale $M_{NN}$
pion effects become non-perturbative.
If $M_{NN}$ is as low as $M_{nuc}$, then 
the theory with perturbative pions is of little use,
since then the pion can be simply replaced by a point.
If $M_{NN}$ is as high as $M_{QCD}$, then non-perturbative pions
are not necessary as long as $Q<M_{QCD}$
\footnote{Currently, the only way to deal with processes with
momenta $Q\sim M_{QCD}$ is by phenomenological models.
Those (few) models which include QCD symmetries correctly
can be roughly thought of as EFTs in which an infinite number
of operators is assumed to be correlated and their parameters
functions of the few model parameters. 
Both model-independence and a systematic expansion are lost.
Truly extending the EFT approach to such ``high'' momenta
presents formidable obstacles.
First, the number of relevant degrees of freedom increases
tremendously. Second, one would need to find a new small
expansion parameter. 
I do not attempt to review efforts in this direction here,
as they have had no clear successes in nuclear physics yet.}.
I will refer to  momenta $Q\saprox  M_{NN}$ as ``low'' 
and to momenta $M_{NN}\saprox Q\saprox  M_{QCD}$ as ``moderate''.
(What the scale  $M_{NN}$ is is a matter of current debate.)

What is unique about the application of EFT to nuclear
physics is this fascinating interplay between perturbative
and non-perturbative phenomena.
EFTs already have a long history,
as its first example likely is the work of Euler and Heisenberg in QED 
\cite{inEul}. 
However, its modern concept evolved only during the late 60's and the
70's, as a confluence of several ideas. First, in $S$-matrix days, 
it was noticed \cite{inwei1} that a chiral Lagrangian used at tree
level was a field theoretical realization of current algebra. 
Second, the spectacular experimental successes of the Standard Model
together with its only partially  unified structure suggested that
it be viewed as the low-energy remainder of a more fundamental
theory \cite{inguts}.
Third, a new view 
of renormalization grew out of renormalization group ideas \cite{inwil}. 
By the end of the 70's it was clearly realized that
the concept of effective Lagrangians provides {\em the\/} rationale behind 
the use of field theory in particle physics phenomenology 
\cite{inwei2,inwei3}.
It was in the 80's that a detailed, systematic analysis
of low-energy hadronic physics based on EFT began \cite{ciGL,ciGSS},
but even then, no non-perturbative effects were dealt with.

Here I will consider mainly the description
of nuclear systems {\it per se} for $Q<M_{QCD}$,
a program that was initiated in Refs. 
\cite{inwei6,chiralfilter,ciOvK,inwei5,taesunphysrep,invk,ciOLvK,civK1,civK2}.
This early work has shown that many of the observed but
otherwise mysterious features of nuclear physics can be naturally
understood in this framework \cite{invk,jimreview}.
More recently, several aspects of this program have received
vigorous attention
---for a sample, see Ref. \cite{book}---
and a reasonable body of successes has been amassed.
A review seems timely. 
However, as many review writers before,
I am compelled to apologize to those who feel
that their work is underrepresented here. 
Spacetime constraints ---the editor's patience notwithstanding---
have biased me towards the ideas that I understand better,
usually (but not always) my own.
For a different perspective on various subsets
of the topics I cover here, the reader is invited to consult
Ref. \cite{reviews}.

After an introduction to the general ideas of the EFT method in
Sect. \ref{sec-eft}, I briefly present some $\chi$PT
highlights in Sect. \ref{sec-01N}.
Sect. \ref{sec-2N} on the two-body system is perhaps the central one.
I consider first the very-low-momentum EFT
as its simplicity makes it worth studying;
then I delve into pions
at low and moderate momentum.
Power counting is the theme that has been played
but whose scope is not yet completely understood.
This sets the stage for the application to few-body
systems in Sect. \ref{sec-3N},
where remarkable universality emerges.
The question we explore is what are the new ingredients
brought in by new bodies.
Several successful applications to processes involving
external probes are mentioned in Sect. \ref{sec-2.5N},
and the first, exploratory studies of the many-body
system are in Sect. \ref{sec-AN}.
The balance is in Sect. \ref{sec-out}.

\section{EFT} \label{sec-eft}

Effective field theories exploit the existence of scales in a system. 
A very brief introduction to the main EFT ideas is presented
Subsect. \ref{subsec-what}.
In Subsect. \ref{subsec-free}
we identify the relevant degrees
of freedom whose propagation should be accounted for explicitly in
order to supply the correct low-energy analytic structure 
of physical amplitudes. 
The symmetries which constrain the low-energy $S$-matrix elements
are discussed in Subsect. \ref{subsec-sym}.
The $S$-matrix can then be generated by
the most general Lagrangian involving the relevant degrees
of freedom and symmetries.
In Subsect. \ref{subsec-pow} one invokes the concept of ``natural'' size
for effective dimensionful parameters
in terms of the scale $M$ of the underlying dynamics, 
in order to establish
an ordering of all possible contributions 
to processes at momenta of order $Q\ll M$ in powers
of the small parameter $Q/M$.
Some features of the method are stressed in 
Subsect. \ref{subsec-prop}.

\subsection{What is effective?}\label{subsec-what}

What is an effective theory? 
In order to have a rough idea
---a fuller answer can be found in several reviews 
\cite{kaplan,popov,polchinsky,lepage,georgi}---
consider a field theory (which I call the 
``underlying'' theory) given by some Lagrangian ${\cal L}_{und}$ 
written in terms 
of some (``elementary'') fields $\Psi$; and suppose that 
this theory adequately describes experiments 
carried out over a certain range of energy
\footnote{ Most likely the underlying theory is itself an effective 
theory to an yet more fundamental theory.}.
This means that $S$-matrix elements can be obtained from the path integral
\begin{equation}
 Z= \int {\cal D}\Psi e^{i \int {\cal L}_{und}(\Psi)}.
\end{equation}
We are interested in how such a theory looks like at scales smaller 
than some scale $\Lambda$ within its range of validity. 
Excitations with
momenta larger than $\Lambda$
cannot be produced directly, so we can profit from
splitting the fields $\Psi$ in 
two components $\Psi_{h}$ (``fast'') and $\Psi_{l}$ (``slow''),
according to whether their momenta 
are greater or smaller than $\Lambda$.
Integrating over $\Psi_{h}$,
\begin{equation}
 Z= \int {\cal D}\Psi_{l} e^{i \int {\cal L}_{eff}(\Psi_{l})},
\end{equation}
\noindent
where the effective Lagrangian ${\cal L}_{eff}$ is given,
in $D$ spacetime dimensions, by
\begin{equation}
 \int d^{D}x {\cal L}_{eff}(\Psi_{l}) =  
     -i \ln \int {\cal D}\Psi_{h} 
     e^{i \int {\cal L}_{und}(\Psi_{h}, \Psi_{l})} 
     = \int d^{D}x \sum_{i} g_{i}(\Lambda) {\cal O}_{i}(\Psi_{l}). 
\label{intro0}
\end{equation}

The operators ${\cal O}_{i}$ may be very complicated but 
they involve only the slowly-varying fields $\Psi_{l}$;
the fields $\Psi_{h}$ are said to have been ``integrated out''.
The ${\cal O}_{i}$'s have two important
properties. First, they are local in the sense that they involve only fields 
at the same spacetime point. Yet,
they contain arbitrary number of derivatives 
of such fields. Indeed, according to the uncertainty principle particles 
with momentum $Q\saprox \Lambda$ can only probe distances 
\linebreak
$1/Q \gaprox 1/\Lambda$ 
and therefore can sense configurations 
with support smaller than $1/\Lambda$ only in an average
sense.
Second, as long as the splitting of fields is done carefully,
the ${\cal O}_{i}$'s transform under various groups according to
the underlying theory.
If the underlying Lagrangian is symmetric under a transformation
that is not anomalous, then so is the effective Lagrangian, although
the symmetry might be realized non-linearly if it is spontaneously
broken. If the symmetry is explicitly broken either at a classical 
or a quantum level, operators will appear at low energies that 
incorporate the breaking accordingly. 
Note, however, that the splitting of fields does require some care.
The Lagrangian ${\cal L}_{eff}$ typically contains interactions
involving the time derivative of $\Psi_{l}$,
$\dot{\Psi}_{l}$, to powers greater than 1.
In this case there is a non-trivial relation between  $\Psi_{l}$ and
its conjugate momentum. As a result the interaction Hamiltonian is
not simply minus the interaction Lagrangian, but contains additional,
Lorentz non-covariant terms. 
It is no problem to include these interactions, 
as other non-covariant pieces arise in covariant 
perturbation theory from contractions
involving derivatives, and in time-ordered formalism from its inherent 
non-covariance. One can show explicitly \cite{inleeyang} that 
the sum of diagrams contributing to any given process is indeed covariant. 
Alternatively, one can add terms $\Delta{\cal L}$ directly to 
${\cal L}_{eff}$. For example, for real scalar $\Psi_{l}$,
if 
\begin{equation}
{\cal L}_{eff}=
\frac{1}{2}\dot{\Psi}_{l}^{T} A(\Psi_{l}) \dot{\Psi}_{l}+\ldots,
\end{equation}
then \cite{insalam}
\begin{equation}
\Delta{\cal L}=\frac{i}{2}\delta^{D}(0)\; Tr [\ln(A^{-1}(\Psi_{l}))] +\ldots
\end{equation}
This Lagrangian is manifestly covariant, but the new non-linear piece
generates self-interactions.
These are certainly ill-defined 
as they stand due to the $\delta^{D}(0)$, but other 
terms in the unrenormalized Lagrangian are equally ill-defined.
The EFT is made well-defined through regularization.
$\delta^{D}(0)$ vanishes in 
dimensional regularization but not when a
mass cutoff is used;
the $\ln$ term is necessary \cite{incharap}
in the renormalization program to get rid of 
loop contributions that could otherwise spoil the assumed symmetries.

As for the coefficients $g_{i}$, they carry 
information about details of the dynamics: 
they are functions of the parameters of the underlying theory, but 
also depend explicitly on the cutoff $\Lambda$
---which lends them the status of ``running coupling constants''.
In diagrams, a change in $\Lambda$ amounts to a reshuffle between contributions
from vertices and from the ultraviolet region of loop integration. 
An obvious requirement is that low-energy observables be insensitive to
changes in $\Lambda$, which places constraints on the $\Lambda$ dependence
of the $g_{i}$'s. The equations that govern this dependence are
the renormalization group equations.

The notion of an effective Lagrangian is particularly useful when
the underlying theory has (at least) one characteristic mass scale $M$.
In this case, the effective degrees of freedom for $\Lambda \leq M$
expressed by the fields $\Psi_{l}$ are in general significantly different 
from the original degrees represented by the fields $\Psi$. 
It frequently happens that that these low-energy phenomena
are ``collective excitations'' as viewed in terms of $\Psi$
\cite{popov}.
The reformulation of the theory in terms of $\Psi_{l}$ 
ends up selecting those effects of the underlying theory 
that are most important at low energies.
The simplest case is the one where $M$ is just the mass of a physical
particle. Production and decay of this particle involve large momenta 
and do not
concern the EFT. Effects of virtual exchange 
(of the particle, or of particle-antiparticle pairs) are of short range
and thus only included indirectly in the 
$g_{i}$'s \cite{inAC}. If, furthermore, the particle is not stable 
in the 
context of the underlying theory, then it does not (explicitly) appear at 
all at low energies; we need associate no field $\Psi_{l}$ to it. 
Another common case is that where $M$ is the scale associated with some
(elementary or composed) scalar field acquiring a non-vanishing vacuum 
expectation value and breaking a (possibly approximate) 
continuous global internal symmetry group, resulting in the appearance 
in the spectrum of either
a (possibly nearly) massless spin-zero particle \cite{inGSW} ---a 
(pseudo)Goldstone boson--- 
or a massive, longitudinal component of a vector boson \cite{inhiggs}.
A third case is more complicated: $M$ is a scale where some
wild non-perturbative phenomena dominate and force asymptotic states
into singlet representations of a local group,
an example being confinement in a non-abelian gauge
theory with the appropriate matter content. 
It seems adequate in this case to restrict the effective
Lagrangian to fields that are singlets under the gauge group.
There are other possibilities (like the existence of a Fermi surface
\cite{popov,polchinsky}), or combinations of various cases, 
but the important point 
is, the mass $M$ provides a measure of the strength of the high-energy 
effects that appear in the coefficients $g_{i}$'s. 
This eventually paves the way to a systematic expansion
in powers of $Q/M$.

If we know and can solve the underlying theory, then we can calculate
the $g_{i}$'s and obtain the complete form of ${\cal L}_{eff}$
to the accuracy of the approximation involved in the solution.
But it frequently happens that we cannot solve the underlying theory
(as it is the case of nuclear physics),
or that we do not know such 
theory at all (as it is for the electroweak theory). Are we then justified in
using an effective theory? A positive answer is given by a ``theorem" due to 
Weinberg \cite{inwei2}: in conciliating 
quantum mechanics and Lorentz invariance in a way consistent with unitarity, 
analyticity and cluster decomposition, we are led to quantum field theory; 
the most general Lagrangian with some assumed 
symmetries will then produce the most general 
$S$-matrix incorporating those 
general 
principles and symmetries, without any other physical content. This is a 
statement based on our own currently accessible energy experience: it has 
only been proved in the particular case of a scalar field with $Z_{2}$ symmetry
in Euclidean space \cite{inball}, 
but no counter-examples are known.

EFT is therefore modern $S$-matrix theory. 
All that any theory can do is to relate observables.
Field theory based on a Lagrangian is the simplest way to do this
given some degrees of freedom and symmetries.
The theory is of course regulated by some sort 
of procedure implementing the cutoff $\Lambda \leq M$ 
that isolates the neglected 
degrees of freedom, so diagrams involve only momenta $Q<\Lambda$. 
Renormalization is carried out as usual, but its aim is {\em not\/} to 
eliminate infinities: because the EFT is not valid at high energies, 
we never have to take the $\Lambda \rightarrow \infty$ limit anyway
\cite{lepage}. 
Rather, the objective is to remove details of the cutoff
procedure from physical amplitudes, 
and to relate Lagrangian parameters and observable
quantities
\footnote{ This can only be done because
we include {\em all\/} allowed interactions. This is in sharp contrast
to what is usually done in models like the Nambu-Jona-Lasinio model of 
low-energy QCD; such models are frequently called ``effective", but
they are not theories in the above sense.}.  
A redefinition
of fields in general changes all amplitudes off shell,
where they cannot be measured. 
Only observables, which appear in on-shell amplitudes, 
are unambiguous in the EFT.

\subsection{Not too much freedom}\label{subsec-free}

The first element of an EFT consists of the identification of 
relevant low-energy degrees of freedom. Here we will be considering
only EFTs for momenta $Q \ll M_{QCD} \sim 1$ GeV. 
This severely restricts the number of degrees of freedom that
need to be included explicitly:
at such low momenta we cannot probe quark degrees directly,
and the number of hadronic states needed is small.

The nucleon mass $m_N$ is not light 
compared to $M_{QCD}$, but
to a very good approximation nucleons cannot decay into lighter states of
high momenta.
Although we can integrate out nucleon-antinucleon pairs 
that pop out of the vacuum,
a number of nucleons exist in asymptotic in- and out-states
of nuclear systems.
Nucleons of momenta 
$Q\ll M_{QCD}$ are thus associated in the EFT with an explicit field $N$,
and nucleon propagation is represented by a line that goes through diagrams. 
The EFT splits into sectors of definite nucleon number $A$:
operators containing more than $2A$ nucleon fields do not
contribute to processes involving only $A$ incoming (and outgoing) nucleons. 
Because $Q \ll m_N$, an incoming slow nucleon
receives only little kicks from other particles. 
Nucleons are thus non-relativistic,
characterized in first approximation by
a non-relativistic dispersion law for the energy,
$E= \vec{p}^{\,2}/2m_N$ as function of the three-momentum $\vec{p}$.
Relativistic corrections are accounted for systematically
in an expansion in $Q/m_N$.

Which other hadronic states need to be included depends
on what range of momenta we are interested in.
Certainly all mesonic states with masses $m_m\gaprox M_{QCD}$ 
can be accounted for indirectly, through their contribution
to coefficients of local interactions.
In complete analogy with the multipole expansion in classical
electrodynamics, we can approximate the
interactions among nucleons originating from the exchange of
these mesons
in a series of contact interactions with an increasing number of
derivatives. In first approximation such contact
interactions are momentum-independent, and corrections
generate an expansion in $Q/m_m$.
Likewise, baryonic states with masses $m_{N^*}\gaprox m_N +M_{QCD}$
cannot be reached by hitting nucleons with 
small momenta.
They can also be integrated out,
generating local interactions 
in an expansion in $Q/(m_{N^*}- m_N)$.

The most relevant state besides the nucleon is the lightest hadron,
the pion of mass $m_\pi \sim M_{nuc}$; next
is the delta isobar, as its mass splitting to the nucleon
is not large, 
$\delta m \equiv m_\Delta-m_N \sim M_{nuc}$.
For $Q\saprox M_{nuc}$ the pion and delta can be integrated out
---unless we are considering external pions,
in which case they could be treated as heavy particles as well.
For $Q\gaprox m_\pi$ the relativistic pion has to be included as an
explicit field $\boldpi$.
For $Q\gaprox m_\Delta-m_N$ a non-relativistic delta field $\Delta$
has to be considered as well.
Because the two thresholds do not exactly coincide,
it is possible ---and possibly efficient---
to integrate
out the delta in a limited region of momenta above the
pion threshold, while maintaining explicit pion fields.
Coefficients of the corresponding effective Lagrangian that 
receive direct contributions from the delta will be relatively
large, being suppressed only by powers of $Q/\delta$.

\subsection{Symmetries} \label{subsec-sym}

Symmetries play a fundamental role in EFTs because they
restrict the possible form of interactions.
The EFT corresponding to the standard electroweak theory at 
an energy scale of a few GeV involves only the lightest
leptons and quarks, gluons, and the photon; 
weak gauge bosons, and heavy leptons and quarks can be
integrated out in favor of non-renormalizable 
interactions.  
Here I am interested mostly in interactions involving
the lightest $u$ and $d$ quarks. They can be arranged in
a flavor doublet $q=(^u_d)$. 
Denoting by
$G_{\mu}$ ($A_{\mu}$) the gluon (photon) field of strength $G_{\mu\nu}$ 
($F_{\mu\nu}$),
the relevant pieces of the effective Lagrangian at this scale are 
\begin{eqnarray}
 {\cal L} & = & 
    -\bar{q}(\partial\!\!\!/-ig_{s}G\!\!\!\!/-ieQA\!\!\!/ )q 
    -\frac{1}{2}(m_{u}+m_{d})\bar{q}q 
    +\frac{1}{2}(m_{d}-m_{u})\bar{q}\tau_{3}q  \nonumber  \\ 
& & -\frac{1}{2}Tr[G_{\mu\nu}G^{\mu\nu}]
    +\frac{\bar{\theta} g_{s}^{2}}{32 \pi^{2}}\varepsilon_{\mu\nu\rho\sigma}
    Tr[G^{\mu\nu}G^{\rho\sigma}] -\frac{1}{4}F_{\mu\nu}F^{\mu\nu}
    +\ldots \label{intro9}
\end{eqnarray}
\noindent
Here $\tau_3$ is the usual Pauli matrix,
$Q=1/6+\tau_{3}/2$  is the quark charge matrix,
and ``$\ldots$'' denote non-renormalizable terms.   
Non-renormalizable interactions are suppressed by
powers of the masses of the heavy particles that have been integrated out.
I will neglect them in a (very good) first approximation.
The leading Lagrangian (QCD + QED of quarks) has $5$ parameters: the gauge 
couplings for strong ($g_{s}$) and electromagnetic ($e$) interactions,
the masses of the up ($m_{u}$) and the down ($m_{d}$) quarks, and the strong
CP parameter ($\bar{\theta}$). 
The theta term is found to be unnaturally small (the so-called strong CP
problem), and I will neglect it as well.

The set of remaining interactions in Eq. (\ref{intro9}) has a number 
of approximate symmetries.
They are clearly invariant under proper Lorentz 
transformations, parity, and time-reversal.
Local changes of the quark-field phase that are
proportional to unity or generators of the $SU(3)$ color group
leave the Lagrangian invariant.
A global $U(1)$ symmetry corresponds to baryon number, while
another, axial $U(1)$ is anomalous.
In the limit where  $m_{u}$, $m_{d}$, and $e$ are zero
there is further invariance under a chiral symmetry
$SU(2)\times SU(2)$, 
consisting of independent rotations with parameters 
$\boldepsilon_L$ and  $\boldepsilon_R$ of the
left ($q_L=(1+\gamma_5/2)q$) and right ($q_R=(1-\gamma_5/2)q$) quark fields: 
\begin{equation}
\delta q_L=-i\boldepsilon_L\cdot \boldt q_L, 
\;\; \delta q_R=-i\boldepsilon_R\cdot \boldt q_R,
\end{equation}
where $\boldt= \boldtau/2$ are the generators of $SU(2)$
in terms of the Pauli matrices $\boldtau$, 
\begin{equation}
t_{a}t_{b}=\frac{1}{4} \delta_{ab} + \frac{i}{2} \epsilon_{abc} t_{c}.
\end{equation}
Acting on quark bilinears, $SU(2)\times SU(2)\sim SO(4)$.
The absence of degenerate parity doublets but
presence of (approximate) isospin multiplets in the hadron spectrum indicates
that chiral symmetry is broken by the vacuum down to the 
diagonal subgroup 
$SU(2)_{V}\sim SO(3)$ of isospin.
Mass terms and electromagnetic interactions
are not invariant under chiral transformations: they transform 
as non-trivial $SO(4)$ tensors and therefore break chiral symmetry 
explicitly.
The operator $\bar{q}q$ 
breaks chiral symmetry as the fourth component of an $SO(4)$ vector
\begin{equation}
S=(2\bar{q}i\gamma_{5}\boldt q,\bar{q}q), \label{vecS}
\end{equation}
while $-\bar{q}\tau_{3}q$ is the third component of another $SO(4)$ 
vector, 
\begin{equation}
P=(-2\vec{q}\boldt q,\bar{q}i\gamma_{5}q). \label{vecP}
\end{equation}
Photon exchange among quarks produces four-quark 
interactions off which we can read the way
electromagnetic interactions break isospin.
Defining the $SO(4)$ antisymmetric tensor
\begin{equation}
T^\mu=\left( \begin{array}{cc}
  \varepsilon_{ijk}\bar{q}i\gamma^{\mu}\gamma_{5}t_k 
  & \bar{q}i\gamma^{\mu}t_i q \\
  - \bar{q}i\gamma^{\mu}t_j q  & 0 \end{array}   \right), \label{tenf}
\end{equation}
one finds that chiral symmetry is explicitly
broken by terms proportional to $T_{34}^\mu$ and $T_{34}^\mu T_{34}^\nu$.

The EFT has to produce the same low-energy $S$-matrix
elements as the underlying theory. 
Although any choice of fields is equally valid,
it is convenient to choose fields that
transform under the above symmetries in the most transparent way.
Although through field redefinitions other forms of interactions can be 
obtained where symmetries arise from a conspiracy of 
several terms, the particular choices of fields 
described next are the most convenient, as the pattern of symmetries
is respected term-by-term in the resulting Lagrangian.

{\bf Spacetime symmetries}.
Parity and time-reversal can be implemented at low energies in the usual
way, but proper Lorentz invariance deserves some care.
Not surprisingly, the pion is represented by a relativistic pseudoscalar field
with standard transformation properties,
but heavy stable fields introduce a small complication.
If I denote nucleons and deltas by a generic
field $\hat{\psi}$ of mass $m\sim M$, then 
interactions will contain 
$\partial_0 \hat{\psi} \sim -im \hat{\psi} +\ldots$ 
and produce factors of $m/M$ that need to be resummed 
before we can develop an expansion in $Q/M$.  
On the other hand, the large energy $m$ is not really available
for the dynamics,
so we expect the result of the resummation
to be well behaved. In fact, the result can be no other
than a rearrangement of the non-relativistic expansion
of the relativistic theory.
 
The resummation can be done from the start, avoiding
the relativistic theory altogether, in 
what is referred to as the heavy-particle formalism 
\cite{inGeor,lukemanohar1}.
We extract the rest 
energy by a field redefinition, $\hat{\psi}\equiv e^{-im v\cdot x}\psi$
in a frame where the heavy particle moves with velocity $v$.
The most general Lagrangian involving $\psi$ will be free of
explicit factors of $m/M$ but will contain explicit factors of $v_\mu$.
Proper Lorentz invariance requires
that the Lagrangian be invariant under a boost
to a frame of different
velocity $w_\mu=v_\mu + q_\mu/m$, 
with $q\ll m$ satisfying $v \cdot q =- q^2/2m$.
Fields in the two frames are related by a phase,
besides the usual matrix $D(q)$
that accounts for the appropriate representation
of $\psi$ under the Lorentz group: 
$\psi_w= e^{iq\cdot x} D(q) \psi_v$.
Derivatives of $\psi$ transform in a more complicate way; as usual
in such circumstances, it is profitable to introduce
a covariant derivative
\begin{equation}
{\rm d}_\mu \equiv \partial_\mu -im v_\mu,
\end{equation}
such that ${\rm d}_\mu \psi_v$ transform in the same way as $\psi_v$
itself.
Out of  ${\rm d}_\mu \psi$ and 
$({\rm d}_\mu \psi)^\dagger\equiv \psi^\dagger{\rm d}_\mu^\dagger$
we can construct a covariant velocity 
and a symmetric derivative
\begin{equation}
{\cal V}_\mu \equiv \frac{i}{2m} ({\rm d}_\mu -{\rm d}_\mu^\dagger), \;\;
\diamondsuit_\mu \equiv \frac{1}{2m} ({\rm d}_\mu +{\rm d}_\mu^\dagger),
\end{equation}
which are useful building blocks, as in bilinears they produce
Lorentz invariants.
Spin can be dealt with in similar manner. 
If $\psi$ has spin 1/2, then at low energy only two
degrees of freedom are important, which can be enforced by imposing
that the spinor $\psi_v$ obeys $(1-\not v)\psi_v=0$. 
The covariant spin is
\begin{equation}
\Sigma_\mu=\frac{i}{2} \gamma_5 \sigma_{\mu\nu} {\cal V}^\nu,
\end{equation}
as indeed $\Sigma\cdot {\cal V}=0$.
If $\psi$ has spin 3/2, then at low energy only four
degrees of freedom are important; this can be enforced
by demanding the field $(\psi_v)_\mu$ 
(which behaves as the tensor product of a Dirac spinor and 
a four-vector)
obey
$v\cdot\psi_v=0$ and $\gamma_5 \sigma^{\mu\nu}v_\mu(\psi_v)_\nu=0$. 
To account for the vector character, 
the covariant spin-3/2 field has to be
\begin{equation}
(\Psi_v)_\mu=(\psi_v)_\mu-\frac{i}{m}  v_\mu \partial\cdot\psi_v.
\end{equation}

Lorentz invariance is ensured by proper contraction of the indices
of these covariant objects \cite{lukemanohar1}.
These invariants reduce in the rest frame of the heavy particle to a set
of $SO(3)$-invariant operators, all with relative
strengths determined by $1/m$.
The heavy-particle formalism is but a way 
to impose invariance under
rotations plus slow velocity boosts; truncated
at leading order in $1/m$ it is just familiar Galilean invariance.
As an example we can look at kinetic terms for $\psi$,
{\it i.e.} terms containing only two heavy fields.
Any such bilinear involving $\diamondsuit_\mu$ can be eliminated in
favor of total derivatives, so the most general kinetic
term is of the form $\psi_v^\dagger({\cal V}^2-1)^n\psi_v$, $n\ge0$.
By field redefinitions involving $({\cal V}^2-1)$
we can eliminate all but 
$\psi_v^\dagger (iv\cdot \partial-\partial^2/2m)\psi_v$;
by field redefinitions involving $iv \cdot \partial$ and $\partial^2$
we can systematically eliminate powers of $v\cdot \partial$
bigger than one. In the rest frame $v=(1,\vec{0})$
we then arrive at the standard
\begin{equation}
{\cal L} =  \psi^\dagger \left( i\partial_{0}
              +\frac{1}{2m} \vec{\nabla}^{2}
              +\frac{1}{8m^3}\vec{\nabla}^{4}+\ldots\right) \psi. \label{kinL}
\end{equation}
When two heavy fields of mass $m$ and $m+ \delta m$, $\delta m \ll M$,
are considered in the same EFT,
it is simpler to redefine both with
the same phase $e^{im v\cdot x}$. 
In this case the kinetic terms of the heavier field
are as in Eq. (\ref{kinL}) with additional terms 
$-\psi^\dagger \delta m (1+\ldots) \psi$.

A consequence is that we can alternatively work in
the rest frame with
heavy fields that provide definite representations of $SO(3)$
\cite{invk}.
By field redefinitions we can systematically remove time-derivatives
of the heavy fields from interactions.
We take $N$ to be a Pauli spinor that provides
the $1/2$ representation $\frac{1}{2} \vec{\sigma}$ of the spin generators,
\begin{equation}
\sigma_{i}\sigma_{j}=\delta_{ij} + i \epsilon_{ijk} \sigma_{k}. 
\end{equation}
In the rest frame, $\psi=(^1_1)N/\sqrt{2}$.
Likewise, the $\Delta$ provides
the $3/2$ representation of the spin generators.
In the rest frame, $\psi_\mu=(0, \vec{S}\Delta)$,
\begin{equation}
  S_{i}S_{j}^{+} = \frac{1}{3} (2\delta_{ij} - 
              i\varepsilon_{ijk} \sigma_{k}).    \label{ci5}
\end{equation}
The $2 \times 4$ transition matrix $\vec{S}$
is used to couple $\Delta$ to $N$ in spin-$1$ bilinears.

{\bf Internal symmetries}.
Goldstone's theorem \cite{inGSW} tells us that massless Goldstone bosons
arise due to the assumed spontaneous breaking 
$SO(4)\rightarrow SO(3)$, and are naturally 
identified as pions.
Pions are then associated with broken generators 
of $SO(4)$, and their fields live in the 
three-sphere $SO(4)/SO(3) \sim S^{3}$. 
We call the diameter of this sphere
$F_{\pi}$; it is directly related to 
the chiral symmetry breaking scale, $\Lambda_{\chi SB}$, but the
precise factor can only be obtained from the 
(so far elusive) mechanism of 
dynamical symmetry breaking in QCD.
It turns out that this diameter can be determined from pion decay,
and is the pion decay constant, $F_{\pi}\simeq 190$MeV\@.
If we embed $S^{3}$ in the Euclidean $E^{4}$ space, $SO(4)$ 
transformations can be viewed as rotations in various $E^{4}$ hyperplanes. 
For example, $SU(2)_{V}$ of isospin consists of rotations in planes 
orthogonal to the fourth axis, while axial $SU(2)_{A}$ are rotations through 
planes that contain the fourth axis.

The sphere can be parametrized any way you want, say
with four cartesian coordinates $\{\boldphi ,\varphi_{4}\equiv\sigma\}$ 
subject to the constraint
$\sigma^{2} + \boldphi^{2} = F_{\pi}^{2}/4$.
The coordinates $\{\boldphi ,\sigma\}$ transform linearly
under $SO(4)$ rotations; for example,
under an infinitesimal 
$SU(2)_{V}$ transformation with parameter $\boldepsilon_V$,
\begin{equation}
   \delta_V \boldphi = \boldepsilon_V \times \boldphi. \label{deltaphi}
\end{equation}
\noindent
It is more convenient, however, to work with three unconstrained coordinates
$\boldpi$ that can be directly associated with the pions.
Any point on the sphere can be obtained by applying a 
four-rotation $R(\boldpi)$, 
$R(\boldpi)R^{T}(\boldpi)=1$, 
to the north pole $({\mbox{\boldmath $0$}}, \frac{1}{2} F_{\pi})$:
\begin{equation}
 \varphi_{\alpha}(\boldpi)= R_{\alpha 4}(\boldpi) \frac{F_{\pi}}{2}.  
                                                                 \label{iv3}
\end{equation}
Different parametrizations of $R$ define different pion fields $\boldpi$.
The $\boldpi$'s just rotate,
\begin{equation}
   \delta_V \boldpi = \boldepsilon_V \times \boldpi,
\end{equation}
\noindent
under $SU(2)_{V}$, but
transform non-linearly under $SU(2)_{A}$. 
{}From a particular choice of $R(\boldpi)$ one can read the
non-linear transformation of $\boldpi$.
The transformation law for $\partial_\mu \boldpi$ is invariably complicated.  

The general theory of non-linear realizations of a group is 
discussed in Ref. \cite{inCCWZ}.
The main idea is to construct covariant objects
that transform under the broken generators
as under the unbroken subgroup, but with a field-dependent ---thus
local--- parameter. 
The transformation we want to imitate
is the rotation (\ref{deltaphi}) and the appropriate
field-dependent parameter for an axial transformation with parameter
$\boldepsilon_A$ is
$\boldepsilon_A \times\boldpi/F_{\pi}$. 
We demand that the covariant derivative of the pion field
\begin{equation}
\boldD_{\mu} =
\frac{\partial_{\mu} \boldpi}{F_{\pi}} (1+O(\boldpi^2/F_{\pi}^2)) 
\end{equation}
transform as
an isospin-$1$ object,
\begin{equation}
   \delta_V \boldD_{\mu} = \boldepsilon_V \times \boldD_{\mu}, 
\;\;   \delta_A \boldD_{\mu} = (\boldepsilon_A \times
       \frac{\boldpi}{F_{\pi}}) \times \boldD_{\mu}. \label{ci37}
\end{equation}
Given the local look of $\delta_A \boldD_{\mu}$, it is not surprising
that the
covariant derivative of $\boldD_{\mu}$ is formed
from a vector field 
\begin{equation}
\boldE_{\mu}= \frac{2i}{F_{\pi}^2}\boldpi 
              \times \partial_{\mu} \boldpi \,
              (1+O(\boldpi^2/F_{\pi}^2)) 
\end{equation}
as
\begin{equation}
{\cal D}_{\mu}\boldD_{\nu} = \partial_{\mu}\boldD_{\nu} +
          i\boldE_{\mu}\times\boldD_{\nu}.    \label{ci7} 
\end{equation}
The pion fields being associated with broken generators,
one can amass them in a matrix
$u(\boldpi\cdot\boldt)
\equiv a(\boldpi^2) + i b(\boldpi^2)\boldpi\cdot\boldt$
---with $a(\boldpi^2)$ and $b(\boldpi^2)$ two functions
that depend on $R(\boldpi)$---
defined so that
\begin{equation}
\boldD_{\mu}\cdot\boldt =
 -\frac{i}{4}(u^\dagger \partial_\mu u -u\partial_\mu u^\dagger),
\;\; \boldE_{\mu}\cdot\boldt = 
\frac{1}{2}(u^\dagger \partial_\mu u +u\partial_\mu u^\dagger).
\end{equation}

For example, one can  
use stereographic coordinates \cite{inwei6}
$\boldpi = 2\boldphi/(1 + 2\sigma/F_{\pi})$,
which correspond to
\begin{equation}
 R[\boldpi]=
\left( \begin{array}{cc}
  \delta_{ij}-2D^{-1}\frac{\pi_{i}\pi_{j}}{F_{\pi}^{2}} & 2D^{-1}\frac{\pi_{i}}
                                                             {F_{\pi}}  \\
   -2D^{-1}\frac{\pi_{j}}{F_{\pi}} & D^{-1}\left(1-\frac{\boldpi^{2}}
                                      {F_{\pi}^{2}}\right) 
                                    \end{array}   \right),  \label{iv4}
\end{equation}
\noindent
where
\begin{equation}
 D\equiv 1+\frac{\boldpi^{2}}{F_{\pi}^{2}}  \mbox{.}       \label{iv5}
\end{equation}
In this case, the pion covariant derivative and the vector field take
particularly simple forms
\begin{equation}
  \boldD_{\mu} =
        D^{-1} \frac{\partial_{\mu} \boldpi}{F_{\pi}},
\;\;  \boldE_{\mu}=\frac{2i}{F_{\pi}}\boldpi \times \boldD_{\mu}.
                                            \label{ci10}
\end{equation}
\noindent
Other common choices of pion field $\boldpi$
are the ``sigma-model'' parametrization
\begin{equation}
u^2= 
\sqrt{1- \frac{4\boldpi^2}{F_\pi^2}} + \frac{4i}{F_\pi} \boldpi\cdot\boldt
\end{equation}
and the ``exponential'' parametrization
\begin{equation}
u=e^{\frac{2i}{F_\pi} \boldpi\cdot\boldt}.
\end{equation}

Fields that couple to pions can be introduced easily
once their isospin character is known.
It simplifies to work with fields that transform
under broken rotations with the same field-dependent parameter
as before.
In the case of a $(2n+2)$-component field $\psi^{(n)}$ 
providing a representation $\boldt^{(n+1/2)}$
of isospin $n+1/2$,
\begin{equation}
 \delta_V \psi^{(n)} =  i \boldepsilon_V \cdot \boldt^{(n+1/2)} \psi^{(n)},
\;\; \delta_A \psi^{(n)}  =  i(\boldepsilon_A \times \frac{\boldpi}{F_{\pi}})
                    \cdot \boldt^{(n+1/2)} \psi^{(n)}. \label{ci40} 
\end{equation}
Its covariant derivative is then
\begin{equation}
{\cal D}_{\mu}\psi^{(n)} =
  (\partial_{\mu}+\boldt^{(n+1/2)}\cdot\boldE_{\mu})\psi^{(n)}.  
\label{ci8}
\end{equation}
\noindent
A nucleon field $N$ is a two-component isospinor and
Eqs. (\ref{ci40}) and (\ref{ci8}) hold for $\psi^{(0)}=N$ and 
$\boldt^{(1/2)}=\boldt$.
A delta field is a four-component spinor $\Delta$
and the same formulas apply for $\psi^{(1)}=\Delta$ and $\boldt^{(3/2)}$.
Isospin-1 bilinears of $N$ and $\Delta$ can be formed using
a $2 \times 4$ transition matrix
$\mbox{\boldmath $T$}$ that satisfies
\begin{equation}
  T_{a}T_{b}^{+}  =  \frac{1}{6} (\delta_{ab} - i \varepsilon_{abc}t_{c}).
                      \label{ci6}
\end{equation}

The introduction of fields that transform non-linearly
makes it easy to construct chiral invariant objects:
any isoscalar built out of 
the fields $\boldD_{\mu}$ and $\psi$,
and their
covariant derivatives
will automatically be invariant under the whole $SU(2) \times SU(2)$
group.
Operators $T_{\alpha\beta\ldots}[\boldpi;\boldD_{\mu},\psi]$
that break chiral symmetry as tensor products
of the tensors (\ref{vecS}), (\ref{vecP}), and (\ref{tenf})
must also be constructed so that we can reproduce the correct
symmetry pattern of low-energy $S$-matrix elements.
Analogously to Eq. (\ref{iv3}), where 
an $SO(4)$ vector involving
the pion field $\boldpi$ is constructed by applying a chiral rotation
on another that does not,
we just need to search for
those tensors 
$T_{\alpha\beta\ldots}[{\mbox{\boldmath $0$}};\boldD_{\mu},\psi]$
written in terms covariant objects only, then rotate them
with $R(\boldpi)$:
\begin{eqnarray}
T_{\alpha\beta\ldots}[\boldpi;\boldD_{\mu},\psi] = 
     \sum_{\alpha'\beta'\ldots}R_{\alpha\alpha'}[\boldpi]R_{\beta\beta'}
     [\boldpi]\ldots\;\;T_{\alpha'\beta'\ldots}
     [{\mbox{\boldmath $0$}};\boldD_{\mu},\psi].  \label{iv14}
\end{eqnarray}
For example, the simplest vector we can construct out of numbers only is
$S_{(1)}[{\mbox{\boldmath $0$}};{\mbox{\boldmath $0$}},0]
                 =({\mbox{\boldmath $0$}},1)$
\noindent
so from (\ref{iv14}) and (\ref{iv4}), the fourth component of
\begin{equation}
 S_{(1)}[\boldpi;{\mbox{\boldmath $0$}},0]
=\left(2D^{-1}\frac{\boldpi}{F_{\pi}},1-2D^{-1}
   \frac{\boldpi^{2}}{F_{\pi}^{2}}\right)  \label{iv16}
\end{equation}
\noindent
breaks chiral symmetry in exactly the same way as $\bar{q} q$.
By considering other vectors and tensors we can construct the
infinite number of operators that break chiral symmetry
as quark masses and electromagnetic interactions.

Because the EFT is formulated in terms of the lowest-lying
color singlets, color gauge invariance plays no direct role.
Soft photons
(those with momenta $< M$) are not integrated out of the EFT
and couple to pions, nucleons, and deltas
in the most general way that respects $U(1)_{em}$ gauge invariance. 
The photon field $A_\mu$ enters through {\it (i)}
``minimal substitution'' in covariant derivatives,
\begin{equation}
\partial_\mu \pi_a\rightarrow \partial_\mu \pi_a-eA_\mu \epsilon_{3ab} \pi_b,
\;\; \partial_\mu \psi^{(n)}
\rightarrow \partial_\mu \psi-ieA_\mu  Q_\psi \psi^{(n)},
\end{equation}
where $Q_\psi$ is the charge matrix of the field
$\psi^{(n)}$ ---for example, $Q_N= (1+\tau_3)/2$;
and {\it (ii)} gauge invariant objects constructed out of the field
strength $F_{\mu \nu}$. 
There is, however, one subtlety regarding the role of gauge invariance:
chiral symmetry is broken explicitly by the electromagnetic
interaction because there is an anomaly in the third (isospin) component 
of the axial current. Terms have to be included in the effective
Lagrangian in such a way
that under a chiral transformation they reproduce the anomaly.
This abnormal intrinsic parity sector is reviewed in Ref. \cite{bij}.    
Finally, further breaking of the above symmetries
by the theta term \cite{pira} and non-renormalizable interactions
\cite{donoghue}
in the Lagrangian (\ref{intro9})
can be included as above, 
{\it mutatis mutandis}.

\subsection{The power of counting} \label{subsec-pow}

Although symmetries restrict the number of interactions
in the EFT, we still have to deal with an infinite number of them. 
The only hope of any predictive power relies on
finding some ordering of contributions.
Since it is not likely that the infinite number of coupling
constants will be all small enough for a coupling-constant 
expansion to converge, we turn to the only obvious
small quantity, the typical momentum $Q$ itself.

Powers of $Q$ of any particular Feynman diagram can be counted
in ways analogous to the one used in finding the superficial degree of
divergence (see, {\it e.g.}, Ref. \cite{itzykson}).
Each space derivative in an interaction produces a three-momentum
in a vertex and therefore counts as $Q$.
Likewise, each time derivative of a light field brings a $Q$ as
well.
In any loop, integration over the zeroth component of the 
four-momentum
will generate two types of poles according to the scales appearing
in the propagators:
{\it (i)} standard poles at $\sim Q$ corresponding to
external three-momenta and to 
the masses of the lightest excitations;
{\it (ii)} shallow poles at $\sim Q^2/2m_N$
corresponding to initial energies.

Processes that involve at most one heavy particle line are the simplest
because we can always close the contour avoiding shallow poles. 
As a consequence,
each four-momentum integration brings a factor $Q^{4}$.
A nucleon propagator is $Q^{-1}$ from the first term
in the kinetic Lagrangian (\ref{kinL}),
the other terms being treated as corrections:
the nucleon is a nearly static source propagating forward
in time, as can be seen from the Fourier transform  
$\theta (t) \delta^{3} (\vec{r})+\ldots$
of its propagator.
If $Q$ is sufficiently high so that pions and deltas are included
explicitly, a pion propagator is  $Q^{-2}$,
and a delta propagator is $Q^{-1}$.

\begin{figure}[t]
\centerline{\psfig{figure=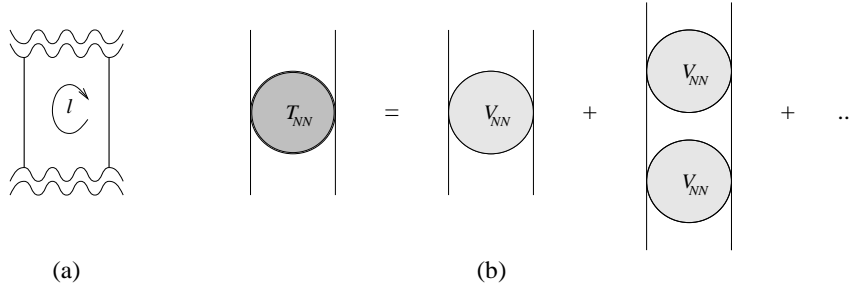,height=1.85in,width=4.50in}}
\vspace{-0.5cm}
\caption{(a) A
two-nucleon intermediate state between two unspecified
interactions, associated with a loop four-momentum $l$; 
(b) the full two-nucleon amplitude $T_{NN}$ as an iteration of a
two-nucleon potential $V_{NN}$.
\label{fig:TV}}
\end{figure}

Processes that involve more than one stable heavy particle present
a complication: a failure of perturbation theory that can
potentially lead to bound states \cite{inwei6}.
To see this, consider the scattering of two heavy particles
with center-of-mass 
momentum $\vec{p}$ and total energy $E\equiv k^2/m +\ldots$. 
In-between two interactions, the heavy particles propagate
forward in time. Such a piece of an arbitrary diagram is shown in 
Fig. \ref{fig:TV}(a).
Neglecting smaller relativistic corrections,
it will contribute 
\begin{equation}
 \int \frac{dl^{0}}{2\pi}\: 
 \frac{1}{l^0+ \frac{E}{2}- \frac{(\vec{l}+\vec{p})^2}{2m_N} +i\epsilon} \;
 \frac{1}{-l^0+ \frac{E}{2}- \frac{(\vec{l}+\vec{p})^2}{2m_N} +i\epsilon}
\end{equation}
\noindent
to the result of the diagram.
The problem is that this integral is infrared enhanced, 
for it picks a large residue from one of the shallow
poles. A naive momentum counting would have suggested a
contribution of $O(1/Q)$, yet we end up with something of $O(m_N/Q^2)$.
This is quite a general phenomenon: such an enhancement of $O(M/Q)$
appears
whenever two or more particles of mass $O(M)$ but momenta $O(Q)$
collide.
As a consequence, for those contributions that come from
shallow poles
each four-momentum integration effectively brings a factor $Q^{5}$.
A nucleon propagator is counted as $Q^{-2}$,
and the first two terms in the kinetic Lagrangian (\ref{kinL})
are important.
A pion propagator still counts as $Q^{-2}$,
but the pion can be taken in first approximation as
static, and it is sometimes referred to as a ``potential'' pion.
Contributions that come from
standard poles naively scale as in processes with none or one heavy particle.
Pions there are non-static or ``radiative''.

After integration over energies, each graph in covariant
perturbation theory produces a number of terms.
One can associate these with graphs in time-ordered perturbation theory,
where loops represent
three-momentum integrations and internal lines stand for denominators 
which contain energy differences between initial and intermediate states. 
In general, we can distinguish two classes of time-ordered 
diagrams. 
(Ir)reducible 
diagrams are those 
that can(not) be split in two pieces if in an intermediate
state we cut all ---and only--- the lines corresponding to initial or final 
particles; their
denominators will (not) contain energy differences of 
$O(Q^{2}/m_N)$.

In a system of heavy particles,
we call the sum of irreducible diagrams the potential $V$. 
Note that the potential being a set of subgraphs,
it can be defined in alternative ways.
All definitions that exclude the infrared enhancement
contributions but differ by a smaller amount are equally good,
as long as no double-counting or omissions are made.
The important point is that the only
scale appearing explicitly in the potential is $Q$, 
so that the power counting 
proceeds as in the case of diagrams with at most one heavy particle.

Reducible diagrams can be obtained by sewing together irreducible diagrams.
The full amplitude  $T$ is a sum of the potential and its 
iterations; schematically, with $G_0$ denoting the intermediate
state containing
only initial or final particles,
\begin{equation}
T = -(V+ VG_0V+ VG_0VG_0V+\ldots)= -V+ VG_0T.
\end{equation}
\noindent
This is just the Lippmann-Schwinger equation,
which is {\it formally} equivalent to the Schr\"{o}dinger equation
with the potential $V$.
Fig. \ref{fig:TV}(b) illustrates the case
of two nucleons, which
formally evolve according to the familiar non-relativistic
Schr\"odinger propagator
\begin{equation}
G_0(l;k)= -\frac{m}{l^2 -k^2 -i\epsilon},         
\;\; G_0(r;k)= -m \frac{e^{ikr}}{4\pi r}.           \label{spropcoord}
\end{equation}
\noindent
Relativistic corrections are accounted for by a two-legged vertex
\begin{equation}
u(l;k) = -\frac{l^4-k^4}{8m^3}+ \dots,
\end{equation}
and can be lumped in the potential.

With these ingredients one can write the 
contribution of any diagram to the amplitude as
\begin{equation}
T \propto Q^\nu {\cal F}(Q/\Lambda), \label{powercount}
\end{equation}
where $\Lambda$ is a renormalization scale, 
and $\nu$ is a counting index. 
The exact formula for $\nu$ 
in terms of the elements of a diagram
depends on the $Q$ we are interested in.
In addition, electromagnetic interactions can be considered
through a simultaneous expansion in 
\linebreak
$\alpha=e^2/4\pi$.
We will discuss the power counting in more detail below.

Counting powers of $Q$ is not in itself sufficient 
for an ordering of interactions. 
We need to find which other scales accompany $Q$. 
As we have mentioned in the introduction, there are
at least three physical scales in the problem: $\aleph$, $M_{nuc}$, 
and $M_{QCD}$. 
The largest possible ratio is $Q/\aleph$.
{}From dimensional analysis alone, then, we expect that 
for $Q\ll \aleph$ amplitudes for scattering
of free nucleons will be perturbative in $Q/\aleph$.
Such systems are labeled dilute.
As $Q$ increases, we have to perform resummations:
first summing all orders in $Q/\aleph$ but
expanding in $Q/M_{nuc}$; then summing all orders in $Q/M_{nuc}$
but expanding in $Q/M_{QCD}$; and so on.
As we resum certain class of diagrams,
bound-states may appear.

Given that we do not yet possess a full solution of QCD, it is
obvious that some assumptions have to be made about the
parameters of the EFT.
The minimal assumption is that of naturalness \cite{in'thooft}.
Naturalness means simply that for each
operator ${\cal O}_{i}$ of dimension $\delta_{i}$ in Eq. (\ref{intro0}), 
the parameter
$g_{i}(\Lambda)$ will be of form $c_{i}(\Lambda/M)/M^{\delta_{i}}$
with $c_{i}(1)=O(1)$
\footnote{Using a mass cutoff $\Lambda \ll M$
introduces a large dimensionless ratio $M/\Lambda$
and can unnecessarily complicate the power counting.}, unless there
exists a 
symmetry that forces $c_{i}$ to be small. 
As we are going to see, the presence in the nuclear problem of  
scales lighter than $M_{QCD}$
requires slightly stronger
assumptions about the scaling of the EFT parameters.
These new assumptions are still fairly general and are
motivated by qualitative features of nuclear phenomenology;
they are much less restrictive than usual model assumptions.
In any case, these assumptions are necessary for
the {\it a priori} ordering of contributions, but they are
subject to test {\it a posteriori}.
If a parameter is found much larger than assumed initially,
the EFT can be improved to better account for its effects.

\subsection{Propaganda} \label{subsec-prop}

The EFT approach is:

({\it i}) systematic. 
Once a power counting is established, 
there is an expansion in a small parameter $\propto Q$.

({\it ii}) consistent with QCD. The only (very important) QCD inputs are
confinement, (chiral, ...) symmetries, and
some version of naturalness. QCD can be represented by a point in the 
space of renormalized parameters (at some renormalization scale) of
the effective theory. 
An explicit solution of QCD (such as from simulations on a lattice)
would provide knowledge of the exact position of this point, 
and then the EFT would be completely predictive. 
Until such a solution is found ---or as a test of QCD after it is found--- 
we can recourse to fitting of low-energy experiments in order to determine 
the region in parameter space allowed on phenomenological grounds. 
Even in this case the theory is predictive, because
to any given order the space of parameters is finite-dimensional.
After a finite number of experimental results are used, an infinite number 
of others can be predicted up to an accuracy depending on the order of the 
expansion. In practice, because the number of parameters grows rapidly 
with the order, model-dependent estimates of parameters based on 
specific dynamical ideas 
---such as saturation by tree-level resonance exchange \cite{mesonsat}---
are sometimes used.

({\it iii}) a consistent quantum field theory.
Renormalization can be carried out to 
relate parameters appearing in different processes and to remove
cutoff dependence at each order in the expansion.
As a consequence, it is applicable in principle to all low-energy 
phenomena. The literature on applications
to mesons and one-nucleon processes is vast; after a lightning
review I will show how we can also gain
insight into nuclear physics {\it per se}.

\section{One- or no-nucleon processes} \label{sec-01N}

The EFT at momenta $Q \ll M_{nuc}$ contains only nucleons,
pions treated as heavy particles, and photons.
All interactions are of contact type,
and form an expansion in $Q/M_{nuc}$.
The relevant symmetries are Lorentz and $U(1)_{em}$ local invariance.
Processes involving two or more pions, or one nucleon and at least one pion
are of the type considered in the next sections.
Processes involving one pion or one nucleon and a number of photons
give rise to the various well-known electromagnetic parameters.
Particle form factors take a multipole
form, that is,
an expansion in charge and anomalous magnetic moment, 
radii, {\it etc.}
Compton scattering has a Thomson-seagull interaction
and a magnetic moment piece, 
followed by polarizabilities, and so on.
The analytic nature of these expansions in $Q/M_{nuc}$ 
and the limited symmetry constraints make this EFT
not very interesting.
We expect, of course, that a sizable fraction of the value
of the parameters
comes from the lightest state that was integrated out, the pion.
As $Q$ is increased, the approximation of zero-range propagation
for pions becomes less and less reliable.

At momenta $Q\sim m_\pi$, pions have to be included explicitly in the
theory. 
Because numerically $\delta m$ is $\sim 2 m_\pi$,
convergence of the EFT for $Q\sim M_{nuc}$ should be optimized by
the concomitant inclusion of an explicit delta degree of freedom.
Now, we are in luck because pion interactions are not arbitrary.
Once explicit pion fields are considered, approximate chiral symmetry
imposes important restrictions on the way pions couple among themselves
and to other degrees of freedom.
The delta, too, can be included without too much hassle, since
at these momenta it is, like the nucleon,
a non-relativistic object, to which we can apply the heavy-particle formalism. 
In the preceding section all the ingredients to construct this EFT
were presented.

The power counting (\ref{powercount}) presents no problems as the
only explicit scales are $M_{nuc}$ and $Q$,
contributions from momenta of order $M_{QCD}$ being lumped
in the vertices.
Assuming $Q\sim M_{nuc}$,
it follows that for a diagram
involving $A=0,1$ nucleon and any number of pions
\begin{equation}
\nu=2- A + 2L+{\sum _i}{V_i}{\Delta _i}, 
\;\;{\Delta _i}\equiv {d_i}+{f_i}/2-2,
\label{index}
\end{equation}
where $L$ is the number of loops and
$V_i$ is the number of vertices of type $i$, which contains $d_i$ 
derivatives or powers of $m_\pi$ and $f_i$ fermion fields
\cite{inwei2,inwei6}.
This formula is important because chiral symmetry places a lower
bound on the interaction index ${\Delta _i}\geq 0$.
Since $L$ is bounded from below ($L\ge 0$),
$\nu\ge \nu_{min}=2-A$ for strong interactions.
An expansion in  $Q/M_{QCD}$ results.
It starts at $\nu= \nu_{min}$ with tree ($L=0$) diagrams
built out of vertices of index 0 ($\sum V_i \Delta_i =0$),
then proceeds at $\nu= \nu_{min}+1$ with further tree diagrams,
now with one vertex of index 1, the remaining
having index 0 ($\sum V_i \Delta_i =1$). 
These first two orders are equivalent
to the current algebra of the 60's, but now 
unitarity corrections can be accounted for systematically.
At $\nu= \nu_{min}+2$, for example, besides tree
diagrams with one index-2 interaction or two 
index-1 interactions ($\sum V_i \Delta_i =2$),
there are also one-loop ($L=1$) diagrams built out of
index-0 vertices ($\sum V_i \Delta_i =0$).
This is generalized  to higher orders in obvious fashion.
We discuss isospin breaking soon. 

It is convenient then to split the chiral Lagrangian in pieces
${\cal L}^{(\Delta)}$
labeled by the index $\Delta$.
For example, the lower-order Lagrangians are 
\cite{invk,ciOLvK,vkolck:cohen,bkm}
\begin{eqnarray}
{\cal L}^{(0)} & = & -\frac{F_\pi^2}{2}\boldD^2
         - \frac{1}{2}m_{\pi}^{2}D^{-1}\boldpi^{2} 
          +N^\dagger i{\cal D}_{0} N 
         -2g_{A}\boldt\cdot\vec{\sigma}\cdot\vec{\boldD} N  \nonumber  \\
   &   & +\Delta^\dagger (i{\cal D}_{0}-\delta m)\Delta 
         -2h_{A} \left[N^\dagger \boldT\cdot\vec{S}\cdot\vec{\boldD}
         \Delta+h.c.\right]+\ldots, \label{L0}
\end{eqnarray}
\begin{eqnarray}
 {\cal L}^{(1)} & = & \frac{1}{2m_N} N^\dagger \vec{\cal D}^{\, 2} N
                     -\frac{g_A}{m_N F_\pi} 
  \left[iN^\dagger \boldt\cdot \boldD_0 \vec{\sigma}\cdot \vec{\cal D}N
        + h.c.\right] \nonumber  \\
  & & + \frac{1}{2m_N} \Delta^\dagger \vec{\cal D}^{\, 2} \Delta
  -\frac{h_A}{m_N F_\pi}
 \left[iN^\dagger \boldT\cdot \boldD_0 \vec{S}\cdot \vec{\cal D}\Delta
         + h.c.\right] \nonumber  \\
  & &   -B_{1}\boldD^2  N^\dagger N 
      -B_{2} (\vec{\boldD}\times\vec{\boldD})
            \cdot N^\dagger \boldt\vec{\sigma}N 
   -\frac{B_{3}
        m_{\pi}^{2}}{F_{\pi}^{2}}D^{-1}\boldpi^{2}N^\dagger N
     -B_4 \boldD_0^2  N^\dagger N  
     +\ldots \label{L1}
\end{eqnarray}
\noindent
Here $g_{A}, h_A =O(1)$ and $B_{i}=O(1/M_{QCD})$
are undetermined
constants, to be obtained either by solving QCD or by fitting data;
``\ldots'' stand for other terms involving the delta isobar
which do not appear explicitly below.
Higher-index interactions can be constructed similarly
Note that pion self-interactions appear only at even index because
of Lorentz invariance.

The most general isospin-violating Lagrangian \cite{invk,civK2}
is constructed out 
of operators that break isospin in three different ways.
First, there are gauge invariant couplings to soft photons.
For example, from minimal substitution in the index-0 Lagrangian,
\begin{eqnarray}
{\cal L}_{sp}^{(-1)} &=& 
- e A^{\mu} (\boldpi \times \partial_{\mu} \boldpi)_3 
    +  \frac{1}{2} e^2 A^2 (\boldpi^2 -\pi_3^2) \nonumber \\
& &   - e A_0 N^\dagger (\frac{1}{2}+t_3) N 
    +\frac{2e g_A}{F_\pi} \vec{A}\cdot  N^\dagger \vec{\sigma}
       (\boldt \times \boldpi)_3 N 
   + \ldots 
\label{iv42}
\end{eqnarray}
Second, there are operators that transform   
as products of the quark-mass-difference term;
they are proportional to $\varepsilon m_\pi^2$, where
$\varepsilon \equiv m_{d}-m_{u}/(m_{d}+m_{u}) \sim 1/3$ \cite{ivwei1}.
For example,
\begin{equation}
 {\cal L}_{qm}^{(1)}=
     \delta m_{N}(-N^\dagger t_{3}N+2D^{-1}\frac{\pi_{3}\boldpi}
                  {F_{\pi}^{2}}\cdot N^\dagger \boldt N) +\ldots,  \label{iv24}
\end{equation}
\begin{equation}
 {\cal L}_{qm}^{(2)} = \frac{1}{2}D^{-2}\delta m_{\pi}^{2}\pi_{3}^{2} 
              + \beta_{1}(\vec{D}_{3}-\frac{2D^{-1}}{F_{\pi}^{2}}\pi_{3}
                     \boldpi\cdot\vec{\boldD})\cdot N^{\dagger}\vec{\sigma}N
              +\ldots,  
\label{iv25}
\end{equation}
with $\delta m_{N}=O(\varepsilon m_{\pi}^{2}/M_{QCD})$,
$\delta m_{\pi}^{2}=O(\varepsilon^{2}m_{\pi}^{4}/M_{QCD}^{2})$,
and $\beta_{1}=O(\varepsilon m_{\pi}^{2}/M_{QCD}^{2})$.
Third, there are operators that transform 
as four-quark interactions generated by hard-photon exchange,
which are proportional to $\alpha$.
For example,
\begin{equation}
 {\cal L}_{hp}^{(-2)} = -\frac{1}{2}D^{-2}\bar{\delta}m_{\pi}^{2}
                        (\boldpi^{2}-\pi_{3}^{2})          \label{iv41}
\end{equation}
\begin{equation}
 {\cal L}_{hp}^{(-1)} = \bar{\delta} m_{N}\left[-N^{\dagger}t_{3}N+
      \frac{2D^{-1}}{F_{\pi}^{2}}
N^{\dagger}(\boldpi^{2}t_{3}-\pi_{3}\boldpi\cdot\boldt)N\right]
       +\ldots \label{iv52}
\end{equation}
with $\bar{\delta}m_{\pi}^{2} \propto \alpha M_{QCD}^{2}$
and $\bar{\delta}m_N=O(\bar{\delta}m_{\pi}^{2}/M_{QCD})$.

The observed smallness of isospin violation at low energies is not 
immediately evident within QCD, as chiral symmetry breaking
effects are not negligible, and $\varepsilon$ is not particularly
small.
Power counting offers an explanation for this phenomenon
\cite{invk,civK2}. 
The first task is to compare the different towers of isospin--violating
operators. This can be accomplished by noticing that photon loops
(containing two electromagnetic interactions)
typically produce factors of $\alpha /\pi$, which is {\em numerically\/}
$\sim \varepsilon (m_{\pi}/m_{\rho})^3$. This immediately
suggests that ${\cal L}_{hp}^{(n)}\sim{\cal L}_{qm}^{(n+3)}$ and
${\cal L}_{sp}^{(n)}\sim{\cal L}_{qm}^{(n+\frac{3}{2})}$.
It can easily be verified that this estimate produces the right order
of magnitude for the various contributions to the pion and nucleon mass
differences,
$\Delta m_\pi^2= \bar{\delta}m_{\pi}^{2}+ \delta m_{\pi}^{2}+ \ldots$
and
$\Delta m_N= \delta m_{N}+\bar{\delta}m_{N}+ \ldots$,
respectively.
One now can see that, because there is no lowest-order isospin-violating
interactions, isospin-violating quantities are typically not only 
$O(\varepsilon)$ smaller than chiral-violating quantities, but are
in general further suppressed by factors of $Q/M_{QCD}$:
{\it isospin is an accidental 
symmetry} \cite{invk,civK2}.
For simplicity, when considering only 
(a fixed number of) external photons, 
I will count a power of $e$ as a contribution of 1
to the index $\Delta$ in Eq. (\ref{index}).

Because $\delta m \sim M_{nuc}$,
delta effects appear 
for $Q\sim M_{nuc}$ typically at the same order as nucleon contributions
\footnote{Some people reserve the name $\chi$PT to the EFT
of the world where 
$Q/\delta m\sim m_\pi/\delta m \ll 1$,
and refer to the EFT of a world where 
$Q/\delta m\sim m_\pi/\delta m \sim 1$
as the ``small scale expansion''.
I find it simpler to refer to them as deltaless and deltaful $\chi$PT,
respectively.}. 
At momenta $Q< \delta m$ we can integrate out the delta,
producing an EFT with nucleons and pions only.
This entails significant reduction in the number of interactions,
as the delta terms disappear and the remaining interactions
among nucleons and pions have the same structure
as those in the ${\cal L}^{(\Delta)}$'s above.
Only their coefficients change, and now contain
$O(1/\delta m)$ terms.
For example, ${\cal L}^{(1)}$ has the same pion-nucleon
seagulls with coefficients traditionally called $c_i$ in the 
literature \cite{bkm}
and related to the $B_i$ via
\[
c_1=\frac{B_3}{8}+\ldots, 
\;\; c_2=-\frac{1}{4} \left(B_4-\frac{g_A^2}{2m_N}\right) 
    +\frac{h_{A}^{2}}{9\, \delta m}+\ldots,
\]
\begin{equation}
c_3=\frac{B_1}{4} -\frac{h_{A}^{2}}{9\, \delta m}+\ldots,
\;\; c_4=\frac{1}{4} \left(B_2-\frac{1}{m_N}\right) 
    +\frac{h_{A}^{2}}{18\, \delta m}+\ldots
\label{c's}
\end{equation}

This machinery can be applied to soft processes involving any number
of pions and photons, and at most one nucleon.
$\chi$PT is particularly relevant for those processes
where the pion plays a dominant role.
Here I briefly present some of the highlights
that are directly relevant for what follows.
For details on 
pion processes see, for example, Ref. \cite{pionreview};
one-nucleon processes in the deltaless theory
have been extensively reviewed in 
Ref. \cite{bkm},
and explicit delta effects are discussed in Ref. \cite{deltapapers}.

Pion-nucleon scattering is perhaps the most important process
where the chiral Lagrangian can be applied,
as it can be used to test the fermionic sector of the theory and
determine some of its parameters.
Contributions start at $\nu=\nu_{min}=1$
from Born graphs with intermediate nucleons and deltas,
and from the $\pi\pi NN$ interaction (``Weinberg-Tomozawa seagull'')
generated
by the chiral covariant derivative in the kinetic nucleon term
of Eq. (\ref{L0}).
At $\nu=\nu_{min}+1$ the $B_i$-seagull diagrams contribute at tree level.
Dropping nucleonic Born terms,
the $\pi_a N\rightarrow \pi_b N$ amplitude is traditionally written 
\cite{TM} as
\begin{equation}
t_{ab} \propto
  \delta_{ab} \left[ a+ b \, \vec{q} \cdot \vec{q}\, '
           + c \, (\vec{q}\, ^2+\vec{q}\, '^{2}) +e \, \omega\omega' \right]
  -d \, \epsilon_{abc} \tau_c \, \vec{\sigma}\cdot \vec{q} \times \vec{q}\, '
  +\ldots
\label{tpiN}
\end{equation}
where $\vec{q}$ ($\vec{q}\, '$) is the incoming (outgoing) pion
momentum 
and $\omega$ ($\omega'$) is the initial (final) pion energy. 
The $a, ..., e$ coefficients are given in terms of the $c_i$'s
as
\begin{equation}
a = \frac{16\, m_\pi^2\, c_1}{F_\pi^2}, \;\;
b = \frac{8\, c_3}{F_\pi^2}, \;\; 
c = \, 0, \;\;  
d = -\frac{4c_4}{F_\pi^2}, \;\;
e = -\frac{8\, (c_2+c_3)}{F_\pi^2}.
\label{abcde}
\end{equation}
At $\nu=\nu_{min}+2$ a host of terms enter, including loop diagrams
and new parameters.
Very close to threshold, where only $S$ waves matter,
chiral symmetry predicts an isovector
scattering length $b^{(1)}\sim m_\pi/F_\pi^2$
from the Weinberg-Tomozawa seagull \cite{ivwei3}
and a smaller isoscalar scattering length 
$b^{(0)}\sim m_\pi^2 /F_\pi^2 M_{QCD}$, related to the sigma-term,
from the subleading $c_i$ seagulls.
The most extensive calculations have been performed to
$\nu=\nu_{min}+2$ in deltaless $\chi$PT.
Different pieces
of $\pi N$ scattering data have been used to determine the 
chiral coefficients. 
In lowest order, $g_{A}\simeq 1.25$;
in sub-subleading order chiral symmetry breaking
corrections appear, and if they are absorbed
in $g_{A}$, then $g_{A}\simeq 1.32$ satisfies 
the Goldberger-Treiman relation without discrepancy.
In Table \ref{tab:ci}
we list some of the determinations of the parameters $c_i$. 
Earlier fits \cite{bkm,bkm2,bkm3,moj} were
made to different sets of threshold and sub-threshold parameters 
obtained from
dispersion analyses of older data. 
Newer fits \cite{fms} were made to different
phase-shift analyses (PSAs), the last two in Table \ref{tab:ci}
including the more modern
meson-factory data. 
The $O(Q^3)$ determinations are consistent with each other
when their error bars (not shown) are considered, except for $c_1$, which
reflects the higher value for the sigma-term in the newer PSAs. 
Note that the
coefficients $c_2$, $c_3$, and $c_4$, which receive contributions from the
$\Delta$ at tree level, are larger than $c_1$, as expected. 
There are not yet very reliable
determinations of $h_{A}$ from $\pi N$ scattering
in the deltaful theory.
The large-${\cal N}_c$-limit value, $h_{A}= 3g_A/\sqrt{2}\simeq 2.7$,
gives a reasonable delta decay width at tree level.

\begin{table}[b]
\caption{Subleading $\chi$PT coefficients in 
GeV$^{-1}$ from several recent fits.
\label{tab:ci}}

\hspace{0.0cm}
\footnotesize
\begin{center}
\begin{tabular}{|c|cccc|}
Fit & $c_1$ & $c_2$ & $c_3$ & $c_4$\\ 
\hline 
$O(Q^2)$\cite{bkm2}                    & $-$0.64 &  1.78 & $-$3.90 &  2.25\\
$O(Q^3)$\cite{bkm}                     & $-$0.87 &  3.30 & $-$5.25 &  4.12\\ 
$O(Q^3)$\cite{bkm3}                    & $-$0.93 &  3.34 & $-$5.29 &  3.63\\ 
$O(Q^3)$\cite{moj}                     & $-$1.06 &  3.40 & $-$5.54 &  3.25\\
$O(Q^3)$\cite{fms}                     & $-$1.27 &  3.23 & $-$5.93 &  3.44\\ 
$O(Q^3)$\cite{fms}                     & $-$1.47 &  3.21 & $-$6.00 &  3.52\\
$O(Q^3)$\cite{fms}                     & $-$1.53 &  3.22 & $-$6.19 &  3.51\\ 
\end{tabular}
\end{center}
\end{table}

The only cases where isospin violation can be large are instances
where the leading isospin-conserving interactions do not
contribute. In such cases we can
encounter an isospin violation of size comparable
to $\varepsilon$.
One example is $S$-wave $\pi N$ scattering 
\cite{ivwei1,invk,civK2,fettesiv}.
The leading isospin-violating
term involving a nucleon is, according to the 
above discussion, given by Eq. (\ref{iv24}). 
It determines not only the most important
quark-mass contribution $\delta m_N$ to the nucleon mass splitting, 
but also isospin-breaking interactions of an even number of pions with
a nucleon. It contributes $\sim \delta m_N/ F_\pi^2$
to the pion-nucleon scattering lengths.
This gives a small contribution 
$\sim \delta m_N/m_{\pi}\sim \varepsilon m_\pi/M_{QCD}\sim 5\%$
to processes that involve at most one $\pi^{0}$, as they receive
a leading-order contribution from the Weinberg-Tomozawa seagull. 
But
for $\pi^{0}N\rightarrow\pi^{0}N$, which starts at
subleading order with the isoscalar seagulls, 
isospin violation can be as large as $\varepsilon \sim 30\%$, thus
revealing the full isospin breaking in the QCD Lagrangian!
Unfortunately, this one case of large isospin
violation cannot be measured easily.

Pion photoproduction
at threshold, where the pion is
mostly in an $S$ wave, has 
received a lot of attention lately
from both theoretical and experimental standpoints.
This process has been studied 
near threshold up to $\nu=\nu_{min}+3$ in the deltaless theory
\cite{vk:bernard1}.
At leading order, $\nu=\nu_{min}=1$, there exist only contributions
to charged channels from 
the fourth term in Eq. (\ref{iv42})
(``Kroll-Ruderman term'') and from the pion pole;
we thus expect the charged channels to be larger.
At $\nu=\nu_{min}+1$, minimal substitution on 
the recoil correction to the pion-nucleon vertex contributes
to photoproduction on the proton.
At $\nu=\nu_{min}+2$ there are tree-level contributions involving
the magnetic moment of the nucleon. Together with the
previous orders, this reproduces an old ``low-energy theorem''.
However, still at $\nu=\nu_{min}+2$ there are large loop corrections
where the photon interacts with a virtual pion which then rescatters
on the nucleon before flying away to the detector.
This quantum effect was missed in the old ``low-energy theorem'' and
in most models. It is determined in terms of known parameters
and thus constitutes a new, corrected low-energy theorem.
At $\nu=\nu_{min}+3$ other loops and new counterterms appear.
Results for the cross-section
at threshold up to $\nu=\nu_{min}+3$
from a fit to data above threshold constrained by resonance saturation 
of parameters \cite{vk:bernard1}
are satisfactory.
To this order there is a prediction for
the near threshold behavior of the
$\gamma n\rightarrow \pi^0 n$ reaction,
which would be important to test.

Compton scattering on the nucleon can also be calculated with this
EFT. To $\nu=\nu_{min}+2$, the amplitude contains non-analytic terms
and is completely predictive \cite{bkm}.
Expanded for small energies, it yields predictions
for the polarizabilities, which are determined by the pion cloud.
There is good
agreement with data on the proton,
and again it would be important to test the corresponding
neutron predictions.

We have seen that there are a number of tantalizing predictions
from $\chi$PT concerning nucleon structure,
particularly involving the neutron and neutral pions.
These are hard to test because neutron
targets and neutral pion beams do not come by easily... 
Nuclear targets and virtual pion effects can compensate
for this, {\it if} we can formulate an EFT
for nuclear systems that is consistent with $\chi$PT.
We are thus naturally led to nuclear physics proper.

\section{The two-nucleon system} \label{sec-2N}

The two-nucleon system is where we hope to understand
the basic modifications in the $\chi$PT power counting
needed to tackle bound states.
I will in turn consider the three regions of momenta
that have been investigated:
$Q<M_{nuc}$ in Subsect. \ref{subsec-2Nverylow},
$Q<M_{NN}$ in Subsect. \ref{subsec-2Nlow}, and
$Q<M_{QCD}$ in Subsect. \ref{subsec-2Nmod},
although this is not a historical order.

\subsection{Very low energies} \label{subsec-2Nverylow}

The lowest range of momenta 
serves to illustrate the method of EFT in the simplest
non-perturbative context.
Because only nucleons appear explicitly, 
we can go a long way following analytical arguments, while
a good fraction of the physics of higher momenta is 
just a generalization of the same ideas to a more complicated situation.
$S$-matrix elements for $NN$ scattering
at momenta $Q\ll M_{nuc}$ can be 
obtained in zeroth order in $\alpha$
from an effective Lagrangian involving arbitrarily complicated 
operators of only $N$ and its derivatives. 
The only symmetry constraints come from
Lorentz transformations of small 
velocity,
parity and time-reversal.
Electromagnetic interactions can be included by adding
all possible $U(1)_{em}$-invariant terms. 
In the two-nucleon system
it will give rise to photon exchange
in addition to dressings already present in one-nucleon systems.
This will be discussed after we consider strong interactions in
the $\alpha \rightarrow 0$ limit.

{\bf The amplitude.} 
The issue of establishing a power counting is largely
independent of complications generated by spin and isospin,
so for now consider \cite{vKolckn} a system of two 
identical, stable spinless particles $\psi$ in their center-of-mass frame
at energy $E= k^2/m- k^4/4m^3+\ldots$ 
much smaller than their mass $m$, their 
internal excitation energy, and
the range of their interaction, which I will denote collectively by $M$.
In addition to the kinetic terms in Eq. (\ref{kinL}), 
we have to consider the most general Lagrangian 
involving four nucleon fields,
which can be written 
as \cite{vKolckn}
\begin{eqnarray}
{\cal L} 
 & = & - \frac{1}{2}C_0 \psi^\dagger \psi\: \psi^\dagger \psi    \nonumber \\
 &  & - \frac{1}{8} (C_2+C'_2) [\psi^\dagger \bothnabla\psi \cdot
                        \psi^\dagger \bothnabla\psi
                        - \psi^\dagger \psi \:
                          \psi^\dagger \bothnabla^2\psi] \nonumber \\
&  & + \frac{1}{4} (C_2-C'_2)
                         \psi^\dagger\psi\:\vec{\nabla}^2(\psi^\dagger \psi) 
       +\ldots,                                   \label{lag}
\end{eqnarray}
\noindent
where the $C_{2n}$'s 
are parameters of mass dimension $-2(1+n)$ that depend on the details 
of the dynamics of range $\sim 1/M$. 
The ``$\ldots$'' stand for operators with more derivatives.

Conservation of particle number
reduces the $T$-matrix to a sum of bubble diagrams. 
The potential is given in momentum space by
\begin{equation}
V(p,p')= C_0+ C_2 
(\vec{p}^{\: 2}+\vec{p}\:'^{\, 2})
+ 2C'_2 \vec{p}\cdot\vec{p}\:'+ \ldots, \label{ver}
\end{equation}
\noindent
$\vec{p}$ ($\vec{p}\:'$) being the relative momentum of
the incoming (outgoing) particles. 
If Fourier-transformed to coordinate space, 
these bare interactions consist of delta functions and 
their derivatives. 
The bubbles require regularization and renormalization,
which introduce a mass scale 
$\Lambda$ through some cutoff function
$F(l^2/\Lambda^2)$ 
\footnote{ Although I refer to this as a cutoff,
I am {\it not} assuming any particular regularization scheme.}.
A bubble involving $n\geq 0$ derivatives
at the vertices gives
\begin{eqnarray}
I_{2n}(k) & = & \int \frac{d^3l}{(2 \pi)^3} \: l^{2n} G_0(l;k) F(l^2/\Lambda^2)
\nonumber \\
& = & -\frac{m}{2\pi^2}
       \left[\sum_{i=0}^{n} k^{2i} \theta_{2(n-i)+1} \Lambda^{2(n-i)+1} 
                                + i\frac{\pi}{2}k^{2n+1}
      +\frac{k^{2(n+1)}}{\Lambda} R(k^2/\Lambda^2)\right], \label{regbubbleany}
\end{eqnarray}
\noindent
where $\theta_n$ is a number that
depends on the regularization scheme chosen,
and $R(x)$ is a regularization-scheme-dependent
function that approaches a finite limit
as $x\rightarrow 0$.
In general $\theta_n$ and $R(x)$ can be non-vanishing with
either sign.
The severe $\Lambda$-dependence comes from the region of high momenta
that cannot be described correctly by the EFT; 
it can be removed
by lumping these terms together with the unknown bare parameters
into ``renormalized'' coefficients $C_{2n}^{(R)}$. 
Only the latter are observable.
The regulator-independent, non-analytic piece in
Eq. (\ref{regbubbleany}), on the other hand, is characteristic of loops:
it cannot be mocked up by re-shuffling contact interactions.
{}From Eq. (\ref{regbubbleany}) we see that we can estimate its effects 
by associating a factor
$m Q/ 4 \pi$ to each loop and a factor $Q$ to each derivative at
the vertices.

\begin{figure}[t]
\centerline{\psfig{figure=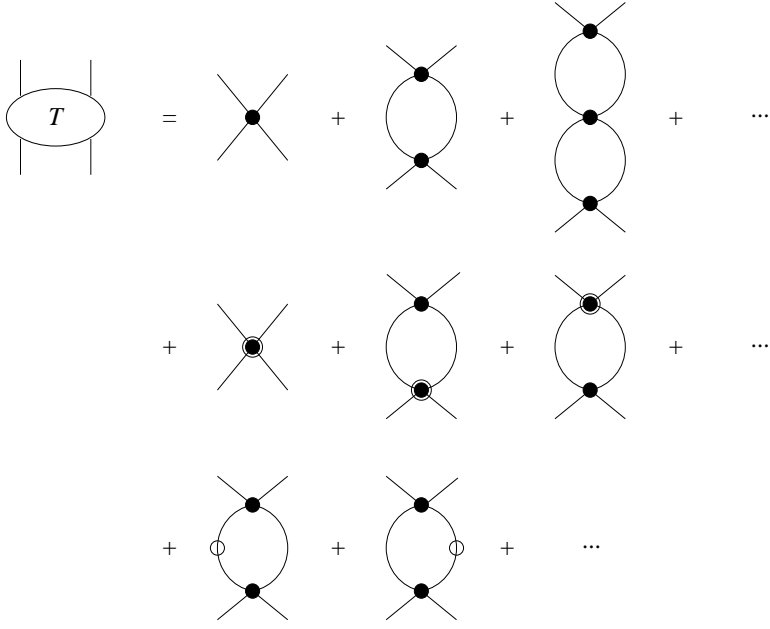,height=3.25in,width=4.00in}}
\vspace{0.5cm}
\caption{First few terms of the 
two-particle amplitude $T$ in a natural EFT.
Two solid lines represent a Schr\"odinger propagator;
a circle on a line represents a $Q^4$ relativistic correction;
a heavy dot stands for a $Q^0$ contact interaction, and
a dot within a circle for a $Q^2$ contact interaction.
\label{fig:natt}}
\end{figure}

The $T$-matrix is given by the diagrams in Fig. \ref{fig:natt}. 
For the first few terms one finds on-shell
\begin{eqnarray}
T_{os}(k, \hat{p}\,'\cdot \hat{p}) 
& = & -C_0^{(R)} \left\{ 1- \left(\frac{mC_0^{(R)}}{4\pi}ik\right)
    +\left(\frac{mC_0^{(R)}}{4\pi}ik\right)^2
    -\left(\frac{mC_0^{(R)}}{4\pi}ik\right)^3  \right. \nonumber \\
  & & 
\ \ \ \ \ \ \ \ \ \ \ + 2\frac{C_2^{(R)}}{C_0^{(R)}} k^2
 \left[1- 2 \left(\frac{mC_0^{(R)}}{4\pi}ik\right) 
\right] 
+2\frac{C_2^{'(R)}}{C_0^{(R)}} k^2 \hat{p}\,'\cdot \hat{p} 
\nonumber \\
  & & \left. \ \ \ \ \ \ \ \ \ \ \ 
  +\left(\frac{mC_0^{(R)}}{4\pi}ik\right)\frac{k^2}{2m^2} 
+ \ldots \right\}.    
                                                        \label{Tpert}
\end{eqnarray}
\noindent
Here dependence on the cut-off was eliminated by 
defining renormalized parameters 
$C_0^{(R)}$, $C_2^{(R)}$, and $C_2^{'(R)}$ 
from the bare parameters 
$C_0(\Lambda)$, $C_2(\Lambda)$, and $C'_2(\Lambda)$:
\[
\frac{1}{C_0^{(R)}} = \frac{1}{C_0} 
+\frac{m}{2\pi^2}\left[ L_1 +\left(\frac{2C_2}{C_0} +\frac{1}{4m^2}\right) 
L_3\right]
    +\ldots,               
\]
\begin{equation}
\frac{C_2^{(R)}}{(C_0^{(R)})^2}
     \equiv \frac{C_2}{C_0^2}
      -\frac{m}{4\pi^2} \left(\frac{R(0)}{\Lambda} +\frac{L_1}{2m^2}\right)
            + \ldots,           
\;\;\;\; C_2^{'(R)}\equiv C_2^{'} +\ldots  \label{C2'R}
\end{equation}
\noindent
Other terms coming from $R(k^2/\Lambda^2)$ are 
$\propto k^4$ or higher powers, and cannot be separated from higher-order
contact interactions that I have not written down explicitly.

It is important to realize that Eq. (\ref{Tpert}) consists of 
{\it two} different expansions: a loop expansion and an expansion 
in the number of insertions of derivatives at the vertices or particle 
lines. The derivative expansion depends on the relative size
of the coefficients $C_{2n}^{(R)}$, 
for example $C_2^{(R)}Q^2/C_0^{(R)}$.
The loop expansion is governed by 
$mC_0^{(R)}Q/4\pi$.
The sizes of terms in both expansions thus depend on the 
absolute size of the  $C_{2n}^{(R)}$'s. 
If one makes the simplest naturalness assumption
that there is only a single mass scale $M$ and no fine-tuning
in the underlying theory, we expect all parameters to scale with $M$
according to their mass dimension.
It is convenient, however, to factor in a $4\pi/m$ and write
$C_{2n}^{(...)(R)}\sim 4\pi/mM^{2n+1}$.
In this case, the derivative expansion is in 
$\sim (Q/M)^2$, while
the loop expansion is in $\sim Q/M$:
the $T$-matrix in Eq. (\ref{Tpert}) is a simple
expansion in powers of $Q/M$.
However, in nuclear or molecular physics we are interested in 
shallow $S$-wave bound states
\footnote{
The remaining case ---that of an unnaturally 
small scattering length--- has also been considered in the literature
\cite{vKolckn}.}. 
A shallow (real or virtual) bound state means that a parameter $\alpha$ 
of the underlying theory is close to its critical value $\alpha_c$
for which a bound state is exactly at threshold.
That is, the underlying theory is so fine-tuned that it has {\it two}
distinct scales: $M$ and a lighter scale 
$\aleph = |\alpha/\alpha_c -1| M \ll M$
induced by the fine-tuning.
I will deal with an $S$-wave shallow bound state
by considering the class of theories where  
\begin{equation}
C_{2n}^{(R)}= \frac{4\pi}{m \aleph (M \aleph)^n}\gamma_{2n}, 
\;\; C_{2n}^{'...(R)}= \frac{4\pi}{m M^{2n+1}}\gamma^{'...}_{2n}
\label{2scaling}
\end{equation}
with 
the $\gamma_{2n}^{(...)}$'s dimensionless
parameters of $O(1)$. 
The natural scenario is recovered when $\alpha$
is tuned out of $\alpha_c$, and $\aleph$ becomes comparable to $M$.
The loop and derivative expansions contained in Eq. (\ref{Tpert}) 
are in 
$\sim Q/\aleph$ and
$\sim Q^2/\aleph M$. 

\begin{figure}[t]
\centerline{\psfig{figure=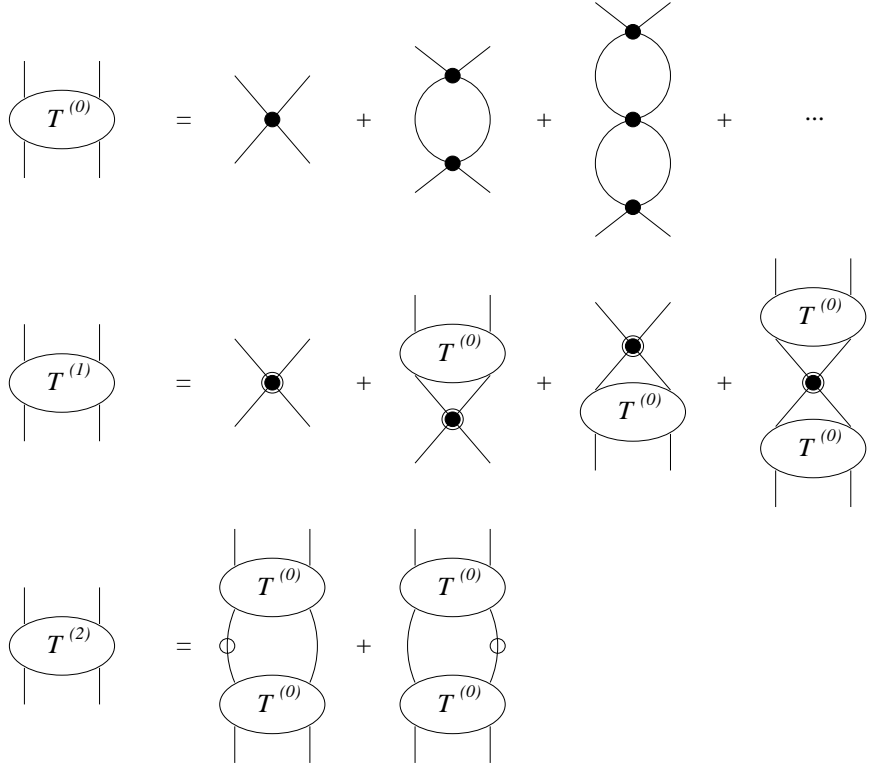,height=4.0in,width=4.50in}}
\vspace{0.5cm}
\caption{First three orders of the
two-particle amplitude $T$ in an EFT with a shallow bound state.
Symbols as in Fig. \protect\ref{fig:natt}.
\label{fig:unnat1t}}
\end{figure}

For $Q\ll \aleph$, the theory is perturbative in $Q/\aleph$.
As $Q$ becomes comparable with $\aleph$,
all the terms in the loop expansion are of the
same order ($O(4\pi/ m \aleph)$) while corrections
from derivatives go in relative powers of
$\aleph/M \ll 1$ and 
can be accounted for systematically, at each order in $\aleph/M$. 
The amplitude is (see Fig. \ref{fig:unnat1t})
\begin{eqnarray}
T_{os}(k, \hat{p}\,'\cdot \hat{p}) & = & 
       T_{os}^{(0)}(k)+  T_{os}^{(1)}(k) +T_{os}^{(2)}(k) +\dots
       +T_{os}^{(3p)}(k, \hat{p}\,'\cdot \hat{p}) +\ldots 
                                                         \nonumber \\
 & = & (T_{os}^{(0)}(k))_0 
       -2C_2^{'(R)} k^2 \hat{p}\,'\cdot \hat{p}
       +O((4\pi/m\aleph)(\aleph/M)^4).              \label{renTonfull}   
\end{eqnarray}
\noindent
The $S$-wave component can be rewritten, up to higher order terms, as
\begin{eqnarray}
(T_{os}(k))_0 & = & [ (T_{os}^{(0)}(k))^{-1} 
  +T_{os}^{(1)}(k) (T_{os}^{(0)}(k))^{-2}
   +T_{os}^{(2)}(k) (T_{os}^{(0)}(k))^{-2} +\ldots]^{-1}
        \nonumber \\
 & = & - \left[ \frac{1}{C_0^{(R)}} -2 \frac{C_2^{(R)}}{(C_0^{(R)})^2} k^2 
         +4\left(\frac{(C_2^{(R)})^2}{(C_0^{(R)})^3}
         -\frac{C_4^{(R)}}{(C_0^{(R)})^2}\right)k^4  \right.    \nonumber \\
 & &   \left.  \ \ \ \ \
       +\frac{imk}{4\pi} \left(1+\frac{k^2}{2m^2}\right)\right]^{-1} 
         \left(1+O(\aleph^4/M^4)\right),
\label{renTon}
\end{eqnarray}
\noindent 
where $C_4$ represents a certain combination of $Q^4$ contact interactions.
By resumming the largest terms,
we have produced a new expansion where the leading amplitude is
$O(4\pi/ m (\aleph+Q))$ and corrections
go as $Q^2/M(\aleph +Q)$ or similar combinations. 
The amplitude (\ref{renTon})
is nominally correct including
$O((4\pi/m\aleph)(\aleph^2/M^2))$ only, but as we will see in a moment it is
actually correct including $O((4\pi/m\aleph)(\aleph^3/M^3))$.
As we consider larger $Q$, $Q\gaprox \aleph$, 
the leading term becomes 
$O(4\pi/ m Q)$  and corrections get relatively more important,
growing in powers of $Q/M$. The new expansion fails at momenta 
$\sim M$, as expected.

The amplitude (\ref{renTonfull}) and (\ref{renTon}) 
is in the form of an effective range expansion
in each partial wave \cite{Bethe,Gold},
\begin{equation}
(T(k))_l =  \frac{4\pi}{mk} 
      \left( 1+ \frac{k^2}{2m^2}+O\left(\frac{k^4}{m^4}\right)\right) ^{-1}
            (\cot \delta_l -i)^{-1},   
         \label{Tondel}
\end{equation}
with
\begin{equation}
k^{2l+1} \cot \delta_l = -\frac{1}{a_2^{(l)}} + \frac{r_2^{(l)}}{2} k^2 
                         +\sum_{n=2}^\infty v_2^{(n)(l)} k^{2n} 
\label{ere}
\end{equation}
In the $S$ wave we obtain for the scattering length and the
effective range
\begin{equation}
a_2^{(0)}= \frac{mC_0^{(R)}}{4\pi},                      
\;\; r_2^{(0)} = \frac{16\pi}{mC_0^{(R)}} 
      \left(\frac{C_2^{(R)}}{C_0^{(R)}}+\frac{1}{4m^2}\right). 
                                                    \label{seffran}
\end{equation}
\noindent
In the $P$ wave, we find
a scattering volume
\begin{equation}
a_2^{(1)}= \frac{mC_2^{'(R)}}{6\pi}.                      \label{pscatlen}
\end{equation}
\noindent
Higher moments $v_2^{(n)(l)}$
can be obtained from the $k^{2n}$ terms that arise in
higher order in the $Q/M$ expansion 
of the amplitude. 
For example, the $S$-wave shape parameter is
\begin{equation}
v_2^{(2)(0)}= -\frac{16\pi}{mC_0^{(R)}} 
          \left(\left(\frac{C_2^{(R)}}{C_0^{(R)}}\right)^2
-\frac{C_4^{(R)}}{C_0^{(R)}}
\right).
                                                    \label{sshaparam}
\end{equation}
An $S$-wave bound state can arise as a pole of 
the amplitude (\ref{renTon})
at $k=i\kappa$,
\begin{equation}
\kappa = \frac{4\pi}{mC_0^{(R)}} 
       \left(1 + 2\frac{C_2^{(R)}}{C_0^{(R)}}
                  \left(\frac{4\pi}{mC_0^{(R)}}\right)^2 
                                  + O((\aleph/M)^2)\right).
\label{boundmom}
\end{equation}
\noindent
$C_0^{(R)} > 0$ ($<0$) implies $\kappa > 0$ ($<0$) and 
represents a real (virtual) bound state. 
(The residue of $i$ times the $S$-matrix
at this pole is indeed positive (negative) if $\kappa > 0$ ($<0$).)
The binding energy $B_2$ 
to this accuracy is, of course, 
just the usual effective range relation among $B_2$, $a_2$ and $r_2$.

We see that an $l(>0)$-wave effective range
parameter of mass dimension $\delta$ 
is 
\linebreak 
$O(1/(2l+1)M^{\delta})$.    
The $S$ wave is trickier.
The scattering length $a_2^{(0)}$ scales with
$\aleph$ rather than $M$, $a_2^{(0)}= O(\aleph^{-1})$.
The $S$-wave bound state is within the range of validity of the EFT,
as $\kappa=O(\aleph)$ and the binding energy
$B_2= \kappa^2/m + O(\aleph^4/M^3)\ll M$.
It is clear that the bound state can be treated in a systematic 
expansion in $\aleph/M$.
The two terms in the 
effective range $r_2^{(0)}$ (\ref{seffran}) come at different orders. 
The main contribution originates in the contact interactions and 
scales with $M$, so that $r_2^{(0)} =O(M^{-1})$. The 
relativistic correction, on the other hand, 
is smaller 
by $\sim M\aleph/m^2\sim \aleph/M$; this 
justifies the neglect of relativistic corrections
in many situations and the usefulness
of a non-relativistic framework in shallow-bound-state problems.
An underlying theory consisting
of a non-relativistic potential can thus serve as a test
for the scalings of $a_2^{(0)}$ and $r_2^{(0)}$. Indeed, it has been explicitly
shown \cite{lekner} that if the underlying theory consists
of a non-relativistic potential of strength $\alpha$ and range $1/M$, 
then near critical binding effective range parameters behave 
precisely in way derived above. 
As for the shape parameter, it appears from Eq. (\ref{sshaparam})
that $v_2^{(2)(0)}\sim 1/\aleph M^2$, but such a scaling is
seen neither in
phenomenological analyses nor in models. This implies that
$\gamma_4= \gamma_2^2/\gamma_0 + O(\aleph/M)$:
although each term in Eq. (\ref{sshaparam})
comes at relative 
$O(Q^4/M^2\aleph^2)$, 
they are correlated and their sum is only of relative 
$O(Q^4/M^3\aleph)$.
We conclude that the shape parameter appears at the same order 
as the first $P$-wave contribution when $Q \gaprox \aleph$. 
Eq. (\ref{renTon}) is therefore good up to 
$O(\aleph^3/M^3)$, as advertised. 
This means that
the first {\it three} orders of the expansion of the amplitude 
---depicted in Fig. \ref{fig:unnat1t}---
are pure $S$ wave and given solely by $C_0^{(R)}$ and $C_2^{(R)}$.
It is easy to verify that,  
in order to avoid $M/\aleph$ enhancements in higher effective
range parameters, there must
be such correlations in higher order coefficients as well:
\begin{equation}
C_{2n}^{(R)}= C_0^{(R)}(C_2^{(R)}/ C_0^{(R)})^n + \mbox{smaller terms}.
\label{corr} 
\end{equation}

We conclude that {\it in the case of short-range interactions,
the EFT approach, when applied consistently, is 
completely equivalent to the effective range expansion}.
The power counting described here and this equivalence
were spelled out in a manifest scheme-independent way in
Ref. \cite{vKolckn}, and in 
specific subtraction schemes in Refs. \cite{Kaplan4,Gegelia1}.

{\bf Dimeron.}
Although not necessary, it is possible to reorganize the EFT in order to 
account explicitly for the dominant correlations among the $C_{2n}^{(R)}$'s.
Because they
form a geometric series in $C_2^{(R)}/C_0^{(R)}$,  
we can sum the dominant correlation terms 
as we have done above with the $C_0^{(R)}$ contributions themselves.
This procedure 
generates a four-$\psi$ interaction of the type
$C_0^{(R)}/(1-2C_2^{(R)}k^2/C_0^{(R)})$;
the first two terms in Eq. (\ref{renTon}) follow immediately. 
Now, this new resummation resembles the exchange of an $s$-channel particle.
This justifies the 
suggestion \cite{david1} 
that one can use a new EFT which involves, besides particles $\psi$,
also a ``dimeron'' field $T$ with the quantum numbers of the shallow
(real or virtual) two-body bound state.  
The leading terms in the 
most general Lagrangian consistent with the same symmetries as before
are
\begin{equation}
{\cal L} =  
    \sigma  T^\dagger\left(i\partial_{0}
                              +\frac{1}{4m}\vec{\nabla}^{2}
                             -\Delta+\ldots\right)T 
    -\frac{g}{\sqrt{2}} (T^\dagger \psi\psi+ h.c.)
        +\ldots                       \label{translag}
\end{equation}
\noindent
Here $g$ and $\Delta$ are parameters and 
$\sigma$ is the sign of the kinetic term.
The Lagrangian (\ref{translag}) is equivalent to (\ref{lag})
if
\begin{equation}
\sigma g^2  = - \frac{C_0^2}{2mC_2},
\;\; \Delta= \frac{C_0}{2mC_2}.
\end{equation}
Thus after renormalization
$(g^{(R)})^2/2\pi =O(M/m^2)$, $\Delta^{(R)}=O(M \aleph/2m)$.
``\ldots'' in Eq. (\ref{translag}) stand for terms with more derivatives,
which are suppressed by powers of $\aleph/M$.
Effects of non-derivative four-$\psi$ contact term
can be absorbed in 
$g^2/\Delta$ and terms with more derivatives.

\begin{figure}[t]
\centerline{\psfig{figure=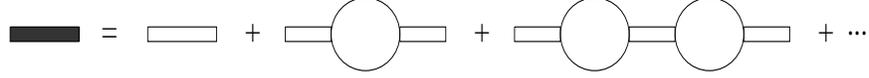,height=0.4in,width=4.50in}}
\vspace{0.5cm}
\caption{The dressed dimeron propagator:
a bar is the bare dimeron propagator, 
while the bubbles represent two-particle propagation.
\label{fig:transprop}}
\end{figure}

The coupling to two-particle states dresses the dimeron propagator.
The dressed propagator contains bubbles as  
in Fig. \ref{fig:transprop}, plus insertions
of relativistic corrections.
This amounts to a self-energy contribution 
proportional to the bubble integral. 
As we know, this integral is ultraviolet 
divergent and requires
renormalization of the parameters of the Lagrangian (\ref{translag}).
Relativistic corrections can be accounted for as in the EFT
without the dimeron.
The dimeron propagator in its rest frame
($p^0\equiv k^2/m -k^4/4m^3+ \ldots$)
has the form
\begin{equation}
S(p^0, \vec{0}\,) = \sigma\frac{(g^{(R)})^2}{g^2} 
      \frac{i}{\frac{k^2}{m}- \Delta^{(R)}
    + \frac{\sigma m (g^{(R)})^2}{4\pi} ik (1+\frac{k^2}{2m^2}) 
    +\ldots +i\epsilon}.
                                   \label{Tprop}
\end{equation}

\begin{figure}[t]
\centerline{\psfig{figure=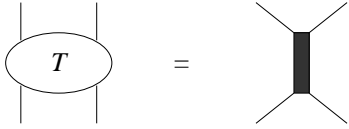,height=0.65in,width=1.8in}}
\vspace{0.5cm}
\caption{The two-particle amplitude $T$ in the EFT with a dimeron field.
\label{fig:transt}}
\end{figure}

The amplitude in this EFT is 
obtained directly from Eq. (\ref{Tprop}) ---see Fig. \ref{fig:transt}.
The result is identical to Eq. (\ref{renTon}) to a given
order.
This is a more direct way to re-derive the
effective range expansion, including the first few orders
at once.
The non-relativistic dimeron includes the first two orders in 
$\aleph/M$ correctly: in leading order only the mass term in the
dimeron kinetic terms needs to be retained, and observables
depend on the combination $(g^{(R)})^2/\Delta^{(R)}$;
in subleading order $g^{(R)2}$ and $\Delta^{(R)}$ appear separately.
The addition of relativistic corrections
regains the first three orders. 
Corrections stemming from higher-derivative operators
can be accounted for perturbatively, and
will contribute to the $\psi \psi$ amplitude starting at
relative $O((\aleph/M)^3)$.
The drawback is a mixing of different orders
in the $\aleph/M$ expansion, with a number
of smaller, higher-order terms included.

The same real or virtual bound state as before exists near zero energy.
The field $T$ corresponds to a light (ghost or normal) 
state of mass $\Delta^{(R)} \sim \aleph$ relative to $2m$.
It serves to introduce through its bare propagator two
bare poles at the $k=\pm \sqrt{m\Delta}$.
This is an undesired pole structure,
but fortunately it changes character as the amplitude gets dressed,
since the poles move 
to the imaginary axis. One ends up at the position
of the shallow bound state, $k=i\kappa_s$, $\kappa_s=\kappa$
given by Eq. (\ref{boundmom}), while the
other pole signals the breakdown of the EFT at $\kappa_d =O(M)$.
The sign of the residue of $-iS$ at each of the poles 
shows that the shallow pole corresponds to a regular (composite) particle,
while the deep pole corresponds to a ghost.
The field $T$ can be used to describe the shallow bound state in processes
other than $\psi\psi$ scattering, if one keeps in mind
that the residue of $-iS$ at the pole is not 1,
but $((g^{(R)})^2/g^2)(\sigma/(1+\sigma m^2 g^2/8\pi\sqrt{mB_2}))$.
Despite the resemblance, this is not an EFT with
an ``elementary'' field for the bound state. 
At $Q\ll \aleph$, the EFT (with or without dimeron) can be matched
onto a lower-energy EFT containing a heavy field for the bound state
which does not couple to $\psi\psi$ states,
but the dimeron field
shows structure at $Q\gaprox \aleph$, which arises from 
$\psi$ loops.

{\bf Schr\"odinger equation.}
Recall that an ordinary potential does not generate 
an $1/r$
behavior of the wave-function close to the origin because action of the 
Laplacian on such a singularity produces a delta function:
$\nabla^2 (1/r) = - 4\pi \delta(\vec{r})$. A local delta-function
potential 
in turn is too singular to admit a solution of this type because then
the potential side of the Schr\"odinger equation becomes
$V \Psi \sim \delta(\vec{r})/r$, more singular than
the kinetic side.
This is the translation into Schr\"odinger language 
of the need for regularization and renormalization of the bubble
diagrams.

The effects of renormalization on the on-shell amplitude are 
two-fold. One is to replace the potential (\ref{ver}) by
\begin{equation}
V^{(R)}(k)= C_0^{(R)}+ 2 C_2^{(R)} k^2 + \dots.             \label{vosR}
\end{equation}
\noindent
The other is to substitute the (free) Schr\"odinger propagator
(\ref{spropcoord}) 
by
\begin{equation}
G_0^{(R)}(r;k)=  -\frac{imk}{4\pi} (1+ O(kr)).            \label{GR}
\end{equation}
It can be shown \cite{vKolckn} that this is equivalent
to the usual Schr\"odinger propagation with an
``effective renormalized potential'' or pseudopotential.
The renormalized Schr\"odinger equation is
\begin{equation}
-\frac{1}{m} (\nabla^2 +k^2) \Psi(\vec{r})= 
 -(C_0^{(R)}+ 2 C_2^{(R)} k^2 +\ldots) \delta(\vec{r})
 \frac{\partial}{\partial r} (r \Psi(\vec{r})).   \label{renSch}
\end{equation}
Recalling the connections 
between EFT and effective range 
parameters, we see that this is nothing but 
a generalization of the well-known pseudopotential of Fermi and Breit 
\cite{Fermi,Breit,Huang}. Therefore,
{\it renormalization of an EFT with only short-range interactions leads 
to a Schr\"odinger equation with a pseudopotential determined by 
renormalized parameters}, rather than 
a Schr\"odinger equation with a local, 
smeared delta-function-type potential determined by bare parameters.

Further insight into the effect of renormalization can be gained
by looking at
the solution of Eq. (\ref{renSch}). 
The effect of the 
$\frac{\partial}{\partial r} r$ operator in the pseudopotential is to
soften the singularity at the origin,  
replacing $\delta(\vec{r})/r$ by $\delta(\vec{r})$. 
Indeed, we can find the solution of Eq. (\ref{renSch}) by expanding
$\Psi(\vec{r}) = \sum_{n=-\infty}^{+\infty} A_n r^n$ at small $r$.
The result is completely determined by $A_{-1}$.
Now, the same solution follows from
\begin{equation}
-\frac{1}{m} (\nabla^2 +k^2) \Psi(\vec{r})= 
    \frac{4\pi}{m} r \delta(\vec{r}) \Psi(\vec{r})
\end{equation}
\noindent
---which for any $\vec{r}\ne 0$ is the free Schr\"odinger equation---
with a peculiar boundary condition at the origin,
\begin{equation}
\left. \frac{\partial}{\partial r} \ln (r \Psi(\vec{r}))\right|_{r=0} =
          -\frac{4\pi}{mC_0^{(R)}} 
      \left(1- 2 \frac{C_2^{(R)}}{(C_0^{(R)})^2} k^2 +\ldots\right). \label{bc}
\end{equation}
We recognize the first term in Eq. (\ref{bc}) as the boundary 
condition of Bethe and Peierls, and Breit 
\cite{Peierls,Breit}. {\it The renormalized EFT of only 
short-range interactions is equivalent to an energy expansion of the most 
general condition on the logarithmic derivative of the wave-function at the 
origin.}
The physics behind these Schr\"odinger formulations is clear if
we look at the resulting radial wave-functions $r\psi(r)=u(r)$,
which obey a free radial equation.
{\it (i)} For $k^2=-\kappa^2<0$,
\begin{equation}
u(r)=c_{(-)} e^{-\kappa r},
\end{equation}
\noindent
with $c_{(-)}$ a normalization factor and $\kappa$ given by 
Eq. (\ref{boundmom}).
{\it (ii)} For $k^2\geq 0$,
\begin{equation}
u(r)=c_{(+)}\left
  (\sin kr -\frac{a_2^{(0)}k}{1-a_2^{(0)}r_2^{(0)} k^2/2+\ldots} 
                             \cos kr\right),
\end{equation}
\noindent
with $c_{(+)}$ a normalization factor and $a_2^{(0)}$ and $r_2^{(0)}$ given in 
Eqs. (\ref{seffran}).
By changing the unspecified dependence of $c_{(+)}$ on $k$ and $\kappa_{s,d}$
we do not change the underlying pole structure of the $\psi\psi$ amplitude,
but we do change the Jost functions. 
This is a reflection of the fact that the EFT cannot distinguish
among short-range potentials which produce the same asymptotic behavior. 
Indeed, it is not difficult to see that with appropriate choices  of $c_{(+)}$
we can cover all situations considered in Ref. \cite{Bargmann}.
The EFT shrinks to the origin the complicated wiggles that the ``underlying''
wave-function might have at small distances, thus replacing
them with a smooth behavior that matches onto the tail of the  underlying
wave-function. The effective wave-function is all tail for the tail
is all one can see from far away.

{\bf Potential and regularization.}
I have so far avoided using specific regularization schemes in order
to emphasize features of the renormalized theory, which is all that
determines observables anyway. 
Now a few words are spent on the related issues of 
the role of potentials and of regularization schemes.
Some of these remarks have been presented in Ref. \cite{vKolckn},
and shown to be supported by a careful numerical
investigation \cite{steele}.

It is clear that iteration of the
potential is superfluous in any partial wave without fine-tuning:
within its region of validity, the EFT
is perturbative,
and any bound states, if they exist, are beyond the range of the
EFT. 
Iteration of the potential includes only part of the higher-order
terms, and this incomplete set cannot {\it a priori} be taken seriously.  
It has been shown explicitly that the iteration
of momentum-independent contact interactions
does not reproduce
the non-analytical behavior of a toy natural underlying theory
\cite{luke2}.
On the other hand, since the error induced by iteration
is small, it does not affect observables much.
The more interesting cases are the ones with fine-tuning
in the underlying theory. 

The appropriate formula for the amplitude, 
Eq. (\ref{renTon}), was obtained
by iterating the $Q^0$ part of the potential to all
orders and by 
using perturbation theory for the corrections.
The same result can 
be obtained more easily by first summing 
the bubbles with the full $V(p',p)$ and then expanding in powers of the 
energy. For example, keeping track only of the $Q^2$ terms 
in the $S$-wave,
\begin{equation}
(T_{os}(k))_0 = -\left(\frac{1}{C_0+2C_2 k^2}- I(k)\right)^{-1},  
\;\; I(k) = \left(1+\frac{k^2}{2m^2}\right) I_0(k) 
           - \left(2\frac{C_2}{C_0}+\frac{1}{4m^2}\right)
             \frac{m L_3}{2\pi^2}
\label{Ton}
\end{equation}
\noindent
This corresponds to the standard procedure of iterating the full potential,
but
it is now obvious that to remove the infinities we {\it have to} expand
\begin{equation}
\frac{1}{C_0+2C_2 k^2}= \frac{1}{C_0}-2 \frac{C_2}{C_0^2} k^2+\ldots, 
\label{exp}
\end{equation}
\noindent
and consistently neglect $Q^4$ terms. 
We are then back to Eq. (\ref{renTon}) truncated at $O(\aleph/M)$.
Terms of $O(Q^4)$ and higher can only be
made regulator-independent by renormalizing higher-order 
parameters in the Lagrangian. Indeed, as we have seen before,
$O(Q^4)$ terms in $V(p,p')$
will contribute $k^4$ terms to Eq. (\ref{Ton}).

Regulator-dependence is allowed insofar as it is small, that is,
of the same magnitude as smaller terms that have been neglected.
Cutoff dependence may arise from two sources.
Some cutoff dependence 
comes from $R(k^2/\Lambda^2)$, for it contains
arbitrarily high powers of $Q/\Lambda$. 
Further cutoff dependence
creeps in through bare parameters if we iterate
the full potential to all orders but neglect to make the
expansion (\ref{exp}). 
For example, to $O((\aleph/M)^2)$, 
by solving the Schr\"odinger equation with a finite cut-off
we induce a shape parameter
\begin{equation}
v_2^{(2)(0)} = -\frac{16\pi C_2^2}{mC_0^3}
            -\frac{2}{\pi\Lambda^3}
             \left( R'(0) +\frac{\Lambda^2}{2m^2}R(0)\right)
            -\frac{r_2^{(0)}}{8m^2}.
\label{fakeshape}
\end{equation}
\noindent
We see that apart from a regularization-independent relativistic correction,
cutoff dependence enters through both 
$R$ and the bare parameter ratio $C_2^2/C_0^3$.
It is natural to ask what are the constraints on regularization
procedures imposed by the requirement that both
sources of cutoff dependence are small,
as in more complicated situations
it might be difficult to carry the step (\ref{exp}) explicitly.
Note that the EFT theory with a dimeron field automatically
performs the step  (\ref{exp}), and therefore
reduces differences among regularization schemes to the $R$ they generate.

In a cutoff regularization, the form of 
$F(l^2/\Lambda^2)$ is chosen explicitly.
The error induced by $R$ is removed order-by-order,
the relative residual error being $O(\aleph/\Lambda)$.
Finite cut-offs have been strongly advocated in Refs. \cite{lepage,Lepage}.
Keeping a cutoff $\Lambda\sim M$ has one useful consequence. 
In this case, the large 
renormalized parameters $C_{2n}^{(R)}$ can be generated by
natural-size bare coefficients
$C_{2n}=O(4\pi /m M^{2n+1})$.
For example, 
\begin{equation}
C_0(\Lambda)=
-\frac{2\pi^2}{\theta_1 m \Lambda} (1+ \pi \aleph/2\gamma_0 \theta_1 \Lambda
+\ldots)
\label{Clambda}
\end{equation}
is of natural size
even though
$C_0^{(R)}=4\pi\gamma_0/m \aleph$ is large.
What a cutoff does is to give a scale for the bare parameters to be 
fine-tuned against.
Before renormalization each loop contributes
a large factor $\propto \Lambda$ but each $C_0$ compensates
with a small factor $\propto \Lambda^{-1}$, resulting a net
$\sim (1+\pi (\aleph+Q)/2\theta_1 \Lambda +\ldots)$.
The non-perturbative character of the theory is evident:
it generates cancellations among the leading terms,
and the relevant physics is contained in the terms that are suppressed
by $\aleph/\Lambda$ or $Q/\aleph$.
These remarks can be couched in a renormalization group analysis
for the potential \cite{birse}.

In this way the ratio of bare parameters can be made of natural
size as well. The difference between a truncation of the rhs of
Eq. (\ref{exp}) and the lhs is then no bigger than other neglected
small effects such as higher-order interactions in the Lagrangian: 
for example, the shape parameter (\ref{fakeshape}) induced
by the cutoff is in this case $O(1/M^3)$.
{\it A power expansion of the bare potential followed
by its iteration to all orders is under these circumstances no worse
than a perturbation treatment of the corrections
to leading order.} 
In this case the power counting (\ref{powercount}) 
applies to the potential
itself: since here the potential is a tree-level object involving
four fermion fields
\begin{equation}
\nu = d,            \label{treenu}
\end{equation}
where $d$ is the number of derivatives at the vertex.
Truncating the potential at certain order
accounts correctly 
for the first few terms in the expansion of the amplitude but
includes also a subset of higher-order terms. 
We can actually invert the previous reasoning and use the freedom
introduced by a finite cutoff to save some work. 
Given a regularization procedure, we can further fine-tune $\Lambda$ 
to its ``optimal" value that improves agreement with data.
If we are dealing with a single channel, 
this is a consistent procedure: it simply means 
that $O(Q^4)$ contributions have been included, but they do not need
to be written explicitly because the corresponding bare parameter is zero.
In general, however, only part of the higher order effects can be accounted 
for in this manner.
As $\Lambda \rightarrow \infty$ the size of implicitly kept higher-order 
terms decreases. For example, all terms in Eq. (\ref{fakeshape}) 
---apart from the finite relativistic correction--- then vanish. 
A consequence of this, noted in Ref. \cite{cohenOK}, is that
the first two terms in $V(p,p')$ are renormalizable
in such non-perturbative context.

To any given order, equations 
relating bare and renormalized parameters will be truncated,
and beyond leading order, be highly non-linear ---see
Eqs. (\ref{C2'R}), for example.
Inverting them, we can find the bare parameters in terms of
the renormalized parameters (known from data) and the cutoff:
$C_{2n}=C_{2n}(C_{2n}^{(R)}; \Lambda)$.
This of course means that the bare parameters are not observables,
depending on the scheme chosen. 
Given $C_{2n}^{(R)}$'s,
whether one finds real solutions $C_{2n}$ for any $\Lambda$ depends on the 
actual values of the $\theta_{n}$'s in Eq. (\ref{regbubbleany}).
In particular, 
to $O((\aleph/M)^2)$ the important term for large $\Lambda$
is $C_2^{(R)}\theta_3\Lambda^3$; 
the sign of $C_2^{(R)}$ being that of
$r_2^{(0)}$, it is the sign of $\theta_3$ which is of concern.
It was pointed out in Ref. \cite{cohen} 
that the most obvious cutoff schemes
(such as a sharp momentum-space cutoff and a coordinate-space
regularization by square wells) imply $\theta_3 >0$.
Imposing real bare parameters in such schemes, it is found that
$r_2^{(0)} \leq 2/\Lambda$ 
for large $\Lambda$, or in other words, that the limit
$\Lambda \rightarrow \infty$ can only be taken if $r_2^{(0)}\leq 0$.
This is a reformulation in  field-theoretical language of 
Wigner's theorem \cite{Wigner}:
$r_2^{(0)}$ cannot be positive if we
take a sequence of ever shorter-range, (possibly non-local) hermitian
potentials in a coordinate space Schr\"odinger equation.
While such a constraint is unusual, it is doubtful that it is of much 
relevance to EFTs applied consistently to a certain order. 
First, this is obviously a regularization-scheme-dependent issue:
it is easy to construct (smooth) cutoff schemes such that 
$\theta_3 \leq 0$ \cite{david3}. 
In the worst case scenario, 
the above constraint is a result of attempting to remove the
cutoff in an inadequate 
regularization scheme. 
Second, this constraint only arises from 
a constraint imposed on unobservable bare parameters. No matter
what process one is considering in the low-momentum regime, 
the bare parameters never appear, except in the right 
combinations with the cutoff that produce renormalized parameters:
an imaginary bare parameter is of no relevance.
Indeed, the Schr\"odinger equation for the 
``renormalized potential'' (\ref{renSch}) 
violates the assumptions made in Ref. \cite{cohen} 
about the nature of the potential. 
Finally, even if one insists on the unfortunate choice of regulator 
and the reality constraint on bare parameters, a positive $r_2^{(0)}$ 
can still 
be achieved by keeping the cutoff finite and of the order of the 
mass scale of the underlying theory, $\Lambda\sim M$. 

An alternative regularization is dimensional regularization.
$F$ is implicit
when we extend the dimension of space to a sufficiently 
small $D-1$ such that the integral converges;
the scale introduced is usually denoted $\mu$.
With minimal subtraction, 
$\theta_n=0$ and $R(x)=0$.
A non-minimal subtraction (``PDS'') scheme was introduced
in Ref. \cite{Kaplan4}, in which a pole in $D=3$ was subtracted.
This particular scheme corresponds to
$\theta_1=\pi/2$, $\theta_{n> 1}=0$, and $R(x)=0$.
For $\mu=0$ this scheme reduces to minimal subtraction.
It is not difficult to relate parameters
in subtraction and cutoff schemes.
If we call the cutoff parameter 
$\Lambda= \bar{\Lambda}(\mu) +\pi\mu/2\theta_1$, we can write
\begin{equation}
I_{2n}(k) = -\frac{m}{2\pi^2}
    \left[\bar{\Lambda}\sum_{i=0}^{n} k^{2i} \varpi_{n-i}(\bar{\Lambda},\mu)
                                + \frac{\pi}{2}k^{2n} (ik+\mu)
      +\frac{k^{2(n+1)}}{\bar{\Lambda}} 
    \bar{R}(k^2/\bar{\Lambda}^2,\mu/\bar{\Lambda})\right], 
\label{runbubbleany}
\end{equation}
\noindent
where $\varpi_{n-i}(\bar{\Lambda},\mu)$ and 
$\bar{R}(k^2/\bar{\Lambda}^2,\mu/\bar{\Lambda})$
are related to the $\theta_n$'s and $R(k^2/\Lambda^2)$.
As we have seen, terms that grow with the cutoff can be cancelled
by the counterterms $C_{2n}(\Lambda)$. If one defines
new parameters  $C_{2n}(\mu)$ as $\bar{\Lambda}$-independent
functions of the $C_{2n}(\Lambda)$ and the $\varpi_{n-i}(\bar{\Lambda},\mu)$,
we can express the renormalized quantities $C_{2n}^{(R)}$
as $\mu$-independent functions of the $C_{2n}(\mu)$ and $\mu$. 
For example, in leading order 
\begin{equation}
\frac{1}{C_{0}(\mu)}=\frac{1}{C_{0}(\Lambda)} 
                     + \frac{m \theta_1 \bar{\Lambda}}{2\pi^2}.
\label{Cmu}
\end{equation}
For $\mu,k\ll \bar{\Lambda}$
contributions from $\bar{R}$ are small;
in the limit $\bar{\Lambda} \rightarrow \infty$ (with $\mu$ fixed)
they vanish and the theory with cutoff can be rewritten in terms of
the $C_{2n}(\mu)$ and
\begin{equation}
I_{2n}(k) = -\frac{m}{4\pi}k^{2n} (ik+\mu), \label{ranbubbleany}
\end{equation}
\noindent
which is equivalent to the theory with 
integrals subtracted at $k^2=-\mu^2$ 
\cite{inwei6,Gegelia1}.
Other subtraction schemes in dimensional regularization
have been discussed in Ref. \cite{scheming}.

Dimensional regularization
suffers from no errors generated by $R$.
Minimal subtraction,
first applied to this problem
in Ref. \cite{Kaplan2}, is special because 
it forces the bare parameters to coincide with
the renormalized ones and be unnaturally large.
Because no cutoff dependence is visible,
if one works to $Q^2$ terms it is not immediately apparent that  
$1/(C_0 + 2C_2 k^2)$ has to be expanded in $k^2$. 
It was one of 
the results of Ref. \cite{Kaplan2} that a fitting of the $^1S_0$ $NN$ phase 
shift based on the effective range expansion worked much better than
one based on the form $1/(C_0 + 2C_2 k^2)$.
One can understand the failure of the $1/(C_0 + 2C_2 k^2)$ fit by noticing that
the induced spurious shape parameter (\ref{fakeshape}) 
is here 
$v_2^{(2)(0)}= a_2^{(0)} r_2^{(0)2}/4 -3r_2^{(0)}/8m^2 +1/4a_2^{(0)}m^4$,
which is $O(1/\aleph M^2)$. 
Since there is no regulator to blame the fine-tuning on, 
the fine-tuning contaminates
all other induced terms as well:
they are all automatically large if $a_2^{(0)}$ is large.
In dimensional regularization with
minimal subtraction one {\it cannot} substitute the perturbative treatment
of corrections by the iteration of the whole potential without
reducing the range of applicability of the theory 
from $M$ to the ratio of parameters in the power expansion
of the contact interactions, which is $\sqrt{\aleph M}$.
The perturbative treatment of corrections can still be carried out
with dimensional regularization,
or equivalently one can use the EFT with a dimeron field. 
This is true of 
minimal subtraction or other subtraction schemes.
Of those, PDS and other schemes with a small subtraction scale
are particularly convenient.
For example, from Eqs. (\ref{Clambda}) and (\ref{Cmu}) we see that
natural-size bare coefficients translate into
\begin{equation}
C_{0}(\mu)=\frac{4\pi}{m(-\mu +\aleph/\gamma_0)}.
\end{equation}
$C_{0}(\aleph)=O(4\pi/m\aleph)$
is a large running coupling.
One can show that other $C_{2n}(\mu)$ are also of the
same size as their renormalized counterparts $C_{2n}^{(R)}$,
and that they satisfy relations analogous to
(\ref{corr}).
We see that the power counting valid for the renormalized
theory is manifest 
when the theory is formulated directly in terms of a 
small mass scale $\mu$ 
and running parameters $C_{2n}(\mu)$ \cite{lutz,Kaplan4}.

{\bf Spin and isospin.}
Generalization to include spin $S$ and isospin $I$
is straightforward. 
In $NN$ scattering, $S,I=0,1$. The Pauli principle requires
$I+S+L$ be odd, where $L$ is the orbital angular momentum.
Channels are labeled by $^{2S+1}L_J$, where $J$ is the total
angular momentum.
There can be mixing between $S=1$ channels with
$L=J-1$ and $L=J+1$, generated by the tensor operator
$S_{12}(q)\equiv 3\vec{q}\cdot\vec{\sigma}_{1}\vec{q}\cdot\vec{\sigma}_{2}/q^2
-\vec{\sigma}_{1}\cdot\vec{\sigma}_{2}$.
The corresponding amplitude can be expressed in terms of 
a mixing angle $\varepsilon_J$,
\begin{equation}
(T(k))_{J-1;J+1} = -i \frac{2\pi}{mk} 
      \left( \begin{array}{cc}
e^{2i\delta_{J-1}}\cos 2\varepsilon_J-1 & 
ie^{i(\delta_{J-1}+\delta_{J+1})}\sin 2\varepsilon_J \\
\;\; ie^{i(\delta_{J-1}+\delta_{J+1})}\sin 2\varepsilon_J & 
\;\; e^{2i\delta_{J+1}}\cos 2\varepsilon_J-1 
             \end{array} \right).
\label{Tondelmix}
\end{equation}

One can use Fierz reordering to reduce the number of independent
contact interactions 
to two $Q^0$ terms \cite{inwei6}, 
seven $Q^2$ terms \cite{ciOvK}, and so on: 
in terms of $\vec{q}\equiv\vec{p}-\vec{p}'$
and $\vec{k}\equiv\frac{1}{2}(\vec{p}+\vec{p}')$, 
one possible choice is \cite{invk,ciOLvK}
\begin{eqnarray}
V(p,p')&=& C_0^{(S)}+C_0^{(T)}\vec{\sigma}_{1}\cdot\vec{\sigma}_{2}
         +C_2^{(1)}\vec{q}\,^{2}+C_2^{(2)}\vec{k}^{2}+
           (C_2^{(3)}\vec{q}\,^{2}+
           C_2^{(4)}\vec{k}^{2})\vec{\sigma}_{1}\cdot\vec{\sigma}_{2} 
                                              \nonumber \\
       &   & +iC_2^{(5)}\frac{\vec{\sigma}_{1}+\vec{\sigma}_{2}}{2}\cdot
             (\vec{q}\times\vec{k})+C_2^{(6)}\vec{q}\cdot\vec{\sigma}_{1}
             \vec{q}\cdot\vec{\sigma}_{2}   
          +C_2^{(7)}\vec{k}\cdot\vec{\sigma}_{1}\vec{k}\cdot\vec{\sigma}_{2}
             +\ldots
                                     \label{ci17}
\end{eqnarray}
The two $Q^0$ terms and two of the $Q^2$ terms 
contribute to the two $S$-wave channels, $^3S_1$ and $^1S_0$.
The remaining five $Q^2$ terms contribute to
$^3S_1-^3D_1$ mixing and to the four $P$-wave channels,
$^1P_1$, $^3P_0$, $^3P_1$, and $^3P_2$.
At higher powers of momenta, 
contributions to $D$ and other waves appear.

Since there are shallow bound states in both $S$-wave channels, 
QCD must belong to the universality class of EFTs
that obey the scaling (\ref{2scaling}) with $M=M_{nuc}$.
The only complication arises from the mixing 
generated by the $C_2^{(6)}$ term.
Analogously to the effective range expansion, we can expand
\begin{equation}
\frac{1}{2}\sin 2\varepsilon_1= \frac{k^3a_2^{(0)}}{\sqrt{1+k^2a_2^{(0)2}}}
                \sum_{n=0}^\infty u_2^{(n)(1)}(ka_2^{(0)}) \; k^{n}, 
\label{eremix}
\end{equation}
where the $u_2^{(n)(1)}(ka_2^{(0)})$'s are certain functions
of the combination $ka_2^{(0)}$ (but not of $k$ or $a_2^{(0)}$ separately).
Models suggest that 
$u_2^{(n)(1)}(ka_2^{(0)})=O(1/M_{nuc}^{(2+n)})$.
This expansion is reproduced by the EFT provided
\begin{equation}
C_2^{(6)(R)}= \frac{4\pi}{m \aleph M^2} \gamma_{2}^{(6)} 
\end{equation}
with $\gamma_{2}^{(6)}$ dimensionless of $O(1)$. 
For example,  
$u_2^{(0)(1)}\propto C_2^{(6)}/(C_0^{(S)}+C_0^{(T)})\sim 1/M_{nuc}^2$.
Such an argument can be generalized to higher orders. 
A local operator that connects a state with angular momentum $l$ to
another with angular momentum $l'$ involves at least
$l+l'$ derivatives and first appears at
$O((Q/M_{nuc})^{l+l'})$ if either $l$ or $l'$ is zero,
or at  $O((Q/M_{nuc})^{l+l'+1})$ if both  $l$ and $l'$ are non-zero,
relative to the lowest
($O(4\pi/m\aleph)$) contribution (from $l=l'=0$). 

Power counting then indicates that the low-energy phase shifts should be larger
in the $S$ waves, with the $Q^0$ terms being treated non-perturbatively.
The other waves, $P$, $D$, $F$, and so on, 
should be 
amenable to perturbation theory up to ever
increasing energy, as we have seen that the suppression factor
is $\sim 1/(2l+1)M_{nuc}$ in an $l$ wave.
We also expect the 
waves to be   
roughly in decreasing order of magnitude as orbital angular
momentum increases, 
as they receive contributions at increasingly higher orders.
These conclusions are qualitatively substantiated by a PSA
\cite{nijmanal}.

\begin{figure}[t]
\centerline{\psfig{figure=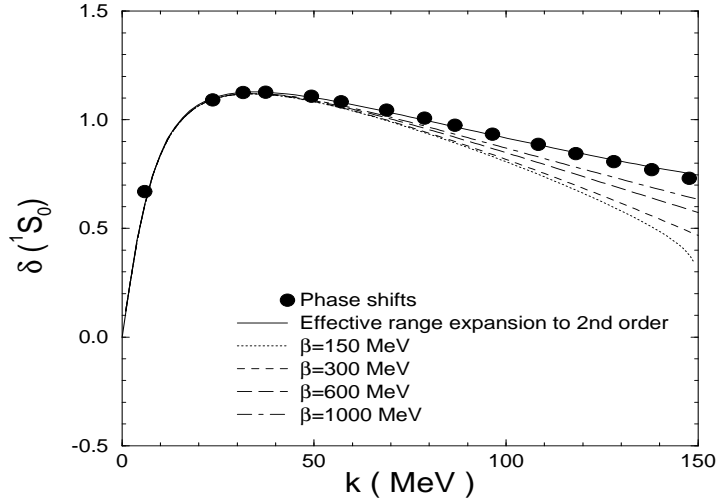,height=3in,width=4in}}
\caption{$^1S_0$ $NN$ phase shift in radians as function
of the center-of-mass momentum in the EFT
to $O(Q/M_{nuc})$ for several values
of the cutoff $\beta$ (curves), compared to the
Nijmegen phase shift analysis (circles). 
Figure courtesy of D. Phillips. 
\label{fig:cutoff}}
\end{figure}

Up to relative $O(Q/M_{nuc})$ only $S$ waves are present.
In each $S$-wave channel we have one non-derivative interaction
at leading order and one two-derivative interaction at subleading order, 
which we can fit to the scattering length
and to the effective range, respectively.
{}From the singlet parameters 
$a_{pp}^{(^1S_0)}=-17.3$ fm, $a_{np}^{(^1S_0)}=-23.75$ fm, 
$a_{nn}^{(^1S_0)}=-18.8$ fm,
$r_{pp}^{(^1S_0)}=2.85$ fm, $r_{np}^{(^1S_0)}=2.75$ fm, 
and $r_{nn}^{(^1S_0)}=2.75$ fm \cite{jerry} we find the averages 
\begin{equation}
C_0^{(R)(^1S_0)}= 1.3 \cdot 10^{-3} \mbox{MeV}^{-2},
\;\; C_2^{(R)(^1S_0)}= 4.5 \cdot 10^{-7} \mbox{MeV}^{-4}.
\end{equation}
{}From the triplet parameters 
$a_2^{(^3S_1)}= 5.42$ fm and $r_2^{(^3S_1)}=1.75$ fm
\cite{nijm} we find 
\begin{equation}
C_0^{(R)(^3S_1)}= 3.6 \cdot 10^{-4} \mbox{MeV}^{-2}, 
\;\; C_2^{(R)(^3S_1)}= 2.1 \cdot 10^{-8} \mbox{MeV}^{-4}.
\end{equation}
Taking a channel-average $\aleph=30$ MeV and $M_{nuc}=140$ MeV, these
translate into reasonable
$\gamma_0^{(^1S_0)}\simeq 2.9$, 
$\gamma_2^{(^1S_0)}\simeq 4.2$,
$\gamma_0^{(^3S_1)}\simeq 0.8$, and
$\gamma_2^{(^3S_1)}\simeq 0.2$.
One can then predict the phase shifts as functions of energy.
As an example, in Fig. \ref{fig:cutoff} 
we see the resulting $^1S_0$ phase shift
as function of $k$ for a number
of cutoffs $\Lambda=\beta$ \cite{dan}, compared to the
values of the Nijmegen PSA \cite{nijmanal}.
Cutoff dependence is significant only at 
momenta of $O(m_\pi)$ where this EFT should fail.
The effective range expansion, which corresponds
to the $\Lambda\rightarrow \infty$ limit, agrees well with the
PSA up to such momenta.
Similar conclusion holds for the $^3S_1$ phase shifts.
In the case of the $^3S_1$ channel, we can alternatively
use the deuteron binding energy to determine $C_0^{(R)(^3S_1)}$, 
the fitting difference being of higher order. 
This is a more natural strategy for studying processes that
involve the deuteron. 

At relative $O((Q/M_{nuc})^{2})$, mixing 
between $^3S_1$ and $^3D_1$ states appears
thanks to $C_2^{(6)(R)}$, and
relativistic corrections need to be included in $S$ waves
\footnote{The latter are numerically small, however, and
can usually be neglected.}. 
$C_2^{(6)(R)}$ can be determined from the low-energy behavior of
$\epsilon_1$ and then deuteron parameters can be predicted,
or {\it vice-versa}.
Using standard formulas in terms of integrals over wave-functions,
deuteron properties 
found with $C_2^{(6)(R)}$ fitted
to a deuteron asymptotic $D/S$ ratio $\eta=0.025$
(as in the Argonne V18 potential \cite{argonne})
are given in Table \ref{tab:taedparam} \cite{taesun2papers}.
(Electromagnetic quantities refer to the contributions
from lowest-order $\gamma NN$ couplings only,
not to a consistent calculation which would include
subleading one- and two-nucleon effects.)

\begin{table}[tb]
\caption{Results from EFT fits to subleading order
for various cut-offs $\Lambda$
and experimental values 
for the deuteron binding energy ($B$), magnetic dipole moment ($\mu_d$), 
electric 
quadrupole moment ($Q_E$), asymptotic $D/S$ ratio ($\eta$), and $D$-state 
probability ($P_D$).
\label{tab:taedparam}}
\vspace{0cm}
\begin{center}
\footnotesize
\begin{tabular}{ccccc}
Deuteron   & \multicolumn{3}{c}{$\Lambda$ (MeV)} &            \\
\cline{2-4}
quantities & 150 & 216 & 250                   & Experiment \\
\hline
$B$ (MeV)  & 1.799 & 2.211 & 2.389                 & 2.224579(9) \\
$\mu_d$ ($\mu_N$) & 0.868 & 0.846 & 0.828       & 0.857406(1) \\
$Q_E$ (fm$^2$) & 0.231 & 0.288 & 0.305          & 0.2859(3)   \\
$\eta$     & 0.025 (fit) & 0.025 (fit) & 0.025 (fit)     & 0.0271(4)  \\
$P_D$ (\%) & 2.11 & 5.89 & 9.09                 &            \\
\end{tabular}
\end{center}
\end{table}

The theory can be extended to higher orders in obvious fashion.

{\bf Electromagnetic corrections.}
Because of the non-relativistic nature of heavy particle propagation,
electromagnetic interactions are dominated by Coulomb-photon exchange.
One-Coulomb-photon exchange is $O(e^2/Q^2)$, while,
using the power counting developed above,
iterated once it 
is $O(m e^4/4\pi Q^3)$. Therefore, Coulomb-photon exchange is 
(non-) perturbative for $Q\gaprox \alpha m$ ($Q\saprox \alpha m$).

In the case of nucleons, $\alpha m_N\simeq 7$ MeV,
which almost coincides with the light scale in the $S$-wave channel
where $pp$ scatter, $\aleph^{(^1S_0)} \sim 8$ MeV;
that is, numerically, $\alpha\sim \aleph^{(^1S_0)}/m_N$.
For $Q\ll \aleph^{(^1S_0)}$, Coulomb interactions
have to be iterated to all orders, while we can treat 
strong interactions generated by the potential (\ref{ci17})
as perturbative insertions.
Conversely, for $Q \gg \aleph^{(^1S_0)}$,
we can add electromagnetic interactions as perturbations.

At $Q\sim \aleph^{(^1S_0)}$, both one-Coulomb-photon exchange
and the momentum-independent contact interaction 
are of the same order and need to be iterated to all orders.
Corrections can  
most simply be classified taking 
$\alpha\sim \aleph^{(^1S_0)}/M_{nuc}$,
and then recognizing that an extra numerical suppression
affects electromagnetic corrections 
because $M_{nuc}/m_N \ll 1$.
At $O(\aleph^{(^1S_0)}/M_{nuc})$, besides
the two-derivative contact interaction,
the most important electromagnetic correction appears due to
vacuum polarization, as it comes
suppressed by a single power of $\alpha$.
(Actually, the more typical suppression of photon loops is $\alpha/\pi$,
reducing further the impact of this effect.)
At $O((\aleph^{(^1S_0)}/M_{nuc})^2)$, besides 
relativistic corrections to the $S$ wave,
transverse-photon exchange generates magnetic interactions.
More exotic electromagnetic interactions 
---such as two-photon exchange generated by the Thomson seagull---
appear in higher orders.
This ordering of electromagnetic interactions 
\cite{pauloandme} is in qualitative
agreement with the findings of the Nijmegen PSA
\cite{Rob}.

The calculation 
of $pp$ scattering including Coulomb interactions to all orders
consists of dressing the bubble diagrams in 
Fig. \ref{fig:natt} with Coulomb exchanges
\cite{pauloandme,ravndal}. 
This gives rise to integrals of the type (\ref{regbubbleany}) but with
the free Schr\"odinger propagator (\ref{spropcoord}) replaced by the 
Coulomb propagator 
\begin{equation}
G_{0C}(l;k)= - 
\frac{2\pi \alpha m_N/l}{\exp (\alpha m_N/l) -1} \; \;
\frac{m_N}{l^2 -k^2 -i\epsilon}.         
\label{coulombprop}
\end{equation}
Dressing of the external legs gives rise to the usual Sommerfeld
factor.
In leading order, 
the amplitude can still be summed as a geometric series
in terms of a single bubble integral. In addition to the linear
cutoff dependence present even in the absence of electromagnetic 
interactions, a new cutoff dependence 
$\alpha m_N^2 \ln (\Lambda/ \alpha m_N)/\pi$ appears.
This requires a new leading-order, no-derivative four-nucleon interaction
\begin{equation}
C_{0em}^{(R)}= \frac{4\pi\alpha}{\aleph^2}\gamma_{0em}
\end{equation}
to be added to $C_{0}^{(R)(^1S_0)}$, which can be fitted 
to the difference $a_{pp}-a_{nn}$. In this EFT the splitting
in scattering lengths cannot be calculated
without matching to the underlying theory.
Effective-range ---but no vacuum polarization--- effects 
were considered in Refs. \cite{ravndal},
where it was shown that to subleading order 
$r_{pp}=r_{nn}$.
Effects of vacuum polarization and two-photon exchange
are discussed in Refs. \cite{taesunburning,latestrho}.

{\bf Moral.} 
The modern EFT method applied to the problem of 
short-range forces is, {\it after 
renormalization}, completely equivalent to the ancient effective range 
expansion, and to the methods of pseudopotentials and boundary conditions.
Why, then, bother with the EFT method?
The gain is indeed marginal for the problem at hand, but there are
advantages to a field-theoretical framework when other processes in
the same energy scale are considered, and we want to
treat them all consistently, free of off-shell ambiguities. 
(We will see an example in Sect. \ref{sec-3N}.)
Yet the gain is potentially 
much more significant in more general cases where there are other, lighter 
degrees of freedom that generate longer-range forces. 
Symmetries are easily incorporated in Lagrangian field theory.
The problem considered in this Section has very 
weak symmetry constraints. In more 
interesting problems, such as the $NN$ system at energies 
comparable to the pion mass, the role of symmetry is more important,
and the EFT method more convenient than the ancient techniques.
We turn to this case next.

\subsection{Low energies} \label{subsec-2Nlow}

As $Q$ approaches $m_\pi$, it becomes increasingly difficult
to account for pion exchange as a short-range effect.
As we further increase momenta past $Q\sim m_\pi$, 
we have to include an explicit pion field
and build up all its interactions allowed by symmetries.
The important new element is chiral symmetry, as discussed in 
Sect. \ref{subsec-sym}.
Adding pions will generate all sorts of non-analytic contributions
to the amplitude.
For $Q \saprox m_\pi$, pions {\it need not} but {\it can}
be introduced explicitly, 
bringing non-analytic terms to $k \cot \delta$.
However, these pionic terms can be expanded in a power series,
and none of the terms with powers higher than the retained contact 
terms can be taken seriously, as neglected
higher-order contact interactions could produce contributions of 
similar size.
(Similar remarks hold for deuteron properties,
as the deuteron is so dilute that it is even within the realm 
of the pionless EFT.)
Nonetheless, one might hope 
to include at least part of the pion effects in this (uncontrolled) way.
Indeed, low-energy
properties seem to improve and cutoff dependence to decrease if
one-pion exchange (OPE) is added to the contact potential at $O(Q/M_{nuc})$
(\ref{ci17}),
which is then iterated to all orders \cite{latestrho}. 
We want to go further and devise a rationale for a controlled
expansion in the presence of pions, valid for momenta $Q\sim M_{nuc}$.
The issue arises immediately of how to estimate the
size of pion contributions. One needs to find
{\it (i)} the importance of pion exchange relative to short-range
interactions;
{\it (ii)} how iterated OPE compares to OPE.

The OPE contribution to the $NN$ amplitude is,
as we will see in more detail below, 
proportional to $Q^2/M_{nuc}^2 (Q^2+M_{nuc}^2)$.
On the other hand, as we have just seen,
leading short-range effects are proportional to $4\pi/M_{QCD} \aleph$.
Thus, as long as $\aleph \saprox M_{NN}$
with $M_{NN}\sim 4\pi M_{nuc}^2/M_{QCD}$,
OPE should be suppressed compared to the leading contact interaction 
by $\aleph/M_{NN}$.
Using power counting arguments again, once-iterated pion exchange  
can give a contribution
$\sim (Q^2/M_{nuc}^2 (Q^2+M_{nuc}^2))^2 (M_{QCD} Q/4\pi)$,
and it also looks like a small correction to OPE as long as
$Q\saprox M_{NN}$. 
Of course, as in the lower-energy EFT, we expect all contributions to
higher-$l$ partial waves to be perturbative 
at least up to momenta 
\linebreak 
$\sim (2l+1)M_{nuc}$.
With such estimates, we conclude that
at sufficiently small $Q$ pions might be treated perturbatively.
The suggestion that
pions can be profitably treated as perturbative even in $S$ waves
and a power counting to support it
were made in Ref. \cite{lutz}
in connection with a more complicated formalism,
but details have not been published. 
Ref. \cite{Kaplan4} gave the first clear
implementation of these ideas.

It is important to find out whether 
this can be done for momenta above $m_\pi$.
Only if $M_{NN}\gaprox m_\pi$ are perturbative pions
an improvement over the effective range expansion.
Moreover, the full power of chiral symmetry in constraining
pion interactions relies on an ordering of chiral-symmetry-breaking 
effects in powers of $m_\pi^2$.
Including pions perturbatively brings terms that are {\it a priori}
suppressed only by powers of $m_\pi/M_{NN}$. Chiral symmetry will be 
really useful only if this ratio is smaller than 1.
The problem is that 
$M_{NN}$ 
is not obviously larger than $M_{nuc}$;
whether or not it actually is  
depends on dimensionless numbers that cannot be accounted for with
dimensional analysis alone. 
The issue of the range of validity of 
the EFT with perturbative pions can only be settled by explicit calculation
of dimensionless factors and comparison with precise
observed quantities, this being done to sufficiently
high order that a significant number of pion effects is tested. 

Let us thus assume $\aleph < M_{NN}$, $Q < M_{NN}$,
and $m_\pi < M_{NN}$.
A simple power counting can be done by taking  
$Q\sim \aleph \sim m_\pi < M_{NN}$, and counting powers
of the light scale $Q$.
This is a direct extension of the power counting in the pionless EFT
in Sect. \ref{subsec-2Nverylow}. 
In particular, the scaling  of the
contact operators is assumed to be the same as before 
with $M=M_{NN}$, and thus their ordering is unchanged.
Because of chiral symmetry, each insertion of a pion exchange brings
a factor of $Q/M_{NN}$.
Electromagnetic interactions can be treated much as in the previous
section.

If $M_{NN}$ is sufficiently high,
we need to include the delta isobar as well.
Beyond contact interactions
considered previously in Eq. (\ref{ci17}), 
and the pion-nucleon
interactions in the one-nucleon sector shown in
Sect. \ref{sec-01N}, 
there are new interactions involving 
four fermion fields where at least one is a delta,
and involving four fermion fields and pions.
For example \cite{invk,civK1},
\begin{equation}
{\cal L}^{(0)} = -D_{T} N^\dagger\boldt\vec{\sigma}N\cdot
         (N^\dagger\boldT\vec{S}\Delta+ h.c.) +\ldots, 
\label{DTlag}
\end{equation}
\begin{equation}
 {\cal L}^{(1)} =  
    -D_{1}N^\dagger N \, N^\dagger \boldt\cdot\vec{\sigma}\cdot\vec{\boldD}N
    -D_{2}\vec{\boldD}\cdot(N^\dagger\boldt\vec{\sigma}N\times
           N^\dagger\boldt\vec{\sigma}N) +\ldots,
\label{dilag}
\end{equation}
\begin{equation}
 {\cal L}^{(2)} = m_\pi^2 
   \left(1-2D^{-1}\frac{\boldpi^{2}}{F_{\pi}^{2}}\right)
                 \left(C_{2qm}^{(S)}N^\dagger N N^\dagger N 
 +C_{2qm}^{(T)}N^\dagger\vec{\sigma} N \cdot N^\dagger \vec{\sigma}N\right)
         +\ldots
\label{pertpilag}
\end{equation}
Here $D_{T}$, $D_{i}$ and $C_{2qm}^{(S,T)}$ 
are undetermined constants of mass dimensions $-2$,
$-3$ and $-4$ respectively.
``\ldots'' above stand for interactions involving more delta fields.
So far investigations in this EFT have integrated out the
delta isobar. As in the one-nucleon sector, this amounts
to omitting the explicit-delta interactions and replacing the coefficients
of nucleonic interactions by new ones that scale with factors of $1/\delta m$.
For example, Eq. (\ref{dilag}) then involves new coefficients
$d_i$ with 
\begin{equation}
d_{1} =D_{1} -\frac{4D_{T}h_{A}}{9\, \delta m} + \ldots,
\;\; d_{2}= D_{2} +\frac{2D_{T}h_{A}}{9\, \delta m} + \ldots
\label{d's}
\end{equation}

\begin{figure}[t]
\centerline{\psfig{figure=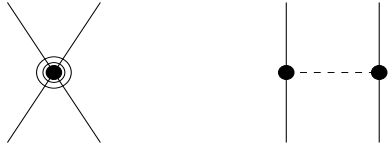,height=0.75in,width=2.0in}}
\vspace{0.5cm}
\caption{Insertions at $O(Q/M_{NN})$
in the perturbative-pion EFT, to be substituted for
the circled dot in 
the second line of Fig. \protect\ref{fig:unnat1t}.
A dashed line represents a pion,
a heavy dot a derivative and
a heavy dot with two circles two derivatives or powers of
the pion mass.
\label{pertpifig}}
\end{figure}

{\bf The amplitude.}
The leading-order $NN$ amplitude
coincides with that in the lower-momentum EFT,
represented by the first line of Fig. \ref{fig:unnat1t}
and by $T_{os}^{(0)}$ (the $C_0$ and $ik$ terms) in Eq. (\ref{renTon}):
it is the iteration of chirally-symmetric,
non-derivative contact interactions. 
At this order there are contributions only in the
two $S$-wave channels.
Subleading terms, of $O(Q/M_{NN})$ relative to leading,
are constructed in a direct
extension of subleading terms of the lower-momentum EFT, 
second line of Fig. \ref{fig:unnat1t} and 
$T_{os}^{(1)}$ (the $C_2$ term) in Eq. (\ref{renTon}). 
The difference is that now besides two-derivative contact interactions
we also insert one-pion exchange and 
a non-derivative contact interaction that breaks chiral symmetry explicitly.
In other words, besides the $C_2$ term in Eq. (\ref{ver})
we need to consider one insertion of
\begin{equation}
 V =  -\left(\frac{2g_{A}}{F_{\pi}}\right)^{2}\boldt_{1}\cdot\boldt_{2}
                \frac{\vec{\sigma}_{1}\cdot\vec{q}\,\vec{\sigma}_{2}\cdot
                \vec{q}}{\vec{q}\,^{2}+m_{\pi}^{2}} 
  +m_\pi^2(C_{2qm}^{(S)}+C_{2qm}^{(T)}\vec{\sigma}_{1}\cdot\vec{\sigma}_{2}).  
\label{pertpipot}
\end{equation}
---see Fig. \ref{pertpifig}.
Here we have neglected the energy transferred in OPE
as it is a higher order effect.
When Eq. (\ref{pertpipot}) is inserted within loops with $T_{os}^{(0)}$,
new divergences arise
that are $\propto m_\pi^2$. These divergences can be absorbed
in the $C_{2qm}$'s, giving rise to renormalized 
$C_{2qm}^{(R)}$'s. 
Both $C_2$ and $C_{2qm}$ contact interactions also only contribute to
$S$ waves. The tensor operator from pion exchange, on the other hand,
introduces mixing between $^3S_1$ and $^3D_1$ waves.
To this order, the $^3S_1-^3D_1$ mixing parameter and the
$^3D_1$ phase shift are 
{\it predicted}
in terms of pion parameters (and the short-ranged parameters
that can be determined from the $^3S_1$ phase shift).

A calculation of the $NN$ system to this order was first carried out
in Ref. \cite{Kaplan4} in the PDS subtraction scheme,
and repeated with a coordinate space
cutoff in Ref. \cite{cohenhansen1}.
One finds
in the $^1S_0$ channel 
\cite{Kaplan4,cohenhansen1,cohenhansen2,cohenhansen3}
\begin{equation}
k \cot \delta^{(^1S_0)}=
 -\frac{1}{a_2^{(^1S_0)}} + \frac{r_2^{(^1S_0)}}{2} k^2 
 + {\cal S}^{(^1S_0)}(k) 
 + O(k^2/\Lambda, m_\pi^2/\Lambda, 1/a_2^{(^1S_0)2}\Lambda), 
\label{pertpiERE}
\end{equation}
where
\begin{equation}
a_2^{(^1S_0)}=\frac{m_NC_0^{(R)}}{4\pi}
              \left[1+\frac{g_A^2m_\pi m_N}{2\pi F_\pi^2}
                   \left(1-\frac{1}{2}\frac{4\pi}{m_\pi m_NC_0^{(R)}}\right) 
                 +\frac{m_\pi^2 C_{2qm}^{(R)}}{C_0^{(R)}} \right],
 \label{pertpiscattlen}
\end{equation}
\begin{equation}
r_2^{(^1S_0)}=\frac{16\pi C_2^{(R)}}{m(C_0^{(R)})^2} 
              +\frac{g_A^2m_N}{2\pi F_\pi^2}
               \left[1-\frac{8}{3} \frac{4\pi}{m_\pi m_NC_0^{(R)}}
                 +2\left(\frac{4\pi}{m_\pi m_NC_0^{(R)}}\right)^2 \right],
\label{pertpieffran}
\end{equation}
and
\begin{eqnarray}
{\cal S}^{(^1S_0)}(k) &=& \frac{g_A^2m_\pi^2m_N}{4\pi F_\pi^2}
      \left[ \frac{1}{4}\ln (1+4\frac{k^2}{m_\pi^2}) -\frac{k^2}{m_\pi^2}
     +2\frac{4\pi}{m_\pi m_NC_0^{(R)}} 
        \left(\frac{m_\pi}{2k}\arctan(\frac{2k}{m_\pi})
        -1+\frac{4k^2}{3m_\pi^2}\right)  \right. \nonumber \\
 & &   
\;\;\;\;\;\;\;\;\;\;\;\;\;\;\;\; \left. 
       -\left(\frac{4\pi}{m_\pi m_NC_0^{(R)}}\right)^2
        \left(\frac{m_\pi^2}{4k^2} \ln (1+4\frac{k^2}{m_\pi^2})
        -1+\frac{2k^2}{m_\pi^2} \right)
      \right].
\label{pertpinonanal}
\end{eqnarray} 
This is in the form of an effective range expansion
supplemented by a non-analytic ``shape function'' generated by pion exchange.
Similar expressions hold for $k \cot \delta^{(^3S_1)}$,
and analytic formulas can also be derived for $\varepsilon_1$
and $\delta^{(^3D_1)}$
\cite{Kaplan4,cohenhansen1,cohenhansen2,cohenhansen3}.

To this order, there are three
unknown parameters in the $^1S_0$ channel 
($C_0^{(^1S_0)}$, $C_2^{(^1S_0)}$, $C_{2qm}^{(^1S_0)}$), 
and three ($C_0^{(^3S_1)}$, $C_2^{(^3S_1)}$, $C_{2qm}^{(^3S_1)}$) in the
$^3S_1$
channel. 
Note, however, that 
the amplitude to this order can be written as a function
of $C_0+m_\pi^2C_{2qm}$ for it is this combination that appears
in the potential  (\ref{ci17})+(\ref{pertpipot}).
Only at higher orders will $C_{2qm}$ contribute
new terms through pion loops.
To subleading order, then, the separate dependence of observables
on  $C_0$ and $C_{2qm}$ is an artifact of perturbation
theory.
The difference between 
treating $C_{2qm}$ perturbatively and non-perturbatively
is an indication of the size of higher-order corrections.

\begin{figure}[t]
\centerline{\psfig{figure=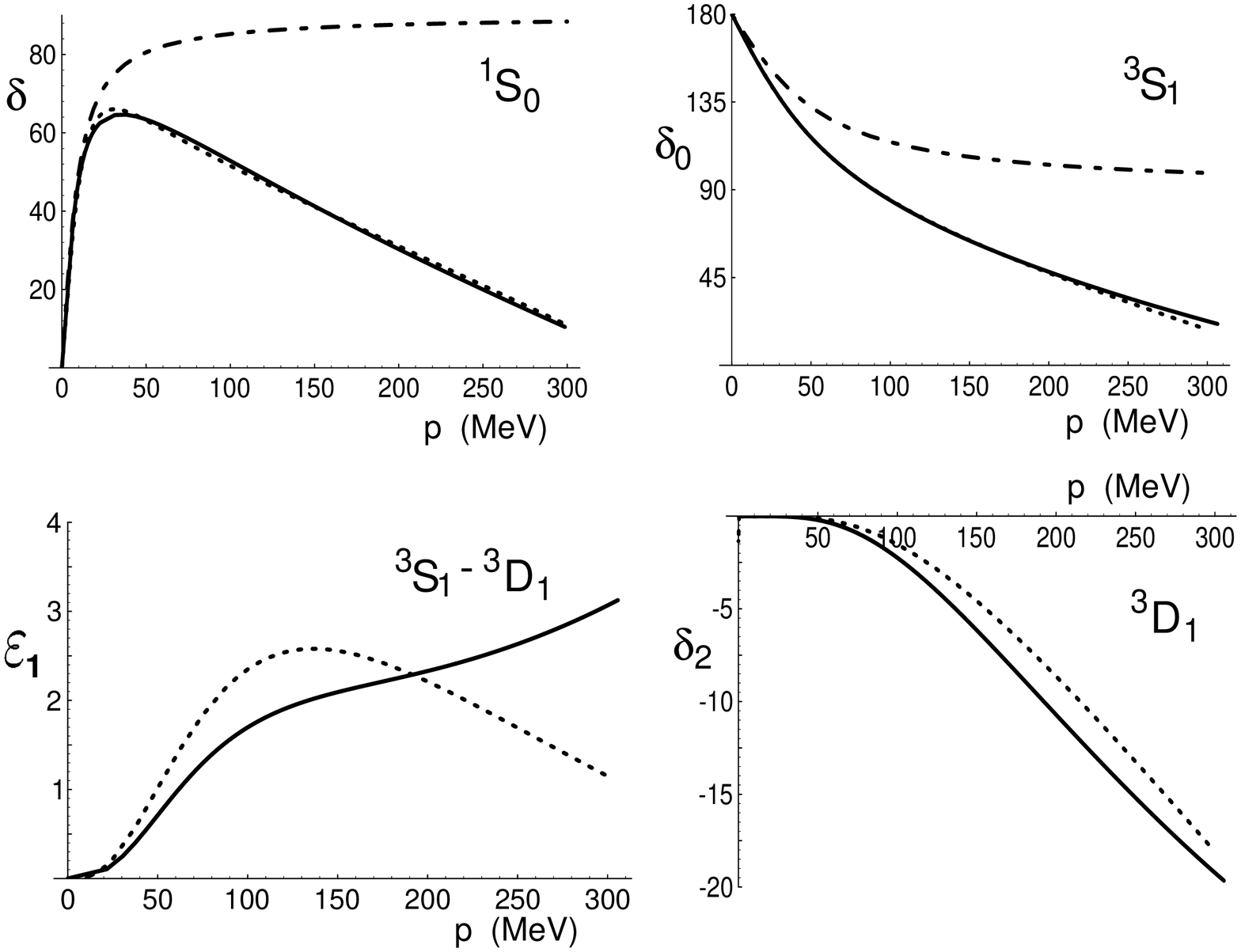,height=3in,width=4in}}
\vspace{0.5cm}
\caption[ ]{ $^1S_0$, $^3S_1$, and $^3D_1$ $NN$ phase shifts and 
$\varepsilon_1$ mixing angle in degrees as functions
of center-of-mass momentum in MeV: 
EFT in leading (dot-dashed) and subleading (dashed line) orders; 
and Nijmegen PSA (solid line).
Figure courtesy of M. Savage.
\label{fig:pertpi}}
\end{figure}

A fit to the Nijmegen $S$-wave phase shifts \cite{nijmanal}
was done in Ref. \cite{Kaplan4}
using the amplitude derived in the PDS scheme. As described in 
Sect. \ref{subsec-2Nverylow}, the theory can then be written in terms
of running couplings such as $C_{2qm}(\mu)$. 
The scattering lengths and effective ranges can be used to determine
$C_0$ and $C_2$ with a choice of $C_{2qm}$. 
Choosing $C_{2qm}(m_\pi)=0$, the predicted $^1S_0$ phase shifts
deviate significantly from the Nijmegen PSA at momenta of
order $m_\pi$. The amplitude with three 
free parameters in each 
$S$-wave gives a much better fit for
momenta below 200 MeV;
the predicted  $^3S_1-^3D_1$ mixing parameter $\varepsilon_1$
and the  $^3D_1$ phase then come out reasonably well
---see Fig. \ref{fig:pertpi} \cite{Kaplan4}.

Note that to this order other waves are determined
as well. 
For example, $P$ waves are given by OPE, and indeed it is
well-known that the sizes of triplet scattering volumes
is roughly reproduced by OPE \cite{ericsonweise}.  
Similarly, it is been part of nuclear folklore that
high partial waves are mostly due to perturbative pion effects 
\cite{ericsonweise}.
The various perturbative contributions from pion exchanges to
$D$ and higher waves was examined in detail 
in Ref. \cite{twokaiserpapers}, 
up to sub-subleading order. 
The $D$ wave cannot not be described
well above laboratory energies of $\sim 50$ MeV
without short-range interactions,
but higher waves get progressively 
better described by perturbative pions, as expected.

{\bf Schr\"odinger formulation.}
The EFT with perturbative pions can be recast
in terms of the Schr\"odinger equation in coordinate
space.
In Ref. \cite{cohenhansen1}
it was shown that the above subleading amplitude
can be obtained by a Born expansion 
on the Yukawa part of the pion potential at distances 
$r\ge R\sim 1/\Lambda$.
The wave-function can be written  as
$u(r)= \sum_{i=0}^\infty u_i(r)$,
where {\it (i)}
the leading term is a solution of the free
Schr\"odinger equation, $u_0=\sin (kr + \delta_0)$, with
$k \cot \delta_0$ determined by the contact interactions;
and {\it (ii)} higher-order terms satisfy 
$u_i(R)= 0$. $u_1(r)$ can be expressed in terms of simple integrals
involving the long-range part of the potential.

{\bf Range of the theory and regularization issues.}
{}From the expressions for the amplitude shown above,
it seems that pion effects are associated with
the scale $M_{NN}\sim 4\pi F_\pi^2/g_A^2 m_N \simeq 300$ MeV
\cite{Kaplan4}.
If so, the expansion parameter is
$\sim m_\pi/M_{NN}\simeq 0.5$.
The scattering length (\ref{pertpiscattlen}) is 
formally of the form
$\sim \aleph + O(Q^2/M_{NN})$.
$C_0^{(R)}$ can be estimated in leading order
using Eq. (\ref{seffran}) and then used
in Eqs. (\ref{pertpieffran}) and (\ref{pertpinonanal})
to estimate the relative size of pion contributions.
For example, OPE 
is responsible for about half of 
the effective range (\ref{pertpieffran}).
Note, however, that this simple-minded procedure is not without
its caveats.
OPE contributions to Eq. (\ref{pertpiscattlen}) are not small.
They include a contribution which is 
non-analytic in the quark mass; this term clearly shows
not only that the expansion is in $m_\pi/ M_{NN}$ even at $k=0$,
but also that dimensionless numerical factors can be dangerous,
since $g_A^2 m_\pi m_N/2\pi F_\pi^2 \simeq 0.9$.  
Nevertheless, $C_{2qm}$ can be adjusted numerically
to compensate, maintaining the lowest-order mostly unchanged.
The reasonable success observed in the low-partial-wave phase shifts
suggests that 
dimensionless numbers do indeed 
conspire to increase $M_{NN}$. 
A number of recent papers have attempted to
provide a more precise estimate of $M_{NN}$.
The issue of a practical test of convergence has been tied
to specific regularization schemes employed in calculations. 

In Ref. \cite{gegelia3} the $^1S_0$ fit has been reexamined
with both PDS and a subtraction at $p^2=-\mu^2$.
Significant differences were 
found in the phases for $Q\gaprox \mu$ 
depending on whether $C_2$ and $C_{2qm}$ were included
perturbatively or non-perturbatively.
This led the author of Ref. \cite{gegelia3} to conclude
that pions can be included perturbatively only
for external momenta well below the pion mass.
In Ref. \cite{secondsteele} a similar conclusion
was reached through a Lepage error analysis.
In Ref. \cite{mehenstewart} it is argued that
this strong sensitivity to $\mu$ arises from the fact that
the subleading amplitude in PDS is not 
$\mu$ independent, so that the parameters that satisfy
renormalization group equations cannot give a good fit to
phase shifts. 
Pion exchange gives contributions
that do not vanish in the chiral limit, and
introduces subleading but potentially significant
dependence on $\mu$.
It was shown that this $\mu$ dependence can be ameliorated
by splitting $C_0$ into two pieces,
$C_0=C_{0[0]}+C_{0[2]}$, where $C_{0[0]}$ is treated non-perturbatively
while $C_{0[2]}\sim m_\pi^2 C_{2qm}\ll C_{0[0]}$ 
is treated as a first-order correction.
The amplitude in subleading order is then a function of
$C_{0[2]}+m_\pi^2C_{2qm}$.
An improved subtraction scheme (``OS'') designed to minimize
the $\mu$ dependence of the perturbative expansion
was also introduced, and the subleading amplitude
computed.
Fits to  $^1S_0$ and $^3S_1$ phase shifts for momenta
between 7 and 100 MeV ---fits that are weighted towards low
momenta--- are then reasonable, at the same
time producing good values for the
scattering lengths.
The freedom of fitting $C_{2qm}$ can enforce
smallness of the full subleading contribution.
The authors of Ref. \cite{mehenstewart} conclude that there are
no obstructions to using perturbative pions
for external momenta larger than the pion mass.

Refs. \cite{cohenhansen2,cohenhansen3} point out 
that the shape functions are completely determined by pion exchange.
For momenta below the pion mass, the shape functions can be expanded
in powers of $k^2$, and
effective range parameters other than scattering
lengths and effective ranges 
are likewise determined.
For example, in either $S$-wave channel
$v_2^{(n)(0)}\sim 1/m_\pi^{n-2} M_{NN}$.
Such relations can be viewed as a one-pion-saturation model \cite{mesonsat}
of the parameters of the very-low-energy EFT of Sect. \ref{subsec-2Nverylow}.
One can see that the perturbative-pion EFT provides no explanation
for the fine-tuning in the scattering lengths,
but does provide higher-order  effective range parameters
which scale approximately as $1/M_{nuc}^{n-1}$
as expected from dimensional analysis.
Note that these parameters blow up in the chiral limit as a consequence of
the long range of pion interactions.
If the perturbative-pion EFT can be shown to hold for $Q\gaprox m_\pi$,
then these predictions can be elevated to {\it bona fide} low-energy theorems.
On the other hand, if the perturbative-pion EFT holds only
for $Q\saprox m_\pi$, 
contact parameters showing up at the same order 
render the pion predictions no better than an 
order-of-magnitude estimate.
The shape function and higher effective range parameters look like good testing
grounds of the convergence of the expansion as 
---contrary to $C_{2qm}$ in $a_2$--- there
are no parameters that can 
be fine-tuned to balance OPE effects. 

In Ref. \cite{cohenhansen2} the shape functions
were expanded in a series $k^2$. The mixing parameter
$\varepsilon_1$ was expanded as well,
\begin{equation}
\varepsilon_1= \sum_{i=1}^\infty g^{(i)} \, k^{2i+1},
\label{epsexp}
\end{equation}
although this is expected to breakdown at very low momenta
$\sim \aleph$; in view of Eq. (\ref{eremix}) it is
no surprise that $g_i\sim 1/\aleph^{2i-1} m_\pi M_{NN}$. 
The result of a comparison between 
perturbative pion predictions and values of effective range (\ref{ere})
and $\varepsilon_1$ (\ref{epsexp}) parameters 
extracted from the Nijmegen PSA \cite{nijmanal}
is in Table \ref{tab:mesonsat} \cite{cohenhansen2}.
Because of the dramatic failure of these predictions 
the authors of Ref. \cite{cohenhansen2} suggest that
the scale of the physics left out from this perturbative-pion EFT
is not much larger than $m_\pi$. 
They further suggest that successes of this EFT
are coming from the expansion in $\aleph/M_{NN}$, with
the failure of the expansion in $m_\pi/M_{NN}$ 
disguised by counterterms.
They attribute
the good fits of Fig. \ref{fig:pertpi} to the fact that any
theory with free parameters consistent with the
effective range expansion can reproduce the rapid rise at low $k$ 
and the subsequent slow decrease of the $S$-wave phase shifts.
However, using the energy-independent Nijmegen analysis,
the authors of Ref. \cite{mehenstewart} have argued 
that the uncertainty in $v_2^{(2)(0)}$
is too large to make a definite test of the OPE predictions.
They claim that uncertainties in the $v_2^{(n)(0)}$'s
from the energy-dependent Nijmegen analysis are unrealistically small,
and question whether
low-energy data alone can be
used to determine these parameters.
Ref. \cite{cohenhansen3} has reexamined these issues
directly in terms of the shape function at finite energy.
It is shown that the shape function in the $^3S_1$ channel
is well-determined by the PSA:
at the deuteron pole, the shape function is
close to zero to a good precision resulting from 
the accurately known values of $B_2$, $a_2$, and $r_2$;
at positive energies the finite-energy phase shifts are also 
need to extract ${\cal S}$.
It is found that OPE saturation fails again way beyond error bars.
Because 
${\cal S}$ is small compared to $k \cot \delta$, 
Ref. \cite{cohenhansen3} points out that one has to
include relativistic and isospin-breaking effects
in the extraction of ${\cal S}$.
This is a little unsettling, however, 
as these are higher-order effects that were not included in the EFT
estimate.

\begin{table}[t]
\caption{Parameters of analytic expansions of
$k \cot \delta$ in the $^1S_0$ and $^3S_1$ channels
and of the $^3S_1-^3D_1$ mixing angle $\varepsilon_1$
predicted by the EFT with perturbative pions compared
to values obtained from the Nijmegen PSA.
\label{tab:mesonsat}}
\vspace{0cm}
\begin{center}
\footnotesize
\begin{tabular}{cccccccccc}
  & \multicolumn{3}{c}{$^1S_0$} &   \multicolumn{3}{c}{$^3S_0$} 
                 &   \multicolumn{3}{c}{$^3S_1-^3D_1$}        \\
\cline{2-4} \cline{5-7} \cline{8-10} 
  & $v_2^{(2)(0)}$ & $v_2^{(3)(0)}$ & $v_2^{(4)(0)}$ 
  & $v_2^{(2)(0)}$ & $v_2^{(3)(0)}$ & $v_2^{(4)(0)}$ 
                & $g^{(1)}$ & $g^{(2)}$  & $g^{(3)}$  \\
\hline
EFT  & $-3.3$ & 17.8 & $-108$ & $-0.95$ & 4.6 & $-25$ 
                 & 3.9 & $-86$ & $1800$ \\
PSA  & $-0.48$ & 3.8 & $-17$ & $0.04$ & 0.67 & $4.0$ 
                 & 1.7 & $-26$ & $220$ \\
\end{tabular}
\end{center}
\end{table}

Although work done so far has gone some distance in assessing the
validity of perturbative pions, this remains one of the most 
important open questions for the EFT approach in nuclear systems.
Work along this line requires the evaluation of higher-order
effects.
At sub-subleading order we encounter new pion exchanges:
both non-static (or radiative) OPE and once-iterated OPE.
Radiative pion diagrams have very recently been studied
in Ref. \cite{mehenstewart3}.
It is suggested that the size of these graphs can be estimated counting powers
of $\sqrt{m_\pi m_N}$, yet
mysterious cancellations have been found that render the 
sum of leading non-vanishing non-static pion diagrams smaller
than expected.
The role of deltas has not yet been examined.
Particularly in diagrams with non-static pions, 
integrating deltas out should significantly limit the range of the EFT. 
So far there is also no explicit calculation of the dimensionless
factors that are involved in once-iterated OPE, and of the sizes
of sub-subleading contact interactions.
Work is in progress in this direction \cite{rupakshoresh}
and a better
determination of the range of validity of this EFT
should be available soon.
As we will see in Sect. \ref{subsec-3Nverylowlow}, this is pivotal 
in assessing the usefulness of this EFT for nuclei
other than the deuteron.
In Sects. \ref{subsec-3Nverylowlow} and
\ref{subsec-2.5Nverylowlow} it will be evident that 
this EFT allows a much simpler
treatment of nuclear systems than afforded by traditional methods.

Finally, note ---in complete analogy to the extensive discussion
in Sect. \ref{subsec-2Nverylow}---
that if pions can be treated perturbatively then little error 
is made in iterating them to all orders.
There might even be something to gain if one is lucky that
short-range effects are small because of unknown dimensionless
factors.
The procedure of including pions in a potential
that is then iterated to all orders
had actually been carried out before any
of the developments reviewed in the present section.
We turn to this next.

\subsection{Moderate energies} \label{subsec-2Nmod}

We want to extend the EFT approach to momenta $Q\gaprox M_{NN}$. 
How this is to be done depends critically on the value of $M_{NN}$. 
At this scale
pions become non-perturbative.
If $M_{NN}$ is as large as $M_{QCD}$
the task is daunting as there is presently no rationale
for a controlled expansion in the presence of
massive degrees of freedom such as the rho.
On the other hand, if  $M_{NN} \ll M_{QCD}$, we might
hope to improve the expansion of the previous section 
by a controlled resummation
of selected terms that go as $Q/M_{NN}$.
If $M_{NN} \saprox 300$ MeV, the lack of known
particle thresholds there suggests that the
resummation could involve {\it only} pions.
For $M_{NN} \gaprox  300$ MeV delta isobars
probably need to be included as well.

In fact, naive dimensional analysis alone suggests that the
leading effect of pion exchange should
be iterated to all orders for $Q\sim M_{nuc}$. 
This is because numerically $M_{QCD}\sim 4\pi M_{nuc}$,
and thus $M_{NN}\sim M_{nuc}$.
The new element is that OPE is $\sim 1/M_{nuc}^2\sim 4\pi/M_{QCD}M_{nuc}$.
Because OPE has short-range components,
it is natural to assume that they set the scale for $\aleph$,
$\aleph \sim M_{nuc} \ll M_{QCD}$.
Part of the unnaturalness of $\aleph$ is thus explained.
(It is still mysterious why $\aleph < M_{nuc}$;
some cancellation between short and long-range
effects is necessary to enforce this.)
This suggests a power counting might be meaningful
in which we take $Q\sim M_{NN}\sim \aleph \sim M_{nuc}\sim M_{QCD}/4\pi$.
Now,
arguments analogous to those in Sect. \ref{subsec-2Nverylow} 
get complicated to the point
that they can no longer be carried out analytically in detail.
Nevertheless they are still valid:
a loop containing a reducible intermediate state
still contributes $M_{QCD} Q/4\pi$, so adding 
no-derivative contact interactions ($\sim 4\pi/M_{QCD}\aleph$)
or static OPE ($\sim 1/M_{nuc}^2$) to
a graph amounts to a power of 
$Q/\aleph\sim Q/ M_{nuc}$. 
In order
to describe $Q\sim M_{nuc}$ we have to sum all graphs
with the leading potential,  
no-derivative contact interactions plus static OPE.
All other contributions would come as corrections.
This power counting is due to Weinberg \cite{inwei6,inwei5},
but many of its elements had been anticipated
by Friar \cite{friarcoon,czechfriar}.

The question has been raised, whether such a power counting
can be carried out consistently.
The problem is that once OPE is iterated to all orders, 
renormalization
of the theory becomes quite different from the perturbative-pion
EFT.
Perturbative arguments have been given 
\cite{Kaplan4,mehenstewart} to support
the claim that Weinberg's power counting is not consistent.
Ref. \cite{Kaplan4} points out that
a diagram with OPE between two $C_0$ interactions
requires the $C_{2qm}$ counterterm, as 
we have seen in the previous section.
Ref. \cite{mehenstewart} finds that thrice-iterated OPE 
and diagrams with twice-iterated OPE plus one $C_0$ interaction
require a $C_2$ counterterm. 
This implies that alleged higher-order contact interactions
might get contaminated by 
powers of $M_{QCD}/M_{nuc}$ in the numerator,
to the point that they are required to appear in leading order.
If this is a feature of an infinite number of contact
interactions, there is no ordering and
this EFT is doomed.
However, it has been known in the context of the
Schr\"odinger equation that perturbative arguments
of this type are not in general reliable for singular
potentials, as is the case for OPE.
The perturbative expansion might have a cut starting at $g_A^2/F_\pi^2=0$;
insistence on a $g_A^2/F_\pi^2$ expansion
would then reflect itself on different orders offering
correlated contributions to counterterms,
each bringing powers of $M_{QCD}/M_{nuc}$, 
yet resulting in a much better behaved sum.
This conjecture ultimately has to be decided on the basis
of actual calculations.
We will assume it to be correct, fit phase shifts and
then examine the cutoff dependence
and convergence of the expansion.
If the conjecture is correct, then we should find
that this EFT has a range of validity greater than $M_{NN}$.
Although less familiar
to those raised with perturbative calculations,
this procedure is quite analogous and no less 
justified than the methods being used to assess the range of validity
of the perturbative-pion EFT.
We are going to see that there is evidence that
renormalization is being performed correctly,
as cutoff dependence turns out to be weak. 

The power counting is most easily done for
the potential $V$.
An arbitrary contribution to the two-nucleon potential 
can be represented by a two-nucleon irreducible Feynman diagram,
that is, with $A=2$ 
continuous nucleon lines
and at least one pion or delta isobar in intermediate states.
Effects from the scale $M_{QCD}$ are lumped
in the vertices, and by restricting ourselves
to the potential we exclude
the scale $Q^2/M_{QCD}$ from energy denominators. 
Since the only explicit scale is $Q$, 
the $\chi$PT counting (\ref{index}) applies
directly if we put $A=2$ \cite{inwei6,inwei5},
\begin{equation}
\nu=2L+{\sum _i}{V_i}{\Delta _i}.
\label{A=2index}
\end{equation}
In this way the benefits of chiral symmetry are extended to
nuclear physics:
since both $L$ and $\Delta_i$ are bounded from below 
($L\ge 0, \Delta_i\geq 0$),
$\nu\ge \nu_{min}=0$.
Note that this counting is  
a direct generalization of the counting (\ref{treenu})
of the pionless EFT,
where the potential has no loops and a single vertex 
of index ${\Delta}=d$.
Isospin violation will be considered shortly.

We assume implicitly
that once $M_{nuc}$ ($F_\pi$, $m_\pi$, possibly $\delta m$,
and, neglecting fine-tuning, $\aleph$) 
have been accounted for, all other dimensionful
quantities are given by $M_{QCD}$.
Since we also assume that no enhancements appear
from the iteration of the potential,
the scaling of contact interactions is determined by
pion exchange in the potential: as in the case of a 
single nucleon, we expect additional $Q^2$ to be accompanied by
$M_{QCD}^2$, so that
\begin{equation}
C_{2n}^{...(R)}= \frac{4\pi}{M_{QCD}^{2n+1} M_{nuc}} \gamma_{2n}^{...}, 
\label{nonpertpiscaling}
\end{equation}
with the $\gamma_{2n}^{(...)}$'s dimensionless factors of $O(1)$.

Thanks to infrared enhancement, reducible diagrams have
to be treated differently than the potential.
The leading-order potential is iterated to all orders. 
Corrections to the leading potential do not need to be iterated to all
orders. Yet, as we have seen in Sect. \ref{subsec-2Nverylow}, they can
be iterated with small error as long as one uses a regularization with a
cutoff mass $\Lambda \sim M_{QCD}$;
for the two-nucleon system, 
this is equivalent to solving the 
Schr\"odinger equation with the potential $V$. 
A potential up to a certain
order ensures that the amplitude is correct to the same
order.
Of course, even if our assumptions about the scaling of
interactions in this EFT are incorrect, results from the iteration
of the potential with a sufficient number of
interactions automatically include all results
of perturbative pions discussed in the previous section. 

Traditionally, potential models have been 
plagued by problems of principle, such as 
the form of meson-nucleon interactions
(for example pseudoscalar {\it vs.} pseudovector pion coupling),
renormalization issues, absence of a small expansion parameter, 
{\it etc.} 
Because the EFT potential includes explicitly the exchange of only pions,
all these problems can be resolved.
As we have seen in Sect. \ref{subsec-sym},
because pions are pseudo-Goldstone bosons, the form of their interactions
with nucleons is determined by the pattern of chiral symmetry and
its breaking. (For example, if the EFT were formulated relativistically, 
pseudovector coupling would be preferred,
as insistence on pseudoscalar form would demand new seagull
vertices to ensure ``pair suppression'' \cite{friarcoon}).
Renormalization can be performed as {\it all} interactions
consistent with symmetries have been included.
And the power counting (\ref{A=2index}) 
for the EFT potential implies that diagrams with an increasing
number of loops $L$ ---and, in particular, with increasing number
of exchanged pions--- should be progressively less important.
Clearly, the EFT potential can be thought of as a low-energy
approximation to standard potential models,
although this can only be taken in an average sense.
The scope of an EFT potential for systems with at least two 
mass scales has been examined in toy models.
Ref. \cite{Lepage} considers the case of a short-range
potential supplemented by a Coulomb potential.
It exemplifies through an efficient error analysis
how the approximation of contact interactions for the short-range 
potential improves the low-energy description systematically.
The small role played by the cutoff $\Lambda$
can also be shown explicitly in this case. 
Similar conclusions hold in toy models
where two nucleons
interact via a light scalar with Yukawa coupling
and a short-range potential \cite{secondsteele,last2epelbaoum}.
It has been shown that, even in the presence of a fine-tuning
that produces a shallow $S$-wave bound state, the
contact interactions obey Eq. (\ref{nonpertpiscaling}),
with $M_{QCD}$ representing the heavy mass scale.
(Ref. \cite{secondsteele} further emphasizes
the advantages of a fit to 
modified effective range parameters.)
The EFT potential thus
provides a QCD justification for some of the phenomenological
successes of potential models.

\begin{figure}[t]
\centerline{\epsfysize=8cm \epsfbox{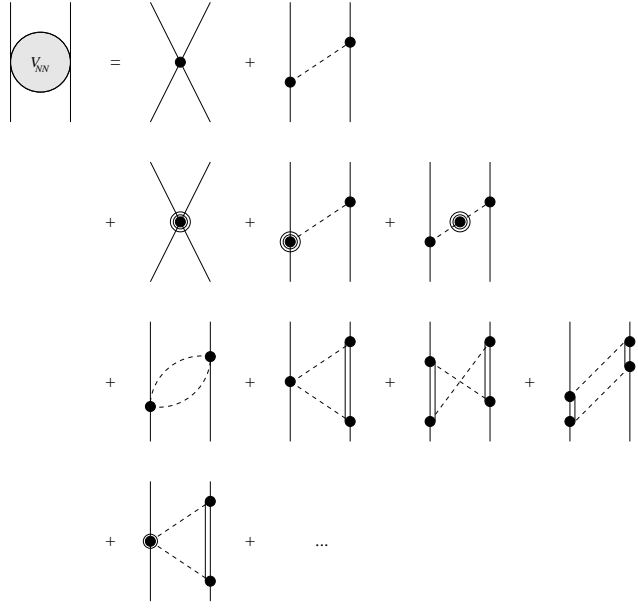}}  
\vspace{0.5cm}
\caption{Some time-ordered diagrams contributing to the two-nucleon potential
$V_{NN}$ in the EFT.
(Double) solid lines represent nucleons (and/or deltas), 
dashed lines pions, a heavy dot an interaction in ${\cal L}^{(0)}$,
a dot within a circle an interaction in ${\cal L}^{(1)}$,
and 
a dot within two circles an interaction in ${\cal L}^{(2)}$.
First line corresponds to $\nu=\nu_{min}$,
second and third lines to $\nu=\nu_{min}+2$, 
fourth line to $\nu=\nu_{min}+3$, and 
``$\ldots$'' denote $\nu\geq \nu_{min}+4$. 
All orderings with at least one pion
or delta in intermediate states are included.
Not shown are 
diagrams contributing only to renormalization of parameters.
\label{F:vkolck:V2N}}
\end{figure}

{\bf Potential and amplitude.}
The Lagrangian for this EFT is the same considered before
in the context of perturbative pions.
A calculation of all contributions to the
two-nucleon potential up to $\nu=\nu_{min}+3$ was carried out
in Refs. \cite{ciOvK,invk,ciOLvK}
using time-ordered perturbation theory. 
Some diagrams are shown in Fig. \ref{F:vkolck:V2N}.

In leading order, $\nu=\nu_{min}$, 
the potential is simply static OPE
and momentum-independent contact terms,
that is, the first term in Eq. (\ref{pertpipot})
and the first two in Eq. (\ref{ci17}) \cite{inwei6}. 
This is obviously a very crude approximation to the $NN$ potential. 
It does allow us to see how energies of $O$(10 MeV) arise from QCD: 
the average of the leading-order potential in coordinate space is
roughly $\langle V_{2N}^{(0)}\rangle \sim
-g_{A}^{2}m_{\pi}^{3}/4\pi F_{\pi}^{2} \sim -10$ MeV.
However, it is known that the nuclear force 
has other sizable components, like a spin-orbit force, a strong short-range
repulsion and an intermediate range attraction.
These are all generated in the next orders:
$\nu=\nu_{min}+1$ corrections vanish due to parity and 
time-reversal invariance, but
$\nu=\nu_{min}+2$ corrections are several.
First, there are short-range corrections;
they come from one-loop pion dressing of the lowest-order contact
interactions, and from
four-nucleon contact interactions with two derivatives
or two powers of the pion mass, displayed in
Eqs. (\ref{ci17}) and (\ref{pertpipot}).
It is easy to show that the result of loop diagrams amount
to a simple shift of the contact parameters.
In Refs. \cite{ciOvK,invk,ciOLvK} the $C_0$'s were redefined
in order to reabsorb both loop contributions and the
chiral-symmetry-breaking $C_{2qm}$'s, as to this order
they cannot be separately determined from $NN$ data.
Second, there are corrections to OPE; these come from vertex dressing
and from recoil upon pion emission:
\begin{equation}
  V^{(2)}_{tree}  =  -\frac{2g_{A}}{F_{\pi}^{2}}\boldt_{1}\cdot\boldt_{2}
        \frac{\vec{\sigma}_{1}\cdot\vec{q}\,\vec{\sigma}_{2}\cdot\vec{q}}
        {\vec{q}\,^{2}+m_{\pi}^{2}} \,
        \left(A_{1}q^{2}+A_{2}k^{2}-2g_{A}
             \frac{E-\frac{1}{4m_{N}}(4\vec{k}^{2}+\vec{q}^{\,2})}
             {\sqrt{\vec{q}\,^{2}+m_{\pi}^{2}}}\right).        \label{ci71}
\end{equation}  
Again, here loop corrections have been absorbed in mass and coupling
constant renormalization. Also,
chiral-symmetry-breaking corrections to the $\pi NN$ coupling
proportional to $m_\pi^2$ can be lumped in $g_A$, which then
satisfies the Goldberger-Treiman relation without discrepancy.
Third, there are two-pion exchange (TPE) diagrams built out 
of lowest-order $\pi NN$ and $\pi N\Delta$ interactions
of Eq. (\ref{L0}).
Denoting
$\omega_{\pm} \equiv \sqrt{(\vec{q}\pm\vec{l})^{2}+4m_{\pi}^{2}}$,
these loop diagrams can be expressed schematically as
\begin{equation}
V^{(2)}_{loop}= \frac{(g_A, h_A)}{F_\pi^4} (1, \boldt_{1}\cdot\boldt_{2})
   \int{\frac{d^{3}l}{(2\pi)^{3}}} 
   \left( \frac{1}{\omega_{\pm}}, 
          \frac{\vec{q}^{\, 2}-\vec{l}^{\, 2}}{(\omega_{\pm}^3, 
                                \omega_{\pm}^2 \delta m)},
          \frac{((\vec{q}^{\, 2}-\vec{l}^{\, 2})^2,
                 \vec{\sigma}_{1}\cdot(\vec{q}\times\vec{l}\, )
                 \vec{\sigma}_{2}\cdot(\vec{q}\times\vec{l}\, ))}
               {(\omega_{\pm}^5, \omega_{\pm}^4\, \delta m,
               \omega_{\pm}^3(\delta m)^2, \omega_{\pm}^2(\delta m)^3)}\right).
\end{equation}
At $\nu=\nu_{min}+3$ a few more TPE diagrams appear,
which involve the $\pi\pi NN$ seagull vertices from Eq. (\ref{L1});
again schematically,
\begin{equation}
V^{(3)}_{loop}= \frac{(g_A^2, h_A^2)B_i}{F_\pi^4}(1, \boldt_{1}\cdot\boldt_{2})
   \int{\frac{d^{3}l}{(2\pi)^{3}}} 
   \frac{(m_\pi^2 (\vec{q}^{\, 2}-\vec{l}^{\, 2}),
               (\vec{q}^{\, 2}-\vec{l}^{\, 2})^2,
               \vec{\sigma}_{1}\cdot(\vec{q}\times\vec{l}\, )
                 \vec{\sigma}_{2}\cdot(\vec{q}\times\vec{l}\, ))}
              {(\omega_{\pm}^4, \omega_{\pm}^3\, \delta m)}.
\end{equation}
To this order there are some relativistic corrections as well, but
they are numerically small and have been neglected.

This potential has all the spin-isospin structure of phenomenological
models, but its profile is determined by explicit degrees of freedom,
symmetries, and power counting. 
The power counting suggests a hierarchy of short-range effects:
$S$ waves should depend strongly on the short-range parameters $C_0$;
contact interactions affect $P$-wave phase shifts 
only in subleading order,
so their effect should be smaller and approximately linear;
$D$ waves are affected by contact interactions only via mixing,
while higher waves should be essentially determined by pion exchange.
Chiral symmetry is particular influential
in the TPE piece. 
The latter includes a particular form of terms previously
considered by Brueckener and Watson \cite{ciBw}, 
Sugawara and von Hippel \cite{ciSH},
and Sugawara and Okubo \cite{ciSO}, plus a few new terms.
Those terms involving the $B_i$'s and the deltas provide the only form of
correlated TPE to this order, as graphs where pions interact in flight appear
only in next order and should thus be relatively small. 
The sum of the $B_1$ term and the corresponding
delta term (which give the $c_3$ of Eq. (\ref{c's}), 
related to the nucleon axial polarizability),
is particularly important in providing an isoscalar central force.
Not surprisingly, in the chiral limit these potentials 
behave at large separations as van der Waals forces.

\begin{figure}[t]
\centerline{
\epsfysize=20cm \epsfbox{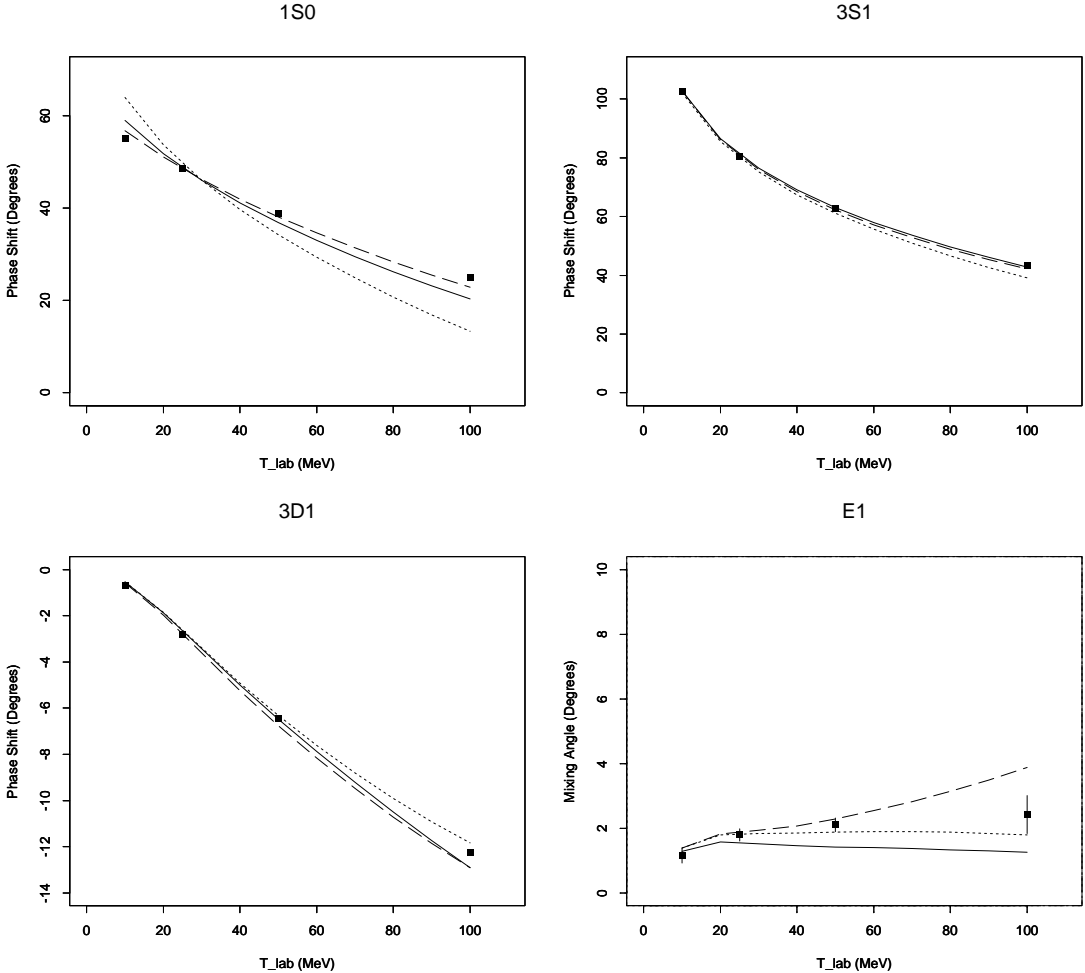}  
}
\vspace{-6.0cm}
\caption[ ]{ $^1S_0$, $^3S_1$, and $^3D_1$ $NN$ phase shifts and 
$\varepsilon_1$ mixing angle in degrees as functions
of the laboratory energy in MeV: 
EFT up to $\nu=3$ for cutoffs of 
500 (dotted), 780 (dashed), and 1000 MeV (solid line); 
and Nijmegen PSA (squares).}
\label{F:vkolck:NN}
\end{figure}

Regularization and renormalization are necessary
not only for the loops in the potential but also
for the loops generated by the solution of the Schr\"odinger
equation. 
For numerical convenience a smooth cutoff of the 
Gaussian type was used, 
and calculations performed with the cutoff parameter
$\Lambda$ taking values 500, 780 and 1000 MeV. 
For each cutoff value a set of 
(bare) parameters was found that fits phase shifts
below 100 MeV laboratory energies. 
There are nine independent parameters stemming from contact
interactions, although for the fit to $NN$ phase shifts 
we have also varied nine other, redundant parameters.
OPE and TPE diagrams are completely determined by
parameters accessible in pion-nucleon reactions,
but because most were not known at the time
they were also searched in the fit.
A sample of the results for the lower, more important partial waves
is presented in Fig. \ref{F:vkolck:NN},
together with phases from the Nijmegen PSA \cite{nijmanal}
---cf. Fig. \ref{fig:pertpi}.
Quality of the fits is typical of other waves.
Waves higher than $F$ were found to be mostly described
well by pion exchange alone, as expected. 
Deuteron quantities 
are shown in Table \ref{tab:dparam}
---cf. Table \ref{tab:taedparam}.
(Electromagnetic quantities refer to the contributions
from lowest-order $\gamma NN$ couplings only,
not to a consistent calculation which would include
sub-leading one- and two-nucleon effects.)
The predicted $S$-wave scattering lengths 
(not used to constrain the fit) were found to be
$a_2^{(^1S_0)}\simeq -15.0$ fm and 
$a_2^{(^3S_1)}\simeq 5.46$ fm.
Important central attraction comes from the $B_i$'s and deltas,
and indeed the central potential does resemble that from 
models that include $\sigma$ and $\omega$ meson exchange explicitly.
Values for the parameters are listed in Ref. \cite{ciOLvK}.
Reasonable values were found for quantities known at the time,
for example 
$g_A=1.33$ (in agreement with the Goldberger-Treiman relation)
and $h_A=2.03$ (smaller but not too far off the large-${\cal N}_c$ value).
However, the values for the $B_i$'s came out different than
those found later in $\pi N$ scattering:
using Eq. (\ref{c's}),
$c_1= 2.2$ GeV$^{-1}$, $c_3=-0.7$ GeV$^{-1}$, and 
$c_4= 2.7$ GeV$^{-1}$, 
while from the leading delta contribution alone 
(because $B_4$ appears here only in higher order)
$c_2\sim 1.8$ GeV$^{-1}$. 
These values are to be compared to those in Table \ref{tab:ci};
one concludes that  $B_3$ is too big and of wrong sign,
while $B_2$ is too small and $B_1$ too big,
and/or the delta contributions are too small.
For a cutoff of $\Lambda=780$ MeV, 
the coefficients $C_{2n}$ of the contact interactions
were found to scale approximately as in Eq. (\ref{nonpertpiscaling}) 
with $M_{QCD} \sim 500$ MeV for $M_{nuc}= F_\pi$.

\begin{table}[t]
\caption{Results from EFT fits at $\nu=3$
for various cut-offs $\Lambda$
and experimental values 
for the deuteron binding energy ($B$), magnetic dipole moment ($\mu_d$), 
electric 
quadrupole moment ($Q_E$), asymptotic $D/S$ ratio ($\eta$), and $D$-state 
probability ($P_D$).\label{tab:dparam}}
\vspace{0cm}
\begin{center}
\footnotesize
\begin{tabular}{ccccc}
Deuteron   & \multicolumn{3}{c}{$\Lambda$ (MeV)} &            \\
\cline{2-4}
quantities & 500 & 780 & 1000                   & Experiment \\
\hline
$B$ (MeV)  & 2.15 & 2.24 & 2.18                 & 2.224579(9) \\
$\mu_d$ ($\mu_N$) & 0.863 & 0.863 & 0.866       & 0.857406(1) \\
$Q_E$ (fm$^2$) & 0.246 & 0.249 & 0.237          & 0.2859(3)   \\
$\eta$     & 0.0229 & 0.0244 & 0.0230           & 0.0271(4)  \\
$P_D$ (\%) & 2.98 & 2.86 & 2.40                 &            \\
\end{tabular}
\end{center}
\end{table}

A number of papers have subsequently examined different
aspects of this EFT.
In Ref. \cite{epelbaoum} it was pointed out that
the parameters $A_i$ in Eq. (\ref{ci71}) needed not be included 
explicitly in the fit, as one combination is fixed
as a numerically small relativistic correction
\cite{lastjim}, while the
$A_1$ contribution can be absorbed in the $C_2^{(6)}$
of Eq. (\ref{ci17}) and in the $\pi NN$ coupling.  
Note that the above EFT potential is energy dependent.
As pointed out in Sect. \ref{subsec-pow}, equivalent
potentials can be obtained through unitary transformations.
An energy-independent potential is more convenient
in many situations, and the corresponding version was derived
in Refs. \cite{jimandsid,epelbaoum,lastjim}.
The effect of the unitary transformation is essentially
to remove the third term in Eq. (\ref{ci71}) and at the same time to modify
the TPE potential. 
It corresponds in what box diagrams are concerned to a change from the 
Brueckener and Watson \cite{ciBw} to the 
TMO \cite{tmo} potential.
The latter has a further advantage: it displays explicitly 
a cancellation of the isoscalar spin-independent central 
and of the isovector spin-dependent interactions, 
which with a energy-dependent choice comes only through 
a combination of terms in the potential and its iterations.  
The components of the EFT TPE potential were studied
in more detail in Ref. \cite{twokaiserpapers}. 
In particular, it is shown explicitly that 
{\it (i)} relativistic corrections are mostly small
(although in the weaker components they can make relatively large
contributions);
{\it (ii)} both isoscalar central and spin-orbit
potentials are numerically similar to $\sigma$ and $\omega$ exchange
in models, if the $B_i$'s take values given by $\pi N$ scattering;
{\it (iii)} the OPE isovector tensor potential is reduced by the
TPE contribution;
{\it (iv)} a subset ---which is invariant under pion field redefinition--- 
of the $\nu=\nu_{min}+4$, two-loop TPE diagrams 
is small and repulsive.
(Other modern studies of chiral aspects of TPE 
include Ref. \cite{shakinetal}.)

The fair agreement of this first calculation and data 
up to center-of-mass momenta $Q\sim 300$ MeV
suggests that this may 
eventually become an alternative to other, more model-dependent approaches 
to the two-nucleon problem.
In particular, it hints that a $\sigma$ exchange might be
unnecessary. 
The relative cutoff insensibility also 
might be indication that renormalization
is being performed correctly.
The large values of the cutoff and the reasonable
value of the contact terms are auspicious
for the range of validity of the theory.
Note that to this order this EFT includes all 
interactions incorporated to date in the perturbative-pion
EFT and then some.
More detailed analyses of the effects of different components of
this potential on the fit have appeared since.
In Ref. \cite{Lepage} the same EFT potential (with the same
type of cutoff) as in Refs. \cite{ciOvK,invk,ciOLvK}
but without TPE (and energy-dependence in OPE) was refitted to
the Nijmegen PSA \cite{nijmanal}
in three channels, $^1S_0$, $^1P_1$, and $^3D_2$,
below center-of-mass energy of $100$ MeV.
An error analysis shows 
explicitly how the second-order corrections
improve over the leading-order fits, particularly
if in a global fit weights are assigned to data.
It is confirmed that pion effects contribute little to the $^1S_0$ phase shift
in this energy range (as expected from fine-tuning),
but play a more important role in higher-wave phases
(as expected by power counting).
The error analysis hints at
a breakdown of the expansion at $M\sim 300$ MeV, and
shows that cutoff effects are minimal when
$\Lambda\sim M$.
It is suggested that the breakdown at  $M\sim 300$ MeV
is related to the neglect of TPE.
Similar conclusions were reached in Ref. \cite{secondsteele}.
More recently, (the tail of the energy-independent version of) the EFT
TPE potential in the limit of a heavy
delta was substituted for 
(the tail of) the one-boson exchange in a 
Nijmegen phase-shift reanalysis of $pp$ data below 350 MeV \cite{robetal}.
A drop in 
$\chi^2$ was observed. 
$c_1$ was found to be poorly determined by the fit;
fixing it from the sigma-term determined by 
the modern $\pi N$ PSA
of Ref. \cite{robpiN}, the best-fit values are 
\begin{equation}
c_1=-0.76(7) \; \mbox{GeV}^{-1},\;\;
c_3=-5.08(28) \; \mbox{GeV}^{-1},\;\;
c_4=4.70(70) \; \mbox{GeV}^{-1}.
\end{equation}
The agreement with $\pi N$ scattering values in Table \ref{tab:ci}
is remarkable and
confirms unequivocally the validity of chiral TPE.
Taken together, these new results suggest
that a 
more extensive search of parameters might 
improve the quality of the EFT potential fit.

{\bf Isospin violation.}
We can order isospin-breaking effects in the potential
in the same way done in Sect. \ref{sec-01N} \cite{invk,civK2}.
At $Q\sim M_{nuc}\gg \aleph$,
photon exchanges are indeed perturbative, as
discussed in Sect. \ref{subsec-2Nverylow}. 
Here I will concentrate on effects beyond those considered
there, that is, 
on those stemming from the quark masses,
from indirect electromagnetic effects, and from simultaneous
pion-photon exchange. 
I follow  the standard nomenclature and refer to an 
isospin-symmetric potential as  ``class I'',
to a potential that breaks charge dependence but not charge symmetry
---defined as a rotation of $\pi$ around the 2-axis in isospin space---
as  ``class II'',
to one that breaks charge symmetry but vanishes in the $np$ system
as  ``class III'',
and to one that breaks charge symmetry but is present in the $np$ system
as  ``class IV''.

No isospin-violating effects appear at
leading order, $\nu = \nu_{min}$, so the 
leading potential is class I.
The first effect appears at 
$\nu = \nu_{min} +1$ in the form of a class II
isospin violation from pion mass splitting 
($\Delta m_\pi^2\propto \alpha M_{QCD}^{2}$) in OPE 
and from Coulomb photon exchange.
One order down, $\nu = \nu_{min} +2$,
a class III force appears
mainly from the quark mass difference
through breaking in the $\pi NN$ coupling 
($\beta_1=O(\varepsilon m_{\pi}^{2}/M_{QCD}^{2})$) in OPE,
from contact terms 
($\gamma_{s,\sigma}=O(\varepsilon m_{\pi}^{2}/M_{QCD}^{4})$),
and from the nucleon mass difference 
($\Delta m_N=O(\varepsilon m_{\pi}^{2}/M_{QCD})$). 
Therefore, to order $\nu = \nu_{min} +2$ the isospin-violating 
nuclear potential is a two-nucleon potential of the form
\begin{equation}
 V=V_{\rm II}[(t_1)_{3}(t_2)_{3}-\boldt_1\cdot\boldt_2]+
        V_{\rm{III}}[(t_1)_{3}+(t_2)_{3}],  \label{iv78}
\end{equation}
\noindent
where 
\begin{equation}
 V_{\rm II}=
   -\left(\frac{2g_{A}}{F_{\pi}}\right)^{2}
     \frac{\vec{q}\cdot\vec{\sigma}_1 \vec{q}\cdot\vec{\sigma}_2}
          {(\vec{q}\,^{2}+m_{\pi^{0}}^{2})
  (\vec{q}\,^{2}+m_{\pi^{\pm}}^{2})}(\Delta m_{\pi}^{2}+\Delta m_{N}^{2}),
                                                     \label{iv79}
\end{equation}
\begin{equation}
 V_{\rm III}=
    \frac{2g_{A}\beta_{1}}{F_{\pi}^{2}}
    \frac{\vec{q}\cdot\vec{\sigma}_1\vec{q}\cdot\vec{\sigma}_2}
         {\vec{q}\,^{2}+m_{\pi}^{2}}
    -(\gamma_{s}+\gamma_{\sigma}\vec{\sigma}_{1}\cdot\vec{\sigma}_{2}). 
\label{iv80}
\end{equation}
\noindent
Finally, class IV forces
appear only at order $\nu = \nu_{min} +3$. 
One concludes that the symmetries of QCD 
naturally suggest a hierarchy of classes in the nuclear 
potential \cite{invk,civK2}:
\begin{equation}
\langle V_{\rm M+1}\rangle /\langle V_{\rm M}\rangle
\sim O(Q/M_{QCD}),
\end{equation}
where $\langle V_{\rm M}\rangle$ denotes the average contribution of 
the leading class ${\rm M}$ potential. 
This qualitatively explains, for example, the observed isospin structure 
of the Coulomb-corrected scattering lengths, 
$a_{np}\simeq 4 \times ((a_{nn}+a_{pp})/2 - a_{np}) 
       \simeq 4^2 \times (a_{pp}-a_{nn})$. 

\begin{figure}[tb]
\centerline{\epsfig{file=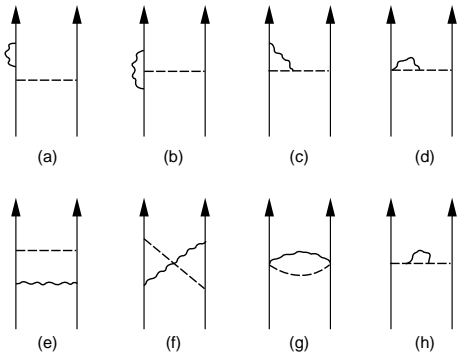,height=2.0in}}
\vspace{0.5cm}
\caption{Electromagnetic loops contributing to isospin violation in the
           one-pion-range 
           nuclear force in first 
           non-vanishing order. Solid lines are nucleons, 
           dashed lines are pions, and wavy lines are photons.
           Other orderings are also included,
           but iteration of static OPE and Coulomb exchange
           is excluded from (e). 
\label{emloops}}
\end{figure}

One can use the above formalism to do consistent and systematic
calculations of isospin violation.
The 
isospin-violating potential of range $\sim 1/m_{\pi}$
up to $\nu = \nu_{min} +3$ was computed in 
Refs. \cite{prlwithnijm,mejimterry}.
It consists of diagrams where one pion flies between the two nucleons,
constructed out of 
tree-level diagrams with interactions of index
up to 3,
plus 16 (not counting trivial particle permutations)
non-vanishing one-loop diagrams,
comprising pion and nucleon self-energy corrections, pion-nucleon
vertex corrections, and simultaneous pion-photon exchange
---see Fig. \ref{emloops}. 
Most of these diagrams are ultraviolet divergent; dimensional 
regularization was used and 
$1/(D-4)$ terms cancelled against bare coefficients.
All diagrams with photons are also individually infrared divergent;
a photon mass is introduced and removed at the end,
the sum being infrared finite. 
The result is invariant under both gauge transformation and
pion-field redefinition: 
\begin{equation}
V_{\pi \gamma}(\vec{q})= 
   \frac{2\alpha g_{A}^{2}} {\pi F_{\pi}^{2}}
   (\boldt_1 \cdot \boldt_2 - (t_1)_3 \, (t_2)_3)  
   \frac{\vec{\sigma}_1 \cdot \vec{q} \, \vec{\sigma}_2 \cdot \vec{q}}{q^2} 
   \left[\frac{(1- q^2/m_\pi^2)^2}{(1+ q^2/m_\pi^2)q^2/m_\pi^2} 
             \ln (1 + q^2/m_\pi^2) 
         -1\right].
\end{equation}
Its isospin structure allows
only charged-pion exchange and therefore affects only $np$ scattering. 
This $\pi\gamma$ potential has been incorporated in a
Nijmegen phase shift reanalysis of 
$np$ data below 350 MeV \cite{prlwithnijm}.
We can use the values for the $\pi NN$ coupling constants
determined by the analysis to find that their isospin breaking 
is consistent with zero,
with an uncertainty comparable to our expectation
from 
dimensional
analysis and from 
$\pi-\eta-\eta'$ mixing \cite{mejimandterry}.
Similarly, $\gamma_{s}$ 
might be viewed 
as originating in $\rho-\omega$ mixing, 
while pseudovector-meson exchange 
(in particular close-lying doublets such as $a_{1}-f_{1}$)
exchange contribute to the $\gamma_{\sigma}$ spin-spin force 
\cite{mejimandterry,coon}.

\section{Few-nucleon systems} \label{sec-3N}

Few-body
systems provide non-trivial testground for ideas developed
in two-body dynamics. 
As we have seen, the deuteron is sufficiently
dilute that an EFT for $Q\sim \aleph$ can be used
to describe it well. Such an EFT is essentially equivalent to
an effective range expansion. The latter 
has been successfully applied to reactions involving the deuteron
and photons, but
difficulties have been encountered in the three-body system.
In Sect. \ref{subsec-3Nverylowlow} we will see how the
EFT can be extended to the three-body system.
In this case
some fairly unusual renormalization takes place, which makes
the EFT method much more challenging and thus interesting than in
its more standard perturbative applications. 
The EFT for higher momenta where pions are treated perturbatively
shares the same leading order as the pionless EFT, differing
only in that pion exchange appears explicitly together with
a contact correction in subleading order.
The important question to be answered is whether few-nucleon bound states
can be dealt with in the approximation where leading-order interactions
are taken as short ranged.
In Sect. \ref{subsec-3Nmod} still higher momenta are considered,
and the connection to potential model approaches discussed.
We can use this EFT to organize the various few-body-force structures.

\subsection{Very low and low energies} \label{subsec-3Nverylowlow}

Symmetries do not forbid three-nucleon contact interactions.
In order to extend the EFT to the three-nucleon system 
we need to find the size of three-nucleon forces
relative to two-nucleon forces.
As in Sect. \ref{subsec-2Nverylow}, I omit at first spin and isospin variables.
I also neglect electromagnetic interactions.
Requiring invariance under
small-velocity Lorentz, parity, and time-reversal transformations,
at sufficiently low momenta, the 
Lagrangian involving six fields $\psi$ is
\begin{equation}
\label{lag3}
{\cal L}  =  
 - \frac{D_0}{6} (\psi^\dagger\psi)^3 + \ldots ,\nonumber
\end{equation}
where ``\ldots'' stand for terms with more derivatives,
which are suppressed at low momenta.
Note that an $S$-wave contact three-body
interaction is allowed for bosons. 
The same is true in the case of 
three nucleons in a spin $S=1/2$ state,
but the Pauli
principle forbids an $S$-wave interaction for 
three nucleons in a $S=3/2$ state.
The first possible three-body interaction in the latter channel
contains an extra $Q^2$, and should be relatively less
important.

A comparison between the tree-level connected three-body diagrams
allows us to guess the relative size of two- and three-body interactions.
We saw in Sect. \ref{subsec-2Nverylow}
that operators involving four nucleon fields contain the small
scale $\aleph$.
At momenta $Q\sim \aleph$,
two sequential two-body $C_0$ interactions
contribute $O((4\pi/m\aleph)^2 m/\aleph^2)$.
If we assume that the coefficients
$D_{2n}$ of the terms with $2n$ derivatives in Eq. (\ref{lag3}) 
are of natural size,
$D_{2n} \sim (4\pi)^2 /mM^{4+2n}$, then we expect 
three-body-force effects to be of relative 
$O((\aleph/M)^{4+2n})$.
But does the small scale $\aleph$ not contaminate operators
involving six nucleon fields?
This question is intimately related to the renormalization
group flow of three-body interactions with the mass scale
introduced in the regularization procedure. 
This flow, in turn, depends on
the behavior of the sum of two-body contact contributions
to three-body amplitude as function of the renormalization scale,
or equivalently, as function of the ultraviolet cutoff $\Lambda$. 

It is convenient
to rewrite this theory
by introducing the dimeron field $T$ of Sect. {\ref{subsec-2Nverylow}
It is straightforward to show 
---for example, by a Gaussian path integration---
that the Lagrangian (\ref{lag3}) is equivalent to 
Eq. (\ref{translag})
supplemented with
\begin{equation}
\label{lagtpsi}
{\cal L}  = h \, T^\dagger T \, \psi^\dagger\psi
 +\ldots
\end{equation}
\noindent
In leading order, since the kinetic term of the dimeron can be neglected,
observables will depend on
the parameter appearing explicitly
in Eq. (\ref{lagtpsi}) only through the combination
$-3hg^2/\Delta^2\equiv D_0$.

Let us consider elastic particle/bound-state scattering at $Q\sim \aleph$
\cite{2stooges,3stooges,3stoogesbosons}. 
(Three-body bound states manifest themselves as poles
at negative energy.
Both the inelastic channel and three-particle scattering 
involve the same type of diagrams considered here and can be obtained by
a  straightforward extension of the following arguments.)
Diagrams contributing to the amplitude are depicted in 
Fig. \ref{fig:part/bs}.
We have already seen that at these momenta all terms in
the full dimeron propagator of Fig. \ref{fig:transprop} are equally important.
It is easy to see that an $L$-loop diagram involving only two-body
interactions (``pinball'' diagram) is
$O((m g^2 /Q^2) \times (m g^2 Q/4\pi (\Delta+m g^2Q/4\pi))^L)$.
Pinball diagrams are thus (superficially) ultraviolet
finite, and are all of the same
order for $m g^2 Q/4\pi\Delta = Q/\aleph \sim 1$.
The three-body amplitude is then the solution of an integral equation.
This equation, including the three-body force, is also depicted in 
Fig. \ref{fig:part/bs}. 

\begin{figure}[t] 
\begin{center}
\epsfxsize=12cm
\centerline{\epsffile{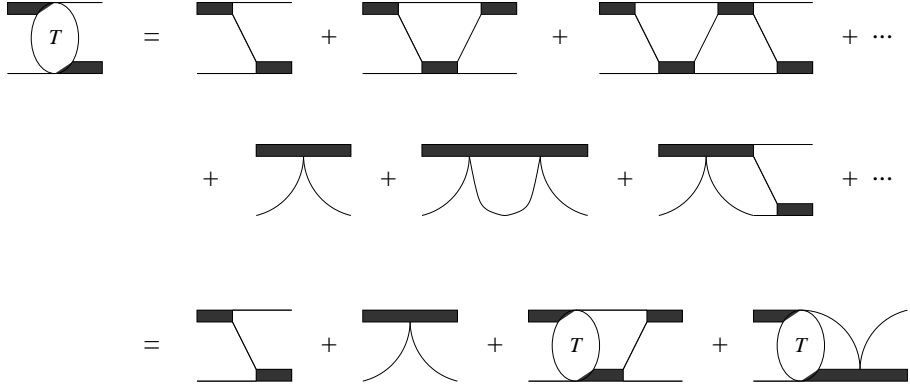}}
\end{center}
\caption{The amplitude $T$ for particle/bound-state scattering
as a sum of dressed pinball and three-body-force diagrams (first 
and second lines)
and as an integral equation (third line).}
\label{fig:part/bs}
\end{figure}

I choose the following kinematics: the incoming 
particle and bound state
are on-shell with four-momenta $(k^2/2m, -\vec{k}\, )$ and 
$(k^2/4m-B_2, \vec{k}\, )$,  
respectively.
The outgoing particle and bound state
are off-shell with four-momenta 
$(k^2/2m-\varepsilon, -\vec{p}\, )$ and 
$(k^2/4m-B_2+\varepsilon, \vec{p}\, )$; the on-shell point has 
$\varepsilon=k^2/2m - p^2/2m$ and $p=k$.
The total energy is  $E = 3k^2/4m - B_2$.
At very low momenta, we can limit ourselves to the more important
$S$-wave.
Using the boosted dimeron propagator (\ref{Tprop}),
after performing the integration over the fourth component
of the loop momentum,
setting $\epsilon=k^2/2m - p^2/2m$, and projecting onto the $S$-wave,
we find that the amplitude $a(k,p)$ normalized
so that $a(0,0)=-a_3$ is the particle/bound-state scattering length
satisfies \cite{3stoogesbosons}
\begin{equation}
a(k,p)
=M(k,p;k)+\frac{2\lambda}{\pi}\int_0^\Lambda dq\ M(q,p;k)
\frac{q^2}{q^2-k^2-i\epsilon} a(k,q), \label{aeq}
\end{equation}
\noindent
with the kernel
\begin{equation}
M(q,p; k)= \frac{4}{3} 
     \left(\frac{1}{a_2}+\sqrt{\frac{3p^2}{4}-mE}\, \right)
   \left[\frac{1}{pq}{\rm ln}
    \left(\frac{q^2+q p +p^2-mE}
               {q^2-q p +p^2-mE}\right)
    +\frac{h}{mg^2} \right]. \label{kernel} 
\end{equation}
\noindent
Here $\lambda=1$ for bosons.
The same equation is valid
in the case of fermions, with different values of $\lambda$. 
For
three nucleons in a spin $S=3/2$ state, this equation holds
with $\lambda=-1/2$ and $h=0$.
For three nucleons in a spin $S=1/2$ state
a pair of coupled integral equations applies, with properties
similar to the bosonic case.
Eqs. (\ref{aeq}) and (\ref{kernel}) reduce to the expressions found in
Refs. \cite{skorny,2stooges,3stooges}
when $h=0$. 
Note that the perturbative series shown in Fig. \ref{fig:part/bs} corresponds
to a perturbative solution of the integral equation
for small $\lambda$. 
The solution of Eq. (\ref{aeq}) is complex even below the threshold for the
breakup of the two-particle bound state due to the $i\epsilon$ prescription.
To facilitate the discussion one uses the function
$K(k,p)$ that satisfies the same Eq. (\ref{aeq}) as $a(k,p)$ but with the
$i\epsilon$ substituted by a principal value prescription. $K(k,p)$
is real below the breakup threshold. $a(k,k)$ and, consequently, 
the scattering matrix can be obtained from  $K(k,p)$ through
\begin{equation}
a(k,k)=\frac{K(k,k)}{1-i k K(k,k)}.
\end{equation}

In order to understand the renormalization group flow due to two-body
forces,
let us first take $h=0$.
When $p\gg 1/a_2$ (but $k\sim 1/a_2$), the inhomogeneous term
is small ($O(1/pa_2)$), the main contribution to the integral comes from
the region $q\sim p$, and the amplitude satisfies the 
approximate equation
\begin{equation}
K(k,p)= \frac{4\lambda}{\sqrt{3}\pi}
                   \int_0^\L \: \frac{dq}{q}\ 
                   {\rm ln} \left(\frac{q^2+pq+p^2}{q^2-pq+p^2}\right)
                   K(k,q).\label{asphieq} 
\end{equation}
Now, in the limit $\Lambda\rightarrow \infty$ there is no scale left.
Scale invariance suggests solutions of the form  
$K(k,p)= p^s$, which exist only if
$s$ satisfies 
\begin{equation}
1- \frac{8\lambda}{\sqrt{3}s} 
   \frac{\sin\frac{\pi s}{6}}{\cos\frac{\pi s}{2}}=0.
\label{rhoeq}
\end{equation}
\noindent
If $K(k,p)$ is
a solution, $K(k,p_\star^2/p)$ 
is also a solution for arbitrary
$p_\star$. Because of this special conformal
symmetry, the solutions of Eq. (\ref{rhoeq}) come in pairs. 

For $\lambda<\lambda_c=3 \sqrt{3}/ 4 \pi \simeq 0.4135$, Eq.
(\ref{rhoeq}) has only real roots.
For $\lambda=-1/2$, 
there are solutions $s=\pm 2, \pm 2.17, \ldots$ 
In this case
an acceptable solution of Eq. (\ref{asphieq})
decreases in the ultraviolet, 
\begin{equation}
K(k,p\gg 1/a_2)= C p^{-|s|}.
\end{equation}
For finite cutoff, the solution should still have this form 
as long as $p\ll \Lambda$. 
The overall constant $C(\Lambda)$ 
cannot be fixed
from the homogeneous asymptotic equation, but is determined
by matching the asymptotic solution onto
the solution at $p\sim 1/a_2$ of the full Eq. (\ref{aeq}).  
Because of the fast ultraviolet convergence, the full solution to
Eq. (\ref{aeq}) is expected to be insensitive to the choice
of regulator, so that a well-defined
$\lim_{\Lambda\rightarrow \infty} C(\L)$ can be found. 
There is thus no evidence for a contamination of three-body forces
by the small scale $\aleph$ in the $S=3/2$ channel of
$S$-wave nucleon-deuteron ($Nd$) scattering. 
Up to and including
relative $O((Q/M)^5)$,
only two-body forces need to be included. A calculation using
Eq. (\ref{translag}) up to and including relative $O((Q/M)^2)$
was carried out in Refs. \cite{2stooges,3stooges}.
As expected, cutoff dependence was found only for $p\sim \Lambda$.
The result for
the phase shifts for energies up to the break-up point is shown in
Fig. \ref{fig:3/2}: the result of the leading-order ($O(1)$) calculation
\cite{skorny} is given by the dashed line
and the $O((\aleph/M)^2)$ result \cite{3stooges} by the solid line.
These are compared to 
data points at finite energy
from a PSA 
of $Nd$ scattering data \cite{vanOers}, and 
to the much more precise
(nearly) zero-energy point \cite{Dilg}.
We expect errors in our calculation to be of 
$O(\aleph/M_{nuc})^3)$.
At zero energy, we find
$a_{3}^{(3/2)}=6.33\pm 0.10$ fm \cite{2stooges},
compared to the experimental value $a_{3}^{(3/2)}=6.35\pm 0.02$ fm \cite{Dilg}.
{\it This is an extremely accurate low-energy theorem!}
Our results seem to deviate from a simple effective range type expansion
only around the pole at $\sim 0.05$ fm$^{-2}$ .
(A pole in $k\ {\rm cot}\delta$ corresponds to a zero in the
scattering matrix, which does not carry special meaning.)
This pole does not appear in potential model calculations, 
and presumably will be (re)moved
by higher-order terms that we have not yet included.
The calculation of higher-order corrections involves the knowledge of further
counterterms like the ones giving rise to 
$S-D$ mixing.
These parameters can be determined either by fitting 
$NN$ data
or by matching with another effective theory
---involving explicit pions--- valid up
to higher energies.
If more precise experimental data ---particularly at
zero-energy--- appear, we would be facing
a unique situation
where high-precision
calculations in strong-interaction physics can be
carried out systematically and tested.

\begin{figure}[tb]
\begin{center}
\epsfxsize=6cm
\centerline{\rotate[r]{
   \rotate[r]{
             \rotate[r]{\epsffile{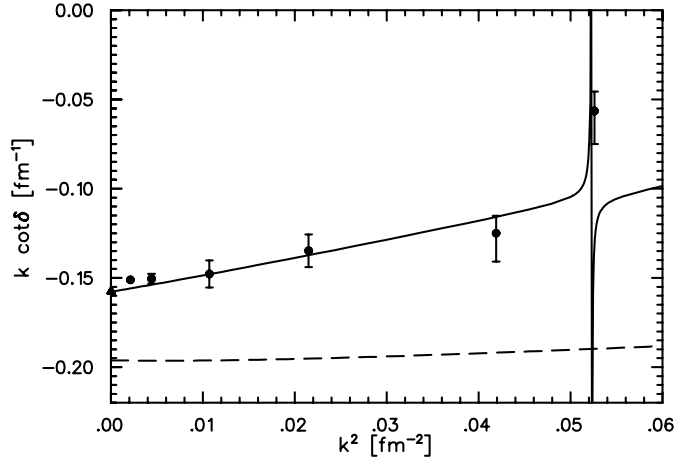}}
             }
          }}
\end{center}
\caption{$k\ {\rm cot}\delta$ in fm$^{-1}$
for $S$-wave $Nd$ scattering
in the $S=3/2$ channel as a function of $k^2$ in fm$^{-2}$ 
in the EFT to 
$O(1)$ (dashed line) and $O((\aleph/M_{nuc})^2)$
(solid line). Circles  are from the PSA in
Ref. \protect\cite{vanOers} and the triangle is from Ref. \protect\cite{Dilg}.}
\label{fig:3/2}
\end{figure}

For $\lambda=1$, on the other hand, there
are purely imaginary solutions: $s=\pm is_0$, where $s_0 \simeq 1.0064$.
The solution of Eq. (\ref{asphieq}) for $p\ll \Lambda$
is 
\begin{equation}
K(k,p\gg 1/a_2)= C \: 
        \cos \left(s_0{\rm ln} \frac{p}{\Lambda_\star}\right), \label{assol}
\end{equation}
\noindent
where, on dimensional grounds, $\Lambda_\star (\L)\propto \Lambda$.
This solution has {\it two} constants 
$C(\Lambda)$ and $\L_\star(\Lambda)$, and implies that 
the solution of Eq. (\ref{aeq}) is not 
unique in the limit $\Lambda\rightarrow \infty$ \cite{danlebed}.
The undetermined phase $\L_\star$ arises from special conformal
symmetry, as there is no way here to select a preferred
oscillatory solution. 
Solutions of Eq. (\ref{aeq}) for $\lambda=1$ and
finite $\Lambda$ (but $h=0$) were obtained in Ref. \cite{3stoogesbosons}.
We have verified that in the region $1/a_2\ll p\ll \Lambda$ 
the solutions indeed
have the form (\ref{assol}), with
$\L_\star$ approximately linear in the cutoff.
The amplitude $C$ was found to be approximately proportional to
$\cos (s_0 \ln (\L_\star a_2))$.
This dependence generates some points 
where the amplitude is cutoff independent, but
apart from them, the amplitude does not
show signs of converging as $\L\rightarrow \infty$.
Since the solution for small $p$ has to match onto
the large-$p$ solution, the cutoff dependence
leaks into the low-momentum region. 
Small differences in
the asymptotic phase lead to large differences in,
for example, the particle/bound-state
scattering length.
This leakage of high-momentum behavior into the low-momentum physics 
is indication that we are not performing renormalization
consistently with our expansion.
Note that if one were to truncate the series
of diagrams in Fig. \ref{fig:part/bs} at some finite number of loops
one would miss 
the asymptotic behavior of $K(k,p)$ that
generates this cutoff dependence.
This is because $s_0$ (and its expansion in powers of $\lambda$)
vanish in a neighborhood of $\lambda=0$.
The truncation of the series in  Fig. \ref{fig:part/bs}
is equivalent to perturbation theory in $\lambda$,
and cannot produce a non-vanishing $s_0$.
We are here facing truly non-perturbative aspects 
of renormalization, which resemble a phase transition. 

Faced with regulator dependence,
one needs to 
modify the leading-order calculation, that is,
to change the bookkeeping of the
higher-energy behavior of the theory
through the addition of at least one new interaction.
In an EFT, one adds local counterterms.
Here the only alternative is to introduce a three-body force, but
the present case is complicated by the fact that the cutoff
dependence of the amplitude is non-analytic around $p=0$.
This, however, does not mean that the renormalization program
in this low-energy EFT is doomed: a three-body force term of 
sufficient strength
contributes not only at tree level, but also in loops
dressed by any number
of two-particle interactions. 
It is convenient to rewrite the
three-body force as 
\begin{equation}
h(\Lambda)=\frac{2mg^2 H(\Lambda)}{\Lambda^2}.
\end{equation}
$H(\Lambda)$ has to be at least big enough to give a non-negligible
contribution in the $p\sim k\sim \Lambda$ region. 
This means that the dimensionless quantity
$H(\Lambda)$ has to be at least of $O(1)$. 
Assuming minimal strength, the three-body force has the feature that its
contribution to the inhomogeneous term is small compared
to the contribution from the two-body interaction, as it is at most 
$p/\Lambda$ of the latter.
When $p\gg 1/a_2$ (but $k\sim 1/a_2$), the amplitude satisfies a new 
approximate equation, with a three-body interaction added
to the rhs of Eq. (\ref{asphieq}).
The solution $K(k,p\sim \L)$ has a complicated form.
But, because in the range $1/a_2\ll p\ll \Lambda$ the three-body force term
is suppressed by $p/\Lambda$ compared to the logarithm, the
behavior
(\ref{assol}) is still correct in the intermediate region.
Now the phase is function of both $\L$ and $H(\L)$,
so if $H(\L)$ is chosen appropriately, we can make $\Lambda_\star$
cutoff independent. 
Matching with the $p\sim 1/a_2$ solution should then determine
the scattering length $a_3=a_3(\Lambda_\star)$ and the low-energy dependence
of the amplitude. 
For $p\sim 1/a_2$, Eq. (\ref{aeq}) becomes 
\begin{eqnarray}
\frac{3}{4}  
\frac{K(k,p)}{\left(\frac{1}{a_2}+\sqrt{\frac{3p^2}{4}-mE}\right)}
& = &\frac{1}{pk}{\rm ln}\left(\frac{p^2+pk+k^2-ME}{p^2-pk+k^2-ME }\right)
\nonumber \\& & 
+\frac{2\lambda}{\pi p}\int_0^\mu \: dq\ 
{\rm ln}
    \left(\frac{p^2+pq+q^2-ME}
               {p^2-pq+q^2-ME}\right) \frac{q K(k,q)}{q^2-k^2}
\nonumber \\& & 
+\frac{4\lambda}{\pi} \int_\mu^\L \: dq 
    \left[\frac{1}{q^2}
    +\frac{H(\L)}{\L^2}\right] K(k,q), \label{lowphieq} 
\end{eqnarray}
\noindent
where $\mu$ is an arbitrary scale such that $\mu\ll \L$,
and we have dropped terms which are smaller by powers of $p/\L$
or $\mu/\L$.
One can use this equation to obtain an approximate expression for $H=H(\L)$.
By integrating out modes between $\L'>\L$ and $\L$ and demanding that
the low-energy amplitude be unchanged, we find \cite{3stoogesbosons} 
\begin{equation}
\label{h}
H(\L)= -\frac{\sin(s_0\ln(\frac{\L}{\L_\star})-
                   {\rm arctg}(\frac{1}{s_0}))}
                 {\sin(s_0 \ln(\frac{\L}{\L_\star})+
                   {\rm arctg}(\frac{1}{s_0}))}. \label{H}
\end{equation}
\noindent
The last diagram in Fig. \ref{fig:part/bs} can absorb
the high-momentum modes of the diagram that precedes it. 

The full solution of Eq. (\ref{aeq}) now depends, apart from
two-body parameters, on  the new, physical parameter $\L_\star$
(but not on the regulator).
This equation was solved  numerically with a non-vanishing $H(\L)$
for several cutoffs.
The three-body force
that is necessary to keep  $a_3$ 
unchanged was found in good accord with the analytical approximation
(\ref{H}).
In Fig. \ref{fig:0} we see the results for 
the corresponding $K(k,k)^{-1}=k \cot\delta$,
where $\delta$ is the $S$-wave phase shift for particle/bound-state
scattering,
for different cutoffs
in the case $a_3=1.56 a_2$.
The effective range, for example, is predicted as $r_3= 0.57 a_2$.

\begin{figure}[t]
\begin{center}
\epsfxsize=8cm
\centerline{\epsffile{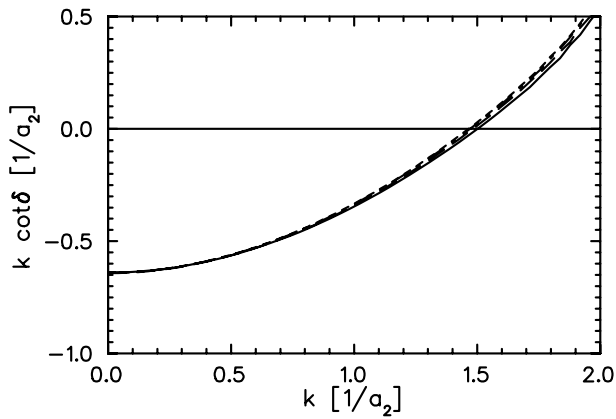}}
\end{center}
\caption{$k\ {\rm cot}\delta$ in units of $1/a_2$
for $S$-wave particle/bound-state scattering
in the $S=0$ case as a function of $k$ in units of $1/a_2$
in the EFT to leading order,
for $\L_\star =19.5 a_2^{-1}$ and different 
cutoffs ($\Lambda=42.6,\,100.0,\,230.0,\,959.0\times a_2^{-1}$).}
\label{fig:0}
\end{figure}

We can extend the preceding analysis to the three-body bound-state problem.
Since the inhomogeneous term 
did not play an important role in the above ultraviolet arguments,
the latter hold for any bound
state with binding energy $B_3$ comparable with $1/ma_2^2$.
Such bound states are shallow, 
having a size $\sim 1/\sqrt{m B_3}\sim a_2$,
and thus should be within range of the EFT.
In principle, all bound states with size larger than
$\sim r_2$ should be amenable to this EFT description.
Their properties will be determined in first approximation
by only $C_0$ and $\Lambda_\star$,
while more precise information can be obtained in an
expansion in $\aleph/M$.
In Fig. \ref{fig:B3} binding energies are shown 
for a range of cutoffs, with the three-body force adjusted to
give a fixed scattering length $a_3=1.56 a_2$.
(For particular values of $a_3$ and $\L$,
these results reduce to those in Ref. \cite{kharchenko}.)
As we can see, for this value of $\L_\star$
there exists a shallow bound state at $B_3 \simeq 2/ma_2^2$
whose binding energy is independent of the cutoff.
This bound state has size $\sim 0.7 a_2$ and can thus be 
studied with the EFT.
As we increase the cutoff, deeper bound states appear
at $\Lambda_n= \Lambda_0 \exp(n\pi/s_0)$ with $n$ an integer and 
$\Lambda_0\simeq 10$,
so that for $\Lambda_{n-1}\le \Lambda\le \Lambda_n$ 
there are $n+1$ bound states.
Similar results can be obtained for other values of $a_3$.
By repeating this procedure we can in fact determine $a_3=a_3(\Lambda_\star)$
(or {\it vice-versa}) and 
$B_3=B_3(\Lambda_\star)$ (or {\it vice-versa}). 
Eliminating $\Lambda_\star$, a universal curve $B_3=B_3(a_3)$ can
be obtained. In the case of the shallowest bound state,
this is seen in Fig. \ref{fig:phil}. 
This curve is known in the three-nucleon case as
the Phillips line \cite{phillips}. It has been derived before in the
context of models for the two-nucleon potential that
differ in their high-momentum behavior.
Varying among two-particle potential models
one could expect to fill up the $B_3\times a_3$ plane.
In the EFT, the high-momentum behavior of the two-particle potential
is butchered, which causes no trouble in describing two-particle scattering
at low-energies, but in the peculiar way described here, does require
a three-body force. 
Varying among two-particle potential models
is thus equivalent to varying the one parameter $\Lambda_\star$
of the three-body force. This spans a single curve $B_3=B_3(a_3)$
in the $B_3\times a_3$ plane. 
Our argument suggests that the Phillips line is a generic consequence
of renormalization group invariance.

\begin{figure}[t]
\begin{center}
\epsfxsize=8cm
\centerline{\epsffile{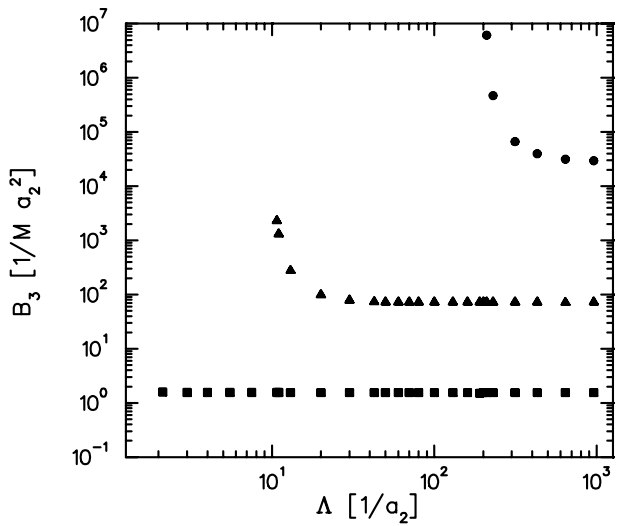}}
\end{center}
\caption{Three-body binding energies $B_3$ in units of $1/m a_2^2$
as functions of the cutoff 
$\Lambda$ in units of $1/a_2$ in the EFT
to leading order in the $S=0$ case, for $\Lambda_\star=19.5 a_2^{-1}$.}
\label{fig:B3}
\end{figure}

\begin{figure}[tb]
\begin{center}
\epsfxsize=8cm
\centerline{\epsffile{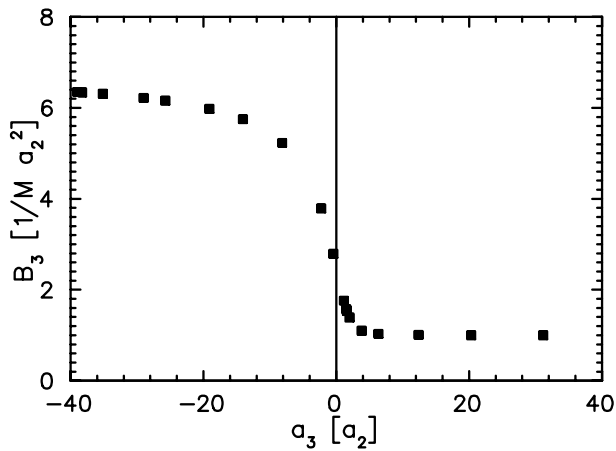}}
\end{center}
\caption{Binding energy $B_3$ of the shallowest bound 
state in units of $1/m a_2^2$
as function of the particle/bound-state
scattering length $a_3$ in units of $a_2$
in the EFT to leading order in the $S=0$ case.}
\label{fig:phil}
\end{figure}

Corrections to this calculation
come from operators with more derivatives, and they are
discussed in Ref. \cite{3stoogesbosons}.
The first correction comes from $C_2$ in Eq. (\ref{lag}).
It can be shown that the large $\L$ dependence it introduces
can be absorbed in $h$ itself, so that a finite, calculable 
contribution remains.
Its effects are being calculated \cite{more3stooges}.
It is reasonable to assume then that more-derivative three-body forces
do {\it not} get further enhanced by inverse powers
of $\aleph$, so that
\begin{equation}
D_{2(n+1)}^{...(R)}= \frac{D_{2n}^{...(R)}}{M^2}, 
\end{equation}
In this case effects of three-body-force corrections
could be accounted for perturbatively.

The above results hold for $^4$He atoms \cite{3stoogesbosons}. In fact,
it has recently been established that the two-$^4$He bound state
(``dimer'') is very shallow, with an average size
$\langle r \rangle= 62\pm 10$ \AA \ \cite{dimer},
more than an order of magnitude larger than the
range of the interatomic potential. 
The low-energy two-$^4$He system should then be describable 
by contact two-body forces. In leading order,
the measured size translates into a scattering length
$a_2=124$ \AA, which determines 
the strength of the contact interaction. 
Unfortunately, although the three-$^4$He (``trimer'') has been observed
\cite{trimer}, there seems to be no low-energy
information on its properties nor on $^4$He/dimer scattering.
At least one three-body datum is needed 
to determine $\Lambda_\star$.
Until such datum becomes available, we can only
illustrate the method by using
a phenomenological $^4$He-$^4$He potential
as a model of a microscopic theory.
We select a potential \cite{heliumpot} 
that is consistent with the recent measurement of the dimer binding energy.
It gives for the two-body system
$a_2=124.7$ \AA \ and $r_2\simeq 7.4$ \AA.
Three-body calculations are much more difficult to perform with such
a phenomenological potential.
An estimate for the $^4$He/dimer scattering length
is $a_3=195$ \AA. Ground and excited bound states have been reported;
estimates for the shallowest bound state place it in the
range $B_3= 1.04-1.7$ mK, while a deeper state 
lies around $B_3= 0.082-0.1173$ K.
There exists a prediction for the low-energy $S$-wave phase shift,
albeit for a different potential,
but $r_3$ could not be determined.
Using such model we can estimate the range of
validity of the EFT in momentum to be $\sim 1/r_2 \simeq 0.14$ \AA$^{-1}$, 
and the leading
order to give an accuracy of $\sim r_2/a_2 \simeq 0.06$,
or about 10\%,
at momenta $Q\sim 1/a_2$.
Using  this potential's $a_3/a_2=1.56$ we can determine $\Lambda_\star$,
and predict both the energy dependence in $^4$He/dimer scattering and the 
trimer binding energy. 
In fact, in Fig. \ref{fig:0} 
I have used exactly this value of $a_3$,
so it 
represents the lowest-order prediction
for $k \cot \delta$ for atom/dimer scattering, from which we can 
extract
an effective range $r_3= 71$ \AA.
{}From Fig. \ref{fig:B3} we predict a bound state at
$B_3=1.2$ mK, which is certainly within the EFT. 
The next-to-shallowest bound state is small enough to be at 
the border of EFT applicability. For a sufficiently
large cutoff, we find $B_3=0.057$ K, but in the best case scenario
corrections from higher orders should be very important.

Because of the similarity of the integral equations,
our arguments should apply to systems of three fermions 
with internal quantum numbers as well. 
Preliminary results for the three-nucleon system in the $J=1/2$ channel
confirm this expectation and produce
good $Nd$ scattering and triton results \cite{more3stooges}.
The results of this section hold
in the EFT where the pion has been integrated out,
and can be straightforwardly generalized
to the EFT with perturbative pions.
Work in the latter is now in progress for the $J=3/2$ channel
\cite{paulo+griess}. 
We also plan to extend these calculations to the four- and
more-nucleon systems \cite{more3stooges}.
We have seen in this section how we need to introduce a
three-body force in leading order to prevent the collapse
of the three-body system that would result from purely
attractive two-body contact interactions.
Is this repulsion enough to allow a prediction of other few-body
bound states?

\subsection{Moderate energies} \label{subsec-3Nmod}

Increasing the range of the EFT, we affect the ultraviolet
behavior of the leading-order two-body amplitude.
As we hit $Q\sim M_{NN}$, contributions from pions
become comparable to the leading contact interaction.
The results of the previous subsection do not necessarily apply,
as the two-body interaction is no longer purely attractive,
but might display short-range repulsion besides long-range attraction.
Here I continue in the vein of Sect. \ref{subsec-2Nmod},
assuming that no enhancements
appear in contact interactions.
The power counting of that section can then be extended
straightforwardly to an $A$-nucleon system.

Consider an arbitrary contribution to an irreducible graph involving
$A$ nucleons. This being only part of the full amplitude,
it will in general have $C$ separately connected pieces
($C=1$ for $A=0,1$; $C=1, ..., A-1$ for $A\ge 2$.)
The straightforward generalization of Eq. (\ref{A=2index}) is
\cite{inwei6,inwei5}
\begin{equation}
\nu=4-A+2(L-C)+\sum_i V_i \Delta_i.  \label{E:vkolck:nu}
\end{equation}
Again, because $C$ is bounded from above ($C\le C_{max}$),
the chiral symmetry constraint $\Delta_i\ge 0$ still
implies a lower bound on the power of $Q$,
$\nu\ge \nu_{min}=4-A-2C_{max}$ for strong interactions,
so that systematic calculations can be carried out.

With this power counting we can
get some insight into few-body forces (those diagrams 
with $C=1$). 
The new forces that appear in systems with more than two 
nucleons have been considered in 
Refs. \cite{ciOvK,inwei5,invk,civK1}. 
The dominant potential, at $\nu=6-3A=\nu_{min}$, 
is simply the two-nucleon potential of lowest order that 
we have already encountered in Sect. \ref{subsec-2Nmod}.
We can easily verify that a three-body potential will arise at 
$\nu=\nu_{min}+2$, a four-body potential at $\nu=\nu_{min}+4$, and so on. 
It is (approximate) chiral symmetry therefore that
implies that $n$-nucleon forces $V_{nN}$ are expected to obey a hierarchy
of the type
\begin{equation}
\langle V_{(n+1)N}\rangle/\langle V_{nN}\rangle
\sim O((Q/M_{QCD})^2),
\end{equation}
with $\langle V_{nN}\rangle$ denoting the contribution per $n$-plet.
This is a non-trivial consequence of chiral symmetry,
as there exist non-chiral models that produce large
three-body forces. 
If 
$|\langle V_{2N}\rangle | \sim 10$ MeV 
as in Sect. \ref{subsec-2Nmod},
we can estimate 
$|\langle V_{3N}\rangle | \sim 0.5$ MeV,
$|\langle V_{4N}\rangle | \sim 0.02$ MeV, and so on. 
This is in accord with detailed few-nucleon calculations using 
phenomenological potentials.

It proves instructive to look at the 
form of the first few terms in the
few-nucleon potential \cite{ciOvK,invk,civK1}.
We do so with time-ordered perturbation theory,
as before. 
At $\nu=\nu_{min}+2$, in addition to corrections to the two-nucleon
potential, one also finds diagrams
involving either three nucleons or two pairs of nucleons
connected via leading contact interactions and static pions.
One finds \cite{inwei6} that the various
orderings of Fig. \ref{F:vkolck:cancel}(a) add to zero. 
Diagrams like those in Fig. \ref{F:vkolck:cancel}(b,c)
give non-vanishing contributions
to three- and double-pair potentials.
However, it is easy to prove \cite{inwei5,civK1}  that
these contributions cancel, to this order in the expansion,
against contributions from reducible graphs 
like those in Fig. \ref{F:vkolck:cancel}(d,e)
involving
the leading and the energy-dependent pieces of the two-nucleon
potential.
This happens because a small energy denominator combines with
the small energy-dependent OPE to produce a result
not only of the same order, but exactly opposite to
the three- and double-pair potentials.
This cancellation had been noted before in the case of the TPE $3N$ 
force \cite{14,friarcoon}, but its model independence and 
generality are particularly 
clear in the EFT. Refs. \cite{jimandsid,epelbaoum} have
emphasized that redefining the potential
to eliminate energy dependence leads to
no $3N$ TPE forces of this type at all. 
Remaining, to this order, are only ``true'' three-body forces generated by
the delta isobar, and
at $\nu_{min}+3$ further terms with similar structure
arise \cite{invk,civK1} ---see
Fig. \ref{F:vkolck:V3N}.
Up to this order, there are no four-nucleon forces, and
the three-nucleon potential
has components with three different ranges: neglecting relativistic
corrections,
\begin{eqnarray}
 V_{3}^{(3)}(\vec{q}_{ij},\vec{q}_{jk}) & = & e_{1}\boldt_{i}\cdot\boldt_{k}
 +e_{2}\vec{\sigma}_{i}\cdot\vec{\sigma}_{k}\boldt_{i}\cdot\boldt_{k}+
 e_{3}\vec{\sigma}_{j}\cdot(\vec{\sigma}_{i}\times\vec{\sigma}_{k})
\boldt_{j}\cdot(\boldt_{i}\times\boldt_{k}) \nonumber  \\
 & & -\frac{2g_{A}}{F_{\pi}^{2}}\frac{1}{w_{jk}^{2}}\vec{\sigma}_{k}\cdot
     \vec{q}_{jk}\left[d_{1}(\boldt_{i}\cdot\boldt_{k}\vec{\sigma}_{i}
     +\boldt_{j}\cdot\boldt_{k}\vec{\sigma}_{j})
     -2d_{2}\boldt_{j}\cdot(\boldt_{i}\times\boldt_{k})
     \vec{\sigma}_{i}\times
     \vec{\sigma}_{j}
         \right]\cdot\vec{q}_{jk} \nonumber  \\
 & & +8\left(\frac{2g_{A}}{F_{\pi}^{2}}\right)^{2}
                              \frac{1}{w_{ij}^{2}w_{jk}^{2}}
     \vec{\sigma}_{i}\cdot\vec{q}_{ij}\vec{\sigma}_{k}\cdot\vec{q}_{jk} \times
 \nonumber \\ & &
\makebox[1.0in][r]{}   
     \times \left[\boldt_{i}\cdot\boldt_{k}(c_{3}\vec{q}_{ij}\cdot\vec{q}_{jk}+
     2c_{1}m_{\pi}^{2}) 
     +c_{4}\boldt_{j}\cdot(\boldt_{i}\times\boldt_{k})
     \vec{\sigma}_{j}\cdot(\vec{q}_{ij}\times\vec{q}_{jk})
       \right]
     \nonumber  \\
 & & + \: \mbox{two cyclic permutations of} \:\:(ijk),
\label{3Npotent}
\end{eqnarray}
with 
$\vec{p}_{i}(\vec{p}_{i}^{\,'})$ the initial (final) momentum of nucleon $i$, 
$\vec{q}_{ij}\equiv\vec{p}_{i}-\vec{p}_{i}^{\,'} 
              = \vec{p}_{j}^{\,'}-\vec{p}_{j}$, 
and $w_{ij}^2\equiv \vec{q}_{ij}^{\, 2}+m_{\pi}^{2}$.
The parameters $c_i$ and $d_i$ are given by
Eqs. (\ref{c's}) and (\ref{d's}) in terms 
of the $B_i$ from Eq. (\ref{L1}), 
the $\pi N\Delta$ coupling $h_A$ from Eq. (\ref{L0}),
the $D_i$ from Eq. (\ref{dilag}),
and the $NNN\Delta$ contact interaction $D_T$ from Eq. (\ref{DTlag}).
Likewise, the $e_i$ are parameters that have
delta and shorter-range components $E_i$,
\begin{equation}
e_{1}=E_1, 
\;\; e_{2} =E_2 + \frac{D_{T}^{2}}{9\, \delta m},
\;\; e_{3} =E_3 - \frac{D_{T}^{2}}{18\, \delta m}.
\end{equation}

\begin{figure}[t]
\centerline{\epsfysize=2.5cm \epsfbox{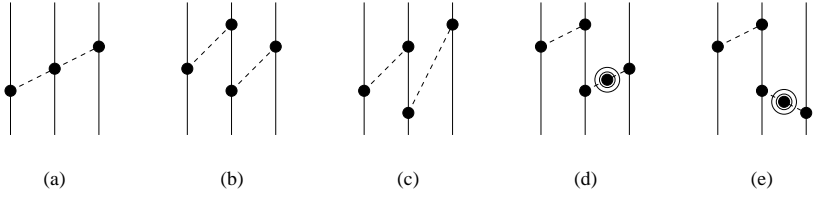}}
\vspace{0.5cm}
\caption{Some time-ordered diagrams whose contributions 
add to nothing in the EFT:
(a) TPE from the Weinberg-Tomozawa seagull, of which only
one ordering is shown;
(b,c) ``uncorrelated'' TPE three-nucleon potential and
(d,e) iteration of energy-dependent OPE two-nucleon potential.  
Solid lines represent nucleons,
dashed lines pions, a heavy dot an interaction in ${\cal L}^{(0)}$,
and a dot within two circles an interaction in ${\cal L}^{(2)}$.
The same cancellation (b-e) occurs in other sets of
three-nucleon diagrams corresponding to other orderings.
It also takes place in diagrams where
short-range interactions are substituted for static OPE,
and in sets of double-pair diagrams.}
\label{F:vkolck:cancel}
\end{figure}

\begin{figure}[t]
\centerline{\epsfysize=10cm \epsfbox{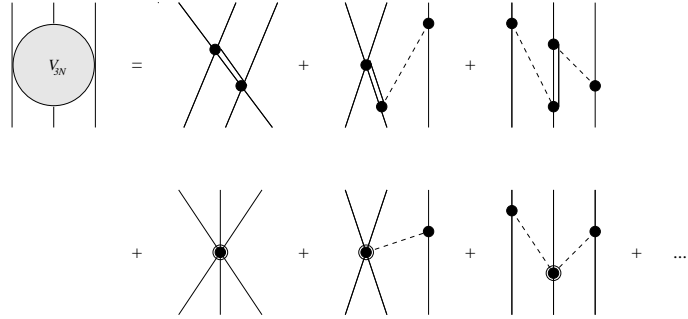}}
\vspace{-5.50cm}
\caption{Some time-ordered diagrams contributing to the three-nucleon potential
$V_{3N}$ in the EFT.
(Double) solid lines represent nucleons (and/or deltas), 
dashed lines pions, a heavy dot an interaction in ${\cal L}^{(0)}$,
and a dot within a circle an interaction in ${\cal L}^{(1)}$.
First line corresponds to $\nu=\nu_{min}+2$,
second line to $\nu=\nu_{min}+3$, 
and ``$\ldots$'' denote $\nu\geq \nu_{min}+4$. 
All nucleon permutations and orderings with at least one pion
or delta in intermediate states are included.}
\label{F:vkolck:V3N}
\end{figure}

The TPE part of the potential (\ref{3Npotent}) is determined
in terms of $\pi N$ scattering observables. 
It is {\it not} identical to the Tucson-Melbourne potential \cite{TM}.
The latter was built from $\pi N$ scattering through a particular
choice of pion fields 
for which chiral symmetry is not respected term-by-term \cite{vk:huber}.
Although the corresponding on-shell $\pi N$ amplitude gives the same
result as Eqs. (\ref{tpiN}) and (\ref{abcde}), they differ off shell.
The corresponding three-body potential involves other interactions 
to guarantee chiral symmetry and to enforce invariance under pion-field
redefinitions, but unfortunately such terms have not 
been included in the original Tucson-Melbourne potential.
After this is done,
all commonly used TPE three-nucleon potentials
agree, except for the numerical values of
parameters that depend on the way $\pi N$ data is fitted 
---see Table \ref{tab:3Nforces} \cite{vk:huber}.
The novel
one-pion/short-range components of the potential (\ref{3Npotent}) can be
related to pion production in $NN$ collisions.
They are starting to be explored in conjunction with
more conventional treatments of few-nucleon systems
\cite{huberInd,carlsonInd}.
The purely short-range components can only be determined from
few-nucleon systems, but to leading order depend only
on one undetermined parameter, $D_T$.
(Recall that $D_T$ could not be isolated in the $\nu=\nu_{min}+3$ EFT fit of
$NN$ phase shifts.)
Relativistic corrections neglected above 
are discussed in Ref. \cite{friarcoon}.

\begin{table}[t]
\centering
\caption{Low-energy $\pi N$ scattering parameters (with Born
terms removed) in a variety of existing TPE three-nucleon forces.
\label{tab:3Nforces}}
\begin{tabular}{c||cccc}
Three-Nucleon Force & $(a-2m_\pi^2c)\, m_\pi $ & $b \, m_\pi^{3}$ & 
$c \, m_\pi^{3}$ & $d \, m_\pi^{3}$\\ 
\hline 
Fujita-Miyazawa \cite{fm}          & $ 0.0$  & $-1.15$ & 0.0  & $-0.29$\\
Tucson-Melbourne \cite{TM,mc}      & $-1.03$ & $-2.62$ & 1.03 & $-0.60$\\ 
Brazil \cite{brazil,mc}            & $-1.05$ & $-2.29$ & 1.05 & $-0.77$\\ 
Urbana-Argonne \cite{ua,li6}       & $ 0.0$  & $-1.20$ & 0.0  & $-0.30$\\ 
Texas \cite{civK1,fms}             & $-1.87$ & $-3.82$ & 0.0  & $-1.12$\\
Ruhr(Pot) \cite{rp}                & $-0.51$ & $-1.82$ & 0.0  & $-0.48$\\ 
\end{tabular}
\end{table}

\section{Probes of few-nucleon systems} \label{sec-2.5N}

Further constraints on our understanding of few-nucleon systems
come from the study of processes involving external probes,
such as pions, photons, electrons and neutrinos. 
In some cases the calculation
in the EFT to a certain order is completely determined
by parameters already known from other processes;
a prediction can be made, accompanied by an educated
guess about the error resulting from neglecting higher orders.
Other cases can be used to determine so-far unknown parameters;
this is particularly useful in the case of neutron
and neutral pion parameters that cannot be easily determined 
in processes with a single-nucleon target ---see Sect. \ref{sec-01N}
for examples.
But even in these cases, 
a prediction for energy dependence is usually possible. 

\subsection{Very low and low energies} \label{subsec-2.5Nverylowlow}

In the EFT with virtual pions integrated out,
scattering of pions on few-nucleon
systems is in itself not very interesting, as pions behave
just as heavy particles. More interesting are processes involving
photons of momenta $Q\sim \aleph$, which we can
use as efficient probes of momentum distributions.
As we have seen, this EFT is equivalent in the two-nucleon sector
to effective range theory. 
An extensive literature exists using this technique in the calculation
of electroweak process involving the deuteron, 
starting with the pioneering work of Refs. \cite{Peierls,bethelongmire}.  
These calculations can be recast in terms of the very-low-momentum
EFT in a straightforward way.
We briefly describe here some of the more modern calculations,
carried out with explicit but perturbative pions, that is, in the
low-momentum EFT.
Calculations involving the two-nucleon system
proceed by separating ``irreducible'' contributions,
which are those that do not split in two when cut at any $C_0$ vertex.
Always contributing is an irreducible two-nucleon function,
intimately related to the two-nucleon amplitude.
Other irreducible components include the external probes.
Each irreducible part is calculated to the same
order in the $Q/M_{NN}$ expansion, they are combined in the
amplitude of the processes of interest,
and only terms up to the desired power of $Q/M_{NN}$ are kept.
These calculations typically contain the result of the lower-scale EFT;
improvement due to the  additional pion effects, if any,
can be taken as evidence for $M_{NN}>m_\pi$.

{\it $\bullet$ Deuteron form factors.}
A calculation of the deuteron form factors in subleading order
was carried out in Ref. \cite{ksw-ff}.
To this order the deuteron is pure $S$ wave, determined by the
$^3S_1$ parameters $C_0+m_\pi^2 C_{2qm}$ and $C_2$.
Interactions with a photon field come, at this order, from known
one-nucleon terms, from minimal substitution in the $C_2$ term,
and from one 
{\it a priori} unknown two-body magnetic-interaction
counterterm, which can only be fitted to data. 
With $C_0+m_\pi^2 C_{2qm}$ fitted to the deuteron binding
and $C_2$ to the phase shifts,
the results for static moments are in 
Table \ref{tab:3muskdparam} \cite{ksw-ff}.
The momentum dependence of the charge and magnetic
form factors can then be predicted.
Good convergence and agreement with data are attained
even at momenta as large as $\sim 400$ MeV;
only the quadrupole moment works poorly.
Unfortunately no improvement over the effective range expansion is seen.
Comparison with effective range theory for the radius and
EFT with non-perturbative pions for the quadrupole
moment
suggest that the most important of the neglect terms 
are from iterated pion exchange \cite{ksw-ff}.
Sub-subleading contributions to the quadrupole
moment are discussed in Ref. \cite{binger}.
The anapole moment of the deuteron was studied in Ref. \cite{anapole}.

\begin{table}[t]
\caption{Results from EFT at leading (LO) and subleading (SLO) order
and experimental values 
for deuteron rms charge radius ($\langle r^2 \rangle^{1/2}_{ch}$),
magnetic dipole moment ($\mu_d$), and
electric quadrupole moment ($Q_E$).
\label{tab:3muskdparam}}
\vspace{0.0cm}
\begin{center}
\footnotesize
\begin{tabular}{c|ccc}
quantity & LO & NL+SLO & expt\\
\hline
$\langle r^2 \rangle^{1/2}_{ch}$ & 1.53 & 1.89  &  2.1303(66)\\
$\mu_d$                          & 0.88 & 0.86 (fit)  & 0.857406(1)\\
$Q_E$                            &  0   & 0.40  & 0.2859(3)\\
\end{tabular}
\end{center}
\end{table}

{\it $\bullet \hspace{.2cm} n p \rightarrow \gamma d$ at threshold.}
To subleading order this reaction proceeds from the $^1S_0$ state;
the same set of $^1S_0$ and $^3S_1$
parameters $C_0+m_\pi^2 C_{2qm}$ and $C_2$ appear
and are determined by phase shifts and deuteron binding.
In leading order only one-body currents from magnetic coupling
appear. At subleading order pion-exchange currents are calculated and
found to be cutoff dependent; a four-nucleon/one-photon
operator appears at the same order and absorbs this dependence
leaving an undetermined finite part behind,
which can then be fitted.
Results \cite{scaldeferri} are presented in
Table \ref{tab:kevradcap}.
The parity-violating asymmetry in the angular distribution
of gamma rays from radiative capture of polarized cold neutrons was
calculated in Ref. \cite{parityasym}.

\begin{table}[b]
\caption{Values for 
the total cross-section $\sigma$ in mb
for radiative neutron-proton capture:
leading (LO) and subleading (SLO) order EFT  
and experiment (expt).
\label{tab:kevradcap}}
\vspace{0.0cm}
\begin{center}
\footnotesize
\begin{tabular}{ccc}
LO & NL+SLO & expt\\
\hline
297.2 & 334.2 (fit) &  334.2$\pm$0.5\\
\end{tabular}
\end{center}
\end{table}

{\it $\bullet \hspace{.2cm} \gamma d \rightarrow \gamma d$.}
Amplitudes with two external photons give rise to interesting
predictions.
In leading non-vanishing order only one-nucleon operators from
subleading interactions 
enter, including the Thomson seagull operator 
and interactions from minimal coupling.
At subleading order pion exchange and $C_2$ appear as well.
Scalar and tensor, electric and magnetic polarizabilities have been computed
in Ref. \cite{deupolar} and are shown in Table \ref{tab:deupol}.
Convergence seems consistent with static electromagnetic moments,
and results are comparable to the effective range expansion.
Nucleon polarizability enters as a higher-order contribution, and
therefore it is unlikely that it can be isolated from the 
deuteron polarizabilities.
Differential cross-sections for 
Compton scattering  at photon energies of
49 and 69 MeV were calculated in Ref. \cite{theircompton}. 
The parameter-free results are in good agreement with existing data,
in particular at the lower energy.
Nucleon polarizabilities from pion loops enter to subleading
order and are important for this agreement.
Tensor polarized scattering was computed in  Ref. \cite{tensorpol}.
At $90^o$ the first non-vanishing contribution
comes only from the pion exchange, and therefore might allow
a sensitive test of the goodness of perturbative pions.

\begin{table}[t]
\caption{Scalar and tensor electric ($\alpha_0$ and $\alpha_2$)
and magnetic ($\beta_0$ and $\beta_2$)
polarizabilities of the deuteron in fm$^3$:
leading (LO) and subleading (SLO) order EFT
\label{tab:deupol}}
\vspace{0.0cm}
\begin{center}
\footnotesize
\begin{tabular}{ccc}
polarizability & LO & NL+SLO  \\
\hline
$\alpha_{0}$ (fm$^3$) & 0.386 & 0.595    \\
$\alpha_{2}$ (fm$^3$) & 0     & $-$0.062 \\
$\beta_0$    (fm$^3$) & 0.067 &  ?  \\
$\beta_2$    (fm$^3$) & 0.195 &  ?  \\
\end{tabular}
\end{center}
\end{table}

\subsection{Moderate energies} \label{subsec-2.5Nmod}

The power counting arguments of
Sect. \ref{subsec-3Nmod} 
can be easily generalized to the case where
external particles with momenta $Q\sim M_{nuc}$
interact with few-nucleon systems.
We define the kernel $K$ as the sum of irreducible diagrams
to which the probes are attached.
A generic diagram contributing to a full amplitude will consist
of irreducible diagrams sewed together by states 
with small energy denominators. 
Interactions among nucleons occurring before or after
scattering can be treated as before: iteration
of the potential gives rise to the wave-function
$|\psi\rangle$ ($|\psi'\rangle$) of the initial (final) nuclear state.
The full scattering amplitude is then 
\begin{equation}
T\sim \langle \psi'|K|\psi\rangle.
\end{equation}
The power counting (\ref{E:vkolck:nu}) applies equally well to the kernel $K$. 
In practice, it is frequently desirable to minimize nuclear
wave-function errors by using a high-precision
phenomenological potential instead of the only
currently available EFT potential \cite{ciOLvK}.
That this is a good approximation is suggested by
a comparison \cite{latestrho}
between a simplified version of the EFT potential 
of Ref. \cite{ciOLvK} and the Argonne V18
potential \cite{argonne}, which show agreement
in most of the wave-function features.

Before plunging into hard results, let me point out the generic results
of this EFT. 
Because of the factor $-2C$ in Eq. (\ref{E:vkolck:nu}), we
see immediately ---in an effect similar to few-nucleon forces--- that
external probes will tend to interact 
predominantly with a single nucleon, simultaneous interactions with more than 
one nucleon being suppressed by powers of $(Q/M_{QCD})^2$. Again, this 
generic dominance of the impulse approximation is a 
well-known result that arises naturally here. 
This is of course what allows extraction, to a certain accuracy, of  
one-nucleon parameters from nuclear experiments. 
A valuable by-product of the EFT is to provide a consistent
framework for one- and few-nucleon dynamics, where
few-nucleon process can be used to infer one-nucleon properties.
More interesting from the purely nuclear-dynamics perspective are, however,
those processes where the leading single-nucleon contribution vanishes by 
a particular choice of experimental conditions, 
for example the threshold region. 
In this case, two-nucleon contributions, especially in the 
relatively large deuteron, can become important. 
Further examination of the structure of the chiral Lagrangian
reveals that two-nucleon contributions tend to be
dominated by pion exchange.
Indeed, photons and pions couple to four-nucleon operators only
at $O(Q/M_{QCD})$ relative to pion-exchange diagrams
constructed out of the leading order Lagrangian.
Thus power counting justifies the ``chiral filter hypothesis''
that has been put forward to summarize some empirical results
on electroweak form factors
\cite{chiralfilter}.
This ``pion dominance'' ensures that
two-body contributions from the EFT in lowest orders
tend to be similar to those in phenomenological models that
include pion-exchange currents.

{\it $\bullet \hspace{.2cm} \pi d\rightarrow \pi d$.}
This is perhaps the most direct way to check the consistency of $\chi$PT in
one- and few-nucleon systems.
For simplicity, consider the region near threshold.
There the lowest-order, $\nu=\nu_{min}=-2$ 
contributions to the kernel vanish because 
the pion is in an $S$ wave and the target is isoscalar.
The $\nu=\nu_{min}+1$ term comes from the 
(small) isoscalar pion-nucleon seagull,
related in lowest-order to the pion-nucleon isoscalar amplitude $b^{(0)}$.
$\nu=\nu_{min}+2, \nu_{min}+3$ contributions 
come from corrections to $\pi N$ scattering
and two-nucleon diagrams, which involve besides $b^{(0)}$ also the much larger 
isovector amplitude $b^{(1)}$. Weinberg \cite{inwei5} has estimated
these various contributions to the $\pi d$ scattering length, finding 
agreement with previous, more phenomenological calculations,
which have been used to extract $b^{(0)}$. 
The impact of the deuteron process on the determination
of $b^{(0)}$ is further discussed in Ref. \cite{vkolck:beanenew}.
The size of two-nucleon, one-loop diagrams that appear 
in next order is estimated to be small
in the related, double-charge-exchange process \cite{koltun}.
Charge symmetry breaking effects are considered in Ref. \cite{rockmore}.

{\it $\bullet \hspace{.2cm} n p \rightarrow \gamma d$.}
This offers a chance of a precise postdiction.
Here it is the transverse nature of the real outgoing photon that is 
responsible for the vanishing of the lowest-order, 
$\nu=\nu_{min}=-2$ contribution
to the kernel. 
The single-nucleon magnetic contributions come 
at $\nu=\nu_{min}+1$ (tree level), $\nu=\nu_{min}+2$ (one loop), 
{\it etc.} 
The first two-nucleon term is one-pion exchange at $\nu=\nu_{min}+2$,
long discovered to give a smaller but non-negligible contribution.
There has been a longstanding discrepancy of a few percent between these
contributions and experiment. 
At $\nu=\nu_{min}+4$ there are further one-pion exchange, two-pion exchange,
and short-range terms. 
In Ref. \cite{vkolck:park1} the two-pion exchange
diagrams in the deltaless theory were 
calculated and resonance saturation used to estimate 
the other $\nu=\nu_{min}+4$ terms. With wave-functions from the Argonne V18 
potential \cite{argonne} and a cut-off $\Lambda=1000$ MeV,
the excellent agreement with experiment shown in 
Table \ref{tab:radcap} was found \cite{vkolck:park1}. 
The total cross-section changes by only 0.3\% 
if the cut-off is decreased to 500 MeV. 

\begin{table}[t]
\caption{Values for various contributions to 
the total cross-section $\sigma$ in mb
for radiative neutron-proton capture:
impulse approximation to $\nu=\nu_{min}+4$ (imp),
impulse plus two-nucleon diagrams at $\nu=\nu_{min}+2$ (imp+tn0),
impulse plus two-nucleon diagrams up to $\nu=\nu_{min}+4$ (imp+tn),
and experiment (expt).\label{tab:radcap}}
\vspace{0.0cm}
\begin{center}
\footnotesize
\begin{tabular}{cccc}
imp & imp+tn0 & imp+tn & expt\\
\hline
305.6 &  321.7 & 336.0  &  334$\pm$0.5\\
\end{tabular}
\end{center}
\end{table}

{\it $\bullet \hspace{.2cm} \gamma d \rightarrow \gamma d$.}
External photons couple to the kernel only in 
subleading order, $\nu=\nu_{min}+1=-1$, via the 
one-body Thomson seagull.
In next order, $\nu=\nu_{min}+2$, more interesting effects 
are present, such
as a pion loop which contributes to the nucleon polarizability. 
There are also two-nucleon contributions from pion exchange,
which can and need to be computed, if one seeks to extract information
about the neutron polarizability.
Work is in progress \cite{compton} to calculate the differential cross-section
at 49, 69, and 95 MeV. Preliminary results are in good agreement
with the existing Illinois data at the lowest two energies,
and a prediction will be made for the highest energy, currently
being measured and analyzed in Saskatoon.

{\it $\bullet \hspace{.2cm} \gamma d\rightarrow \pi^0 d$.}
This reaction offers the possibility to test a
prediction arising from a combination of two-nucleon contributions 
and the neutral-pion single-neutron amplitude. 
The differential cross-section for a photon of momentum $k$ to
produce a pion of momentum $q$ is, at threshold, 
$[(k/q) (d\sigma/d\Omega)]_{q=0} = 8|E_d|^2/3$.
$E_d$ has been studied up to $\nu=\nu_{min}+3$ with the delta integrated out
in Ref. \cite{vk:beane}. 
There are two classes of contributions, according to whether
the external light particles interact with one 
or with both nucleons.
The one-nucleon part of the kernel is given by the same 
$\nu=\nu_{min},\nu_{min}+1,... $
mechanisms described in Sect. \ref{sec-01N}, with due account of $P$ waves and 
Fermi motion inside the deuteron.
The neutrality of the outgoing $S$-wave pion ensures
that the leading $\nu=-2=\nu_{min}$ terms vanish. 
The first two-nucleon part of the kernel 
appears at $\nu=\nu_{min}+2$; it comes from
a virtual charged pion photoproduced on one nucleon
which rescatters on the other nucleon with charge exchange.
These contributions are actually numerically larger than indicated
by power counting due to relatively large deuteron size.
Smaller two-nucleon terms 
appear at $\nu=\nu_{min}+3$ from corrections in either nucleon.  
Results for $E_{d}$ up to $\nu=\nu_{min}+3$ \cite{vk:beane}
are shown in Table \ref{T:vkolck:Ed}.
They correspond to the Argonne V18 potential \cite{argonne} and a cutoff 
$\Lambda=1000$ MeV. Other realistic potentials and cutoffs from 650 to
1500 MeV give the same result within 5\%.
The chiral potential of Sect. \ref{subsec-2Nmod}
is more cumbersome to use, but it has been verified that
it gives $\nu=\nu_{min}+2$ 
results that are similar to other realistic potentials.  
We see that two-nucleon
contributions seem to be converging, although more convincing
evidence would come from next order, where loops appear. 
A model-dependent estimate \cite{vk:wilhelm} of some $\nu=\nu_{min}+4$ terms 
suggests a 10\% or larger error from neglected higher orders 
in the kernel itself. 
The single-scattering amplitude depends on 
amplitude for $\gamma n\rightarrow \pi^0 n$, $E_{0+}(\pi^0 n)$,
in such a way that $E_{d} \sim -1.79 -0.38(2.13-E_{0+}(\pi^0 n))$ 
in units of $10^{-3}/m_{\pi^+}$.
Thus, sensitivity to $E_{0+}(\pi^0 n)$ survives the large two-nucleon
contribution at $\nu=\nu_{min}+2$. 

\begin{table}[b]
\caption{Values for $E_{d}$ in units of $10^{-3}/m_{\pi^+}$
from one-nucleon contributions ($1N$) up to $\nu=\nu_{min}+3$,
two-nucleon kernel ($2N$) at $\nu=\nu_{min}+2$ 
and at $\nu=\nu_{min}+3$,
and their sum ($1N+2N$).\label{T:vkolck:Ed}}
\vspace{0.0cm}
\begin{center}
\footnotesize
\begin{tabular}{cccc}
$1N$   & \multicolumn{2}{c}{$2N$} &    $1N+2N$        \\
\cline{2-3}
$\nu\le \nu_{min}+3$ & $\nu= \nu_{min}+2$ & $\nu= \nu_{min}+3$   
                                          & $\nu\le \nu_{min}+3$  \\
\hline
 0.36  & $-$1.90 & $-$0.25                  & $-$1.79 \\
\end{tabular}
\end{center}
\end{table}

We see that working within the effective theory yields a testable
prediction, $E_d=-(1.8 \pm 0.2)\cdot 10^{-3}/m_{\pi^+}$. 
It is remarkable that for this process $\chi$PT gives
results that are somewhat different from tree-level models of
the type traditionally used in nuclear physics.
For example, the models in Ref. \cite{vk:justusetal} predict the
threshold cross-section about twice as large as $\chi$PT.
Most of the difference comes from one-nucleon loop diagrams:
tree-level models tend to differ from $\chi$PT mostly
by having a smaller $E_{0+}(\pi^0 n)$, which increases $|E_d|$.
A test of this prediction is thus an important 
check of our understanding of the role of QCD at low energies.
Such a test was recently carried out
at Saskatoon \cite{vk:berg}.
The experimental results for 
the pion photoproduction cross-section near threshold are shown 
in Fig. \ref{F:vkolck:piphotodata},
together with 
our $\chi$PT prediction at threshold.
Inelastic contributions have been estimated
in Ref. \cite{vk:berg} and are smaller than 10\% 
throughout the range of energies shown.
At threshold, Ref. \cite{vk:berg} finds 
$E_{d}=-(1.45\pm 0.09)\cdot 10^{-3}/m_{\pi^+}$.
While agreement with $\chi$PT to order $\nu=\nu_{min}+3$ 
is not better than a reasonable estimate of higher-order
terms, it is clearly superior to
tree-level models. This is compelling evidence of chiral loops. 
Further test will come from
an electroproduction experiment currently under analysis in Mainz
\cite{vk:bernstein}.

\begin{figure}[t]
\centerline{
\epsfysize=8cm \epsfbox{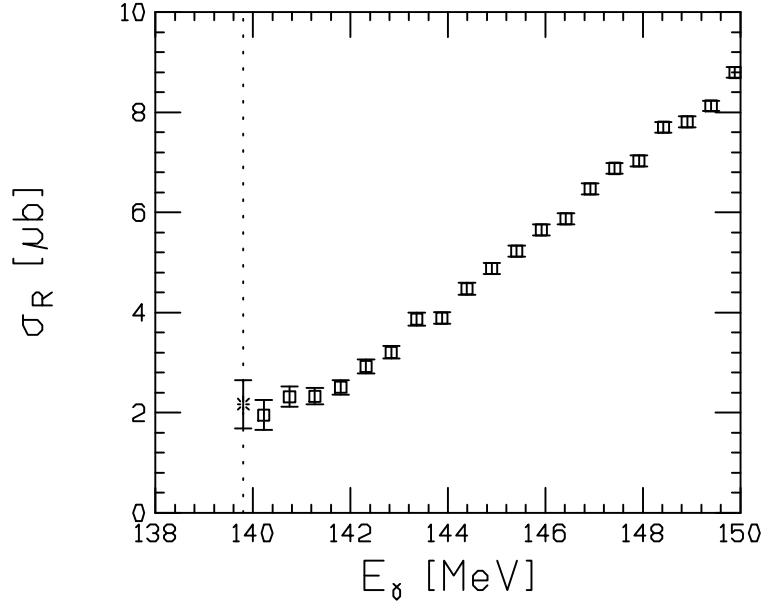}  
}
\vspace{0.05cm}
\caption[ ]{ Reduced cross-section  $\sigma_R=(k/q)\sigma$ in $\mu$b
for neutral pion photoproduction as function of the photon 
energy in MeV.
Threshold is marked by a dotted line.
Squares are data points from Ref. \protect\cite{vk:berg}
and the star at threshold is the 
$\chi$PT {\it pre}diction of Ref. \protect\cite{vk:beane}.
Figure courtesy of U.-G. Mei{\ss}ner.}
\label{F:vkolck:piphotodata}
\end{figure}

{\it $\bullet$ Axial currents.}
The long-range two-nucleon contributions to the axial current 
come from one-pion exchange in first non-vanishing order and 
from two-pion exchange as a first correction.
They have been evaluated in Ref. \cite{taesunphysrep}.
The related proton-burning process $pp \rightarrow d e^+ \nu_e$
is found \cite{taesunburning} in agreement with previous calculations.

{\it $\bullet \hspace{.2cm} pp \rightarrow pp\pi^0$ close to threshold.}
This reaction has attracted a lot of interest because of the failure of 
standard phenomenological mechanisms in reproducing 
the small cross-section observed near threshold. 
It involves larger momenta of $O(\sqrt{m_\pi m_N})$, so 
the relevant expansion parameter here is the not so small
$(m_\pi/m_N)^{\frac{1}{2}}$.
This process is therefore not a good testing ground for the above ideas. 
But $(m_\pi/m_N)^{\frac{1}{2}}$ is still $<1$, so at least in some formal sense
we can perform a low-energy expansion. 
In Ref. \cite{vkolck:cohen}
the chiral expansion was adapted to this reaction and the first few 
contributions estimated. 
(Note that ---contrary to what is stated in Ref. \cite{ulfpespelasmauns}---
momenta $\sim \sqrt{m_\pi m_N}$ do not necessarily
imply a breakdown of the non-relativistic
expansion, as $p^4/m_N^3 \sim  (m_\pi/m_N) (p^2/m_N)$.)
Again, the lowest order terms all vanish. The formally leading non-vanishing 
terms ---an 
impulse term and 
a similar diagram from the delta isobar--- 
are anomalously small and partly cancel. 
The bulk of the cross-section must then arise from contributions that 
are relatively unimportant in other processes. 
One is isoscalar pion rescattering for which two sets of $\chi$PT
parameters were used: 
``ste'' from a sub-threshold expansion of 
the $\pi N$ amplitude \cite{bkm2}
and ``cl'' from an one-loop analysis of threshold 
parameters \cite{bkm}. 
Others are two-pion exchange and short-range $\pi NNNN$ terms, 
which were modeled by heavier-meson exchange: 
pair diagrams with $\sigma$ and $\omega$ exchange, and 
a $\pi\rho\omega$ coupling, among other, smaller terms
\cite{withriska}.    
Two potentials ---Argonne V18 \cite{argonne} and Reid93 \cite{reid93}--- 
were used. 
Results \cite{withriska} are shown in Fig. \ref{F:vkolck:mesexfig3} 
together with IUCF \cite{iucf} and Uppsala \cite{Uppsala} data.
Other $\chi$PT studies of this reaction can be found in 
Ref. \cite{vkolck:pppppizero}. 
The situation here is clearly unsatisfactory, and presents therefore a unique 
window into the nuclear dynamics. Work is in progress, for example, on
a similar analysis for the other, not so suppressed channels
$\rightarrow d\pi^+, \rightarrow pn\pi^+$ \cite{vkolck:carocha}. 

\begin{figure}[tb]
\vspace{-1.5cm}
\includegraphics{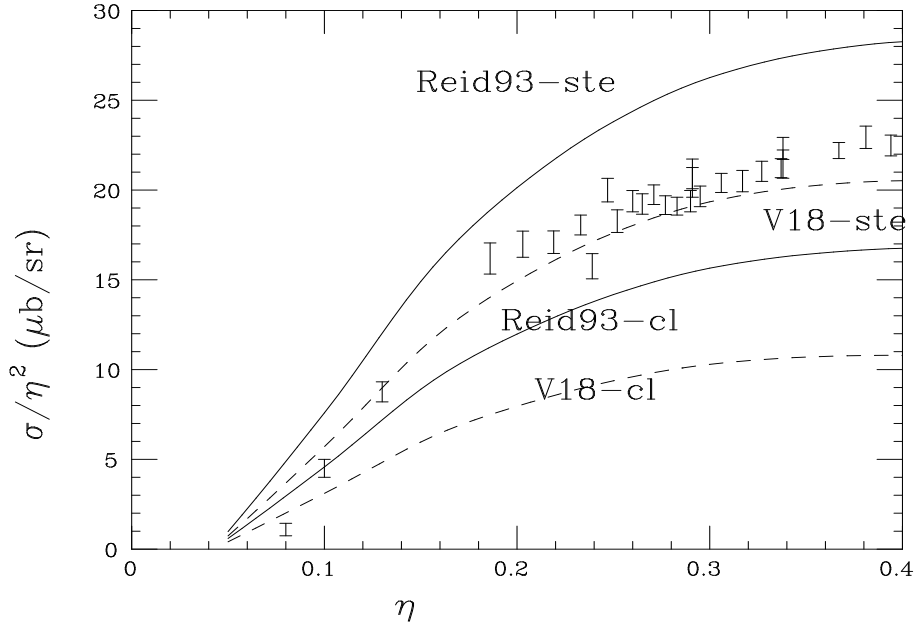}
\vspace{10.2cm}
\caption{Cross-section for $pp \rightarrow pp \pi^0$ in $\mu$b/sr
as function of the pion momentum $\eta$ in units of $m_\pi$ for 
two $NN$ potentials (Argonne V18 and Reid93) 
and two parametrizations of the isoscalar pion-nucleon amplitude (ste and
cl).}
\label{F:vkolck:mesexfig3}
\end{figure}

\section{Many-nucleon systems} \label{sec-AN}

The density of a many-body system sets the scale of typical
particle momenta $Q$. At sufficiently low density such that
$Q <\aleph$, we have seen that the EFT can be formulated
in terms of contact interactions only, and should be perturbative.
There is a vast literature on dilute systems 
(see, {\it e.g.} Ref. \cite{popov}).
Most of its results can be rederived in an EFT language.
For example, Ref. \cite{braaten} has reexamined some
of the higher corrections to the energy density of a dilute
boson gas; and Ref. \cite{papenbrock} has studied pairing
in a dilute Fermi system, producing analytical expressions 
relating the pairing gap, the density, and the energy density to 
the scattering length.
Only exploratory work has so far been carried out regarding
the more challenging, higher densities where $Q/\aleph \gaprox 1$.
This should not be surprising in view of the non-trivial
aspects that even the three-body system bears, as 
discussed in Sect. \ref{subsec-3Nverylowlow}.
Attempts to look at nuclear matter at equilibrium density
and at finite nuclei have been phenomenological;
within its limited scope, this pioneering work has 
nevertheless uncovered remarkable systematics.

It was very quickly realized that chiral Lagrangians could
be related to Walecka-type models.
Under the assumptions that pion and spin effects average to very little
in heavy nuclei, the main interactions should be given by contact
interactions. Under the further approximation of
mean field, two four-nucleon contact interactions can indeed
be related to the original Walecka model,
with parameters of the correct order of magnitude \cite{gelmini}.
More elaborate versions of the Walecka model can be shown
to require further contact interactions. 
The more extensive analysis of Ref. \cite{crazyland} has shown a remarkable
regularity in the size of parameters in a relativistic 
point-coupling model that includes 
several four-, six-, and eight-nucleon
interactions, is 
treated in Dirac-Hartree mean-field, and is fitted
to a large number of nuclear data.
This regularity was confirmed and extended
to more complete versions of the same model \cite{rusnakfurn},
and was shown to also hold in non-relativistic, 
Skyrme-force models  \cite{furnhack}.
It was found that a $2n$-nucleon field operator without derivatives
contributes $\sim (4\rho_0/F_\pi^2 M_{QCD})^{n-1} M_{QCD}$ 
to the energy per particle at equilibrium density $\rho_0$.
This is tantalizingly reminiscent of our previous
naturalness assumptions.
It has also been noted \cite{furnhack} that the
cancellation between large scalar and vector densities that
is characteristic of relativistic point-coupling models
is incorporated automatically in a smaller
contribution in the non-relativistic models.
Although these models do not include all operators
consistent with symmetries to a certain order in a definite
power counting, additions of new interactions 
with natural parameters could change
the precise values of all 
existing parameters but not their orders of magnitude.
This regularity is a nontrivial result because
at mean-field level effective interactions
absorb long distance many-body effects from ladder
and ring diagrams \cite{serotwalecka}.
It implies a convergent density expansion for mean-field
contributions to nuclear matter, as the expansion parameter
is $4\rho_0/F_\pi^2 M_{QCD}\sim 0.2$ \cite{jimreview}.
For an attempt at an EFT expansion around $ \rho_0$, 
see Ref. \cite{matthias}.
This all hints that a controlled EFT expansion for finite density
nuclear systems is possible.

\section{Outlook} \label{sec-out}

We have gone a long way in identifying the essential
ingredients of the EFT appropriate for nuclear physics.

\begin{itemize}

\item 
In the two-nucleon system,
the way fine-tuning surfaces in contact interactions,
the role of a potential and of a dimeron field,
and the virtues and limitations of various regulators
have been
understood. 
As a consequence, systematic calculations can and have been performed,
in some cases producing simple analytical results.
Two-nucleon observables work well at low energies,
but not yet at the level of phenomenological potential models.

\item 
In the three-nucleon system,
we have learned about the subtlety of renormalization
in a non-perturbative context and started exploring
the rich features of phenomenology and
three-body forces. Universal features 
---such as the Phillips line--- arise as consequence
of the renormalization group.

\item
Nuclear processes involving external particles have been treated
consistently with one-nucleon $\chi$PT,
providing a way to access neutron properties.
In most cases results consistent with conventional models 
have been obtained,
but in one case there was
a {\it pre}diction in disagreement with these models. 
This has stimulated experimental efforts
which unequivocally favor the EFT.

\item
Evidence for naturalness 
has been unearthed in many-nucleon systems.

\end{itemize}

When these developments are taken together, 
a simple physics picture of nuclei emerges from QCD.
Low-momentum nucleons can be thought of as point particles
surrounded by 
{\it (i)} 
an inner cloud which is dense but of short range $O(1/M_{QCD})$, and
{\it (ii)} 
an outer cloud made out of pions which has long range $O(1/M_{nuc})$
but is sparse.
If
a few non-relativistic nucleons are put together, 
each nucleon is unable
to distinguish details of the others' inner clouds. 
Interactions in the inner
region
can be expanded in
delta functions and their derivatives,
of progressively smaller importance. 
The outer 
cloud yields interactions which are non-analytic, 
but chiral symmetry guarantees that
the chance of finding $n$ pions in the air at the same time
decreases rapidly with $n$. 
The systematic theory built on this simple picture
has a finite number of interactions at any desired accuracy.
It in fact explains many regularities
that have been discovered in previous phenomenological treatments
but whose origins are otherwise mysterious. 
These include the decreasing importance of 
many-pion exchanges, many-nucleon forces, 
and higher isospin classes. 

This picture is not in contradiction with potential-model experience.
In fact, the method of EFT has already provided quantitative
input for more standard analyses that
could not have been obtained before by other means.
The chiral TPE two-nucleon force and simultaneous
pion-gamma exchange have been successfully incorporated in the
Nijmegen phase shift analysis;
chiral TPE three-nucleon forces have been used 
to correct existing models; and,
at the same time, the effects of the shorter-range 
three-nucleon forces
on discrepancies remaining in conventional approaches
are starting to be explored. 
From a conservative point-of-view,
the EFT method can be considered a refinement of 
other hadronic approaches in that it brings
to an otherwise free-fantasy context the constraints
of QCD symmetries.

There is no denying that much is still to be done in raising 
this pre-teen EFT formulation of nuclear physics
into a fully mature subject.
One major source of malnourishment
has been the so-far elusive determination of the scale
where pion effects cease to be perturbative, and 
the scope of their resummations.
Yet, this is a problem that can and will be resolved.
A coherent description of few- and many-nucleon systems
is still green but 
we can already glimpse the shape it is going to take.

\vspace{1cm}
\noindent
{\large\bf Acknowledgments}

\noindent
The ideas reviewed here evolved from conversations
with a large number of friends, collaborators, and colleagues.
I thank them all, but particularly Silas Beane,
Paulo Bedaque, and Jim Friar for their continuous 
input along the years.
Helpful discussions on specific issues reported here
are also acknowledged: 
Paulo Bedaque on $\aleph$ counting and the dimeron,
Paulo Bedaque and Rob Timmermans on
electromagnetic interactions,
and Tom Cohen on perturbative pions.
I am grateful to 
Silas Beane for helpful comments on the manuscript,
to Ulf Mei{\ss}ner, Dan Phillips, and Martin Savage
for allowing me to use their figures,
and to Jim Friar, Barry Holstein, Tom Mehen, Gautam Rupak,
Iain Stewart, and Rob Timmermans for making their unpublished
papers available to me. 
This research was supported in part by the US National Science Foundation,
grant PHY 94-20470.


\end{document}